\documentclass[aps,preprint]{revtex4}
\pdfoutput=1
\usepackage{amsmath}
\usepackage[english]{babel}
\usepackage{amssymb}
\usepackage{graphicx}
\usepackage{epsfig}
\usepackage{bm}
\usepackage{dcolumn}
\usepackage{xcolor}
\usepackage{mathtools}

\usepackage{bbm}
\usepackage{soul}
\usepackage{verbatim}
\usepackage{ulem}

\definecolor{forestgreen}{rgb}{0.13, 0.55, 0.13}
\definecolor{internationalorange}{rgb}{1.0, 0.31, 0.0}

\def\ffQxbarq{\mathbb{F}^{\scriptscriptstyle{\cal{Q}}}(x, \bar q)}

\def\ffQxbarqprima{\mathbb{F}^{\scriptscriptstyle{\cal{Q}}}(x, \bar q')}
\def\ffmQxbarq{\mathbb{F}^{\scriptscriptstyle{-\cal{Q}}}(x, \bar q)}
\def\ffQybarq{\mathbb{F}^{\scriptscriptstyle{\cal{Q}}}(y, \bar q)}

\def\elle{k}
\def\ellef{k}

\def\ffpxbarq{\mathbb{F}^{+}(x, \bar q)}
\def\ffpxprimabarqprima{\mathbb{F}^{+}(x', \bar q')}
\def\ffpxprimabarq{\mathbb{F}^{+}(x', \bar q)}
\def\ffpybarq{\mathbb{F}^{+}(y, \bar q)}
\def\ffpyprimabarq{\mathbb{F}^{+}(y', \bar q)}
\def\ffpyprimabarqprima{\mathbb{F}^{+}(y', \bar q')}

\def\piprop{\Delta_{\pi^{\scriptscriptstyle{\cal{Q}}}}}
\def\pipropt{\tilde{\Delta}_{\pi^{\scriptscriptstyle{\cal{Q}}}}}
\def\pipropb{{\bar \Delta}_{\pi^{\scriptscriptstyle{\cal{Q}}}}}
\def\pipropq{\hat \Delta_{\pi^{\scriptscriptstyle{\cal{Q}}}}(\elle,q_\parallel)}

\def\pipropqplus{\hat \Delta_{\pi^+}(\elle,q_\parallel)}
\def\pipropqprimaplus{\hat \Delta_{\pi^+}(\elle',q_\parallel')}
\def\pipropqplusLOC{\hat \Delta_{\pi^+}^{\rm (LOC)}(\elle,q_\parallel)}

\def\qprop{S_f}
\def\qpropb{\bar S_f}
\def\qpropq{\hat S_{f}(\elle,q_\parallel)}

\def\picx{\pi^{\scriptscriptstyle{\cal{Q}}}(x)}
\def\pitcx{\tilde{\pi}^{\scriptscriptstyle{\cal{Q}}}(x)}
\def\pimcx{\pi^{-\scriptscriptstyle{\cal{Q}}}(x)}
\def\ach{a_\pi^{\scriptscriptstyle{\cal{Q}}}}
\def\amch{a_\pi^{-\scriptscriptstyle{\cal{Q}}}}
\def\apmch{a_\pi^{\pm\scriptscriptstyle{\cal{Q}}}}

\def\psif{\psi_f}

\def\rhocmunu{\rho^{\scriptscriptstyle{\cal{Q}},\mu\nu}}
\def\rhocmunud{\rho^{\scriptscriptstyle{\cal{Q}}}_{\mu\nu}}
\def\rhocmu{\rho^{\scriptscriptstyle{\cal{Q}},\mu}}
\def\rhocmud{\rho^{\scriptscriptstyle{\cal{Q}}}_{\mu}}

\def\rhocnud{\rho^{\scriptscriptstyle{\cal{Q}}}_{\nu}}

\def\rhopropnula{D_{\rho^{\scriptscriptstyle{\cal{Q}}}}^{\nu\gamma}}

\def\rhopropnulab{\bar D_{\rho^{\scriptscriptstyle{\cal{Q}}}}^{\nu\gamma}}

\def\scrcalQ{\scriptscriptstyle{\cal{Q}}}

\def\rhopropbarmunu{\bar{D}_{\rho^{\scriptscriptstyle{\cal{Q}}}}^{\mu\nu}}
\def\rhopropmunu{D_{\rho^{\scriptscriptstyle{\cal{Q}}}}^{\mu\nu}}
\def\arhoch{a_\rho^{\scriptscriptstyle{\cal{Q}}}}
\def\arhopmch{a_\rho^{\pm\scriptscriptstyle{\cal{Q}}}}
\def\arhomch{a_\rho^{-\scriptscriptstyle{\cal{Q}}}}

\def\wqchmu{W^{\mu}_{\scriptscriptstyle{\cal{Q}}}}

\def\wqmchmu{W^{\mu}_{-\scriptscriptstyle{\cal{Q}}}}

\def\dcmu{{\cal D}^\mu}
\def\dcmud{{\cal D}_\mu}

\def\dcnud{{\cal D}_\nu}
\def\dcga{{\cal D}^\alpha}
\def\dcgad{{\cal D}_\alpha}

\def\QP{Q_{\mbox{{\scriptsize P}}}}

\def\SP{\Phi_{\mbox{{\scriptsize P}}}}
\def\SPtilde{{\tilde \Phi}_{\mbox{{\scriptsize P}}}}

\def\SPpiQ{\Phi_{\pi^{\scriptscriptstyle{\cal{Q}}}}}
\def\SPpiQtilde{{\tilde \Phi}_{\pi^{\scriptscriptstyle{\cal{Q}}}}}

\def\SPpiplus{\Phi_{\pi^+}}
\def\SPrhoplus{\Phi_{\rho^+}}

\def\SPrhoQ{\Phi_{\rho^{\scriptscriptstyle{\cal{Q}}}}}

\def\SPq{\Phi_f}

\def\SPpiplus{\Phi_{\pi^+}}

\def\calF{{\cal F}_{\scriptscriptstyle{Q}}}

\def\mpi{m_\pi}
\def\Bpi{B_\pi}
\def\Epi{E_\pi}

\def\mq{m_f}
\def\Bq{B_f}
\def\Eq{E_f}

\def\mrho{m_\rho}
\def\Brho{B_\rho}
\def\Erho{E_\rho}

\DeclareMathOperator*{\sumint}{%
\mathchoice%
  {\ooalign{$\displaystyle\sum$\cr\hidewidth$\displaystyle\int$\hidewidth\cr}}
  {\ooalign{\raisebox{.14\height}{\scalebox{.7}{$\textstyle\sum$}}\cr\hidewidth$\textstyle\int$\hidewidth\cr}}
  {\ooalign{\raisebox{.2\height}{\scalebox{.6}{$\scriptstyle\sum$}}\cr$\scriptstyle\int$\cr}}
  {\ooalign{\raisebox{.2\height}{\scalebox{.6}{$\scriptstyle\sum$}}\cr$\scriptstyle\int$\cr}}
}

\begin{document}

\title{\sc\Large{Charged meson masses
under strong magnetic fields: gauge invariance and Schwinger
phases}}

\author{D. G\'omez Dumm$^{a,b}$,
S. Noguera$^{c}$ and N.N.\ Scoccola$^{b,d}$}

\affiliation{$^{a}$ IFLP, CONICET $-$ Departamento de F\'{\i}sica, Facultad de Ciencias Exactas,
Universidad Nacional de La Plata, C.C. 67, (1900) La Plata, Argentina}
\affiliation{$^{b}$ CONICET, Rivadavia 1917, (1033) Buenos Aires, Argentina}
\affiliation{$^{c}$ Departamento de F\'{\i}sica Te\'{o}rica and IFIC, Centro Mixto
Universidad de Valencia-CSIC, E-46100 Burjassot (Valencia), Spain}
\affiliation{$^{d}$ Physics Department, Comisi\'{o}n Nacional de Energ\'{\i}a At\'{o}mica,
Av.\ Libertador 8250, (1429) Buenos Aires, Argentina}

\begin{abstract}
We study the role of the Schwinger phase (SP) that appears in the
propagator of a charged particle in the presence of a static and uniform
magnetic field $\vec{B}$. We first note that this phase cannot be removed by
a gauge transformation; far from this, we show that it plays an important
role in the restoration of the symmetries of the system. Next, we analyze
the effect of SPs in the one-loop corrections to charged pion and rho
meson selfenergies. To carry out this analysis we consider first a simple
form for the meson-quark interactions, and then we study the $\pi^+$ and
$\rho^+$ propagators within the Nambu-Jona-Lasinio model, performing a
numerical analysis of the $B$ dependence of meson lowest energy states. For
both $\pi^+$ and $\rho^+$ mesons, we compare the numerical results arising
from the full calculation
---in which SPs are included in the propagators, and meson wavefunctions
correspond to states of definite Landau quantum number--- and those obtained
within alternative schemes in which SPs are neglected (or somehow
eliminated) and meson states are described by plane waves of definite
four-momentum.
\end{abstract}


\maketitle

\renewcommand{\thefootnote}{\arabic{footnote}}
\setcounter{footnote}{0}

\global\long\def\theequation{\thesection.\arabic{equation}}%
\setcounter{equation}{0}

\section{Introduction}

The study of the behavior of charged particles in the presence of an intense
magnetic field within the framework of relativistic quantum field theory has
a long history (see e.g.\ Ref.~\cite{Dittrich:2000zu} and references
therein). In recent years, the interest in this topic has been renewed in
the context of the physics of strong
interactions~\cite{Kharzeev:2012ph,Andersen:2014xxa,Miransky:2015ava}. The
motivation arises mostly from the realization that intense magnetic fields
might play an important role in the study of the early
Universe~\cite{Grasso:2000wj}, in the analysis of high energy noncentral
heavy ion collisions~\cite{Skokov:2009qp,Voronyuk:2011jd} and in the
description of compact stellar objects like the
magnetars~\cite{Duncan:1992hi,Kouveliotou:1998ze}. It is well known that
magnetic fields also induce interesting phenomena like the chiral magnetic
effect~\cite{Kharzeev:2007jp,Fukushima:2008xe,Kharzeev:2015znc}, the
enhancement of the QCD quark-antiquark condensate (``magnetic
catalysis'')~\cite{Gusynin:1995nb} and the decrease of critical
temperatures for chiral restoration and deconfinement QCD
transitions~\cite{Bali:2011qj}.

In the above context, the study of the properties of magnetized light
hadrons shows up as a very relevant task. In fact, this subject has been
addressed by several works in the framework of various approaches to
non-perturbative QCD. These include e.g.\ Nambu-Jona-Lasinio (NJL)-like
models~\cite{Chernodub:2011mc, Fayazbakhsh:2013cha, Fayazbakhsh:2012vr,
Liu:2014uwa, Avancini:2015ady, Zhang:2016qrl, Liu:2016vuw, Mao:2017wmq,
GomezDumm:2017jij, Wang:2017vtn, Liu:2018zag, Coppola:2018vkw, Mao:2018dqe,
Avancini:2018svs, Avancini:2016fgq, Avancini:2018svs,
Coppola:2019uyr,Coppola:2019idh, Cao:2019res, Ghosh:2020qvg,
Avancini:2021pmi, Cao:2021rwx, Carlomagno:2022inu, Carlomagno:2022arc},
quark-meson models~\cite{Kamikado:2013pya,Ayala:2018zat}, chiral
perturbation theory
(ChPT)~\cite{Andersen:2012zc,Agasian:2001ym,Colucci:2013zoa}, hidden local
symmetry~\cite{Kawaguchi:2015gpt}, path integral
Hamiltonians~\cite{Orlovsky:2013gha,Andreichikov:2016ayj} and QCD sum
rules~\cite{Dominguez:2018njv}. In addition, results for the $\pi$ and
$\rho$ meson spectra in the presence of background magnetic fields have been
obtained from lattice QCD (LQCD)
calculations~\cite{Bali:2011qj,Luschevskaya:2015bea,Brandt:2015hnz,
Luschevskaya:2016epp,Bali:2017ian,Ding:2020hxw}.

In models with explicit quark degrees of freedom, like e.g.\ the NJL model
or the meson-quark model, the determination of meson properties demands the
evaluation of quark loops. In the presence of a magnetic field $\vec B$, the
calculation of these loops requires some care due to the appearance of
Schwinger phases~(SPs)~\cite{Schwinger:1951nm} associated with quark
propagators. These phases are not invariant under either translational or
gauge transformations. When all external legs in the quark loop correspond
to neutral particles, SPs cancel out and one can take the usual momentum
basis to diagonalize the corresponding loop correction; this is the case,
for example, of one-loop corrections to neutral meson selfenergies. In
contrast, when some of the external legs correspond to charged particles
---as in the case of the one-loop correction to a charged meson mass---
Schwinger phases do not cancel, leading to a breakdown of translational
invariance that prevents to proceed as in the neutral case. In this
situation, some existing calculations within the NJL
model~\cite{Chernodub:2011mc,Liu:2014uwa,Zhang:2016qrl,Liu:2016vuw,Liu:2018zag,Cao:2019res,Cao:2021rwx}
just neglect Schwinger phases; if this is done, one can set meson transverse
momenta to zero, considering only the translational invariant part of the
quark propagators to determine charged meson masses. In fact, it has even
been argued~\cite{Li:2020hlp} that this way to proceed would be consistent
with gauge invariance. On the other hand, a method that fully takes into
account the translational-breaking effects introduced by SPs has been
presented in Ref.~\cite{Coppola:2018vkw} for the calculation of charged pion
masses, and then it has been subsequently extended for the determination of
the charged pion masses at finite temperature~\cite{Mao:2018dqe}, for the
analysis other charged pion properties~\cite{Coppola:2019uyr}, for the
determination of charged kaon~\cite{Avancini:2021pmi} and rho meson
masses~\cite{Carlomagno:2022arc}, and for the study of diquark and nucleon
masses~\cite{Coppola:2020mon}. This method, based on the use of the
eigenfunctions associated to magnetized relativistic particles, allows one
to diagonalize the charged meson polarization functions in order to obtain
the corresponding meson masses.

The main objective of this article is to clarify the role played by
Schwinger phases in the calculation of quark loops associated to the
determination of charged meson properties in the presence of an external
magnetic field. One important point to be shown is that these phases cannot
be ``gauged away'': if a SP does not vanish in a given gauge, it cannot be
removed by any gauge transformation. In fact, the assumption of neglecting
SPs might be considered at best as some kind of approximation in which the
polarization functions are forced to be gauge invariant, instead of {\it
gauge covariant}, as they should be. To be fully consistent and
self-contained we devote the first few sections of this article to review
some properties of the SP as well as to provide the explicit form of quantum
fields and propagators for particles with spin 0, 1/2 and 1. Then, we
dedicate one section to the determination of one-loop corrections to the
charged pion and rho meson selfenergies in the context of quark meson model,
and another section is devoted to the calculation of charged pion and rho
meson masses in the framework of the NJL model. Throughout these
calculations we focus on the role of SPs and the preservation of gauge
properties of the involved quantities. In this way we uncover the issues
that appear when the SPs are neglected, providing further support to the
method introduced in Ref.~\cite{Coppola:2019uyr}. We also show that the
assumption of neglecting SPs may have a significant qualitative impact on
the theoretical predictions for the behavior of meson masses under a strong
magnetic field.

This work is organized as it follows. In Sec.~II we review the definition of
the SP and state its explicit form in commonly used gauges. Then, we show
how the SP plays an important role in the preservation of the expected
symmetries of the system ---although it is not itself translational and
gauge invariant--- and we discuss the related constraints on the form of the
invariant part of charged particle propagators. In Sec.~III we present the
explicit form of charged particle quantum fields in the presence of an
external magnetic field. The corresponding expressions are given in a quite
general form, in terms of eigenfunctions associated to the more commonly
used gauges. In Sec.~IV we provide the explicit form of the charged particle
propagators; this is done in terms of both the field eigenfunctions and the
product of a SP and a gauge invariant function obtained using the Schwinger
proper time method. Next, in Sec.~V we determine the leading order
correction to the charged pion and to the charged rho meson selfenergies for
some typical quark-meson interaction lagrangian. In particular, we show that
these corrections are diagonal in the basis of the corresponding meson
eigenfunctions. We also show that this implies taking into account some
transverse momentum fluctuations, which would have been neglected by
disregarding the SP (and considering plane wave meson wavefunctions). In
Sec.~VI we extend the analysis to the calculation of the charged pion and
rho meson masses in the framework of the NJL model. To give an idea of the
importance of properly taking into account the SP, we perform a numerical
analysis of the effect of the magnetic field on these masses, comparing the
results obtained from the expressions that include/neglect the SP. Finally,
in Sec.~VII we provide a summary of our work, together with our main
conclusions. We also include Appendixes A, B, C and D to provide some
formulae related with the formalism used throughout our work.

\section{Schwinger phase and charged particle propagators}
\label{section:SP}
\setcounter{equation}{0}

\subsection{Gauge transformations and gauge fixing for a constant magnetic field}

We start by considering the electromagnetic field strength $F^{\mu\nu}$
associated with a general electromagnetic field ${\cal A}^\mu(x)$,
\begin{eqnarray}
F^{\mu\nu} = \partial^\mu \mathcal{A}^{\nu}-\partial^\nu
\mathcal{A}^{\mu}\ .
\end{eqnarray}
Throughout this work we use the Minkowski metric $g^{\mu\nu} =
\mbox{diag}(1, -1, -1, -1)$, while for a space-time coordinate four-vector
$x^\mu$ we adopt the notation $x^\mu = (t, \vec x )$, with $\vec x = (x^1,
x^2, x^3)$.
We also consider the covariant derivative $\dcmu$ that appears in the field
equations associated with an electrically charged particle,
\begin{equation}
\dcmu\ = \ \partial^{\mu} + i\, Q \mathcal{A}^{\mu}\ ,
\label{covdev}
\end{equation}
where $Q$ is the particle electric charge. Now, under a gauge transformation
$\Lambda(x)$ the electromagnetic field transforms as
\begin{eqnarray}
{\cal A}^\mu \ \rightarrow \ {\tilde {\cal A}}^{\mu} = {\cal A}^{\mu}
+ \partial^\mu \Lambda\ .
\label{gautr}
\end{eqnarray}
While the electromagnetic field strength $F^{\mu\nu}$ is invariant under
this transformation, the operator $\dcmu$ transforms in a covariant way,
namely
\begin{equation}
\dcmu \ \rightarrow \ \tilde\dcmu = e^{-iQ \Lambda(x)}\,\dcmu\, e^{iQ
\Lambda(x)}\ .
\label{gaugecovdev}
\end{equation}

So far we have considered a general external electromagnetic field
$\mathcal{A}^{\mu}(x)$. In what follows we concentrate on the case
associated with a static and uniform magnetic field $\vec B$. The
tensor $F^{\mu\nu}$ is given in this case by
\begin{equation}
F^{ij} = F_{ij} = -\epsilon_{ijk} B^k\ , \qquad\qquad F^{0j} = 0\ ,
\end{equation}
with $i,j=1,2,3$, whereas the corresponding electromagnetic field can be
written as
\begin{eqnarray}
{\cal A}^\mu(x) = \frac{1}{2} x_\nu F^{\nu\mu} + \partial^\mu
\Psi(x)\ ,
\label{gengauge}
\end{eqnarray}
where $\Psi(x)$ is, in principle, an arbitrary function. For any form of
this function one obtains a particular gauge. Without loosing generality,
one can now choose the 3-axis to be parallel (or antiparallel) to the
magnetic field, writing $\vec B = (0,0,B)$. In addition, one can take
$\Psi(x)$ to be only a function of the spatial coordinates that are
perpendicular to $\vec B$, i.e., $x^1$ and $x^2$. The reason is that only
the components $F^{12}$ and $F^{21}$ of the field strength tensor are
different from zero, which implies that only $\partial^1 \mathcal{A}^{2}$
and $\partial^2 \mathcal{A}^{1}$ are relevant. In what follows we adopt this
coordinate choice.

For the considered situation, some commonly used gauges are
\begin{align}
\mbox{Symmetric gauge (SG),} & \quad \Psi(x) = 0\ , & &
{\cal A}^\mu(x) = \left(0,-\frac{B}{2}\, x^2, \frac{B}{2}\, x^1,0\right) \ ; \\
\mbox{Landau gauge 1 (LG1),} & \quad \Psi(x) = \frac{B}{2}\, x^1 x^2 \ , & &
{\cal A}^\mu(x) = \left(0, -B\,x^2,0,0\right)\ ; \\
\mbox{Landau gauge 2 (LG2),} & \quad \Psi(x) = -\frac{B}{2}\, x^1 x^2 \ , & &
{\cal A}^\mu(x) = \left(0, 0, B\,x^1,0\right)\ .
\end{align}
In what follows, we refer to them as ``standard gauges''.

According to the above introduced coordinate choice, given a four vector
$V^\mu$ we find it convenient to distinguish between ``parallel''
components, $V^0$ and $V^3$, and ``perpendicular'' components, $V^1$ and
$V^2$. Thus, we introduce the definitions
\begin{equation}
V_{\parallel}^{\mu}\equiv(V^{0}, 0, 0, V^{3})\ ,\qquad\qquad
V_{\perp}^{\mu}\equiv(0, V^{1},V^{2},0)\ .
\end{equation}
In addition, we define the metric tensors
\begin{equation}
g_{\parallel}^{\mu\nu}=\text{diag} (1,0,0,-1)\ ,\qquad\qquad
g_{\perp}^{\mu\nu}=\text{diag} (0,-1,-1,0)\ .
\end{equation}
The scalar products of parallel and perpendicular vectors are thus given by
\begin{eqnarray}
V_{\parallel}^\mu\, W_{\parallel\,\mu} & = & V_{\parallel}\cdot W_\parallel
= V^{0}W^{0}-V^{3}W^{3} \ , \nonumber \\
V^\mu_{\perp}\, W_{\perp\,\mu} & = & - \vec{V}_{\perp}\cdot\vec{W}_{\perp} =
- (V^{1}W^{1}+V^{2}W^{2})\ ,
\nonumber \\
V_{\parallel}^\mu\, W_{\perp\,\mu} & = & 0\ .
\end{eqnarray}

\subsection{Schwinger phase}

We introduce here the so-called Schwinger phase, which will be a relevant
quantity throughout this work. Given a particle $\mbox{P}$ with electric
charge $\QP$, we denote the associated SP by $\SP(x,y)$; its explicit form
is~\cite{Schwinger:1951nm}
\begin{equation}
\SP(x,y) = \QP \int_{x}^y d\xi_\mu \left[{\cal A}^\mu(\xi) +
\frac{1}{2}\, F^{\mu\nu}\, (\xi_\nu-y_\nu)\right]\ ,
\label{sp}
\end{equation}
where $F^{\mu\nu}$ is assumed to be constant, and the integration is
performed along an arbitrary path that connects $x$ with $y$. In general,
the SP is found to be not invariant under either translations or gauge
transformations. On the other hand, the integral in Eq.~(\ref{sp}) is shown
to be path independent; thus, it can be evaluated using a straight line path.
In this way, using Eq.~(\ref{gengauge}) one can obtain a closed expression
for the SP associated to a static and uniform magnetic field in an
arbitrary gauge. It reads
\begin{eqnarray}
\SP(x,y) = \frac{\QP}{2}\, x_\mu \, F^{\mu\nu} \, y_\nu - \QP
\big[\Psi(x)-\Psi(y)\big]\ .
\label{SPgeneral}
\end{eqnarray}
{}From Eqs.~(\ref{gautr}) and (\ref{gengauge}), it is seen that under a gauge
transformation the SP transforms as
\begin{equation}
\SP(x,y) \ \rightarrow \ \SPtilde(x,y) = \SP(x,y) - \QP \left[ \Lambda(x) - \Lambda(y) \right]\ .
\label{spgau}
\end{equation}

The expressions for the SP in the particular gauges introduced above can be
now readily obtained from Eq.~(\ref{SPgeneral}). We have
\begin{align}
& \mbox{SG:} & & \hspace{-1.5cm} \SP(x,y) = -\frac{\QP B}{2} (x^1 y^2 - y^1 x^2) \ ; \hspace{1.5cm} \\
& \mbox{LG1:} & & \hspace{-1.5cm} \SP(x,y) = -\frac{\QP B}{2} (x^2+y^2)(x^1-y^1) \ ; \hspace{1.5cm} \\
& \mbox{LG2:} & & \hspace{-1.5cm} \SP(x,y) = \frac{\QP B}{2} (x^1+y^1)(x^2-y^2) \ . \hspace{1.5cm}
\end{align}
It is worth noticing that in all cases the SP includes products that mix the
coordinates of the points $x^\mu$ and $y^\mu$. Clearly, there is no way in
which these combinations could be expressed in terms of the difference
between a scalar function evaluated at $x^\mu$ and the same function
evaluated at $y^\mu$. Therefore, it follows from Eq.~(\ref{spgau}) that if
the SP does not vanish in a given gauge it will be nonvanishing {\it in any
gauge}. This means that the SP cannot be ``gauged away''.

\subsection{Charged particle propagators in a static and uniform magnetic field}
\label{section:chargedprop}

Let us study now the propagators of charged particles. We start by
considering the propagator of a spin zero meson, e.g.\ a charged pion,
which we denote as $\piprop(x,y)$ with ${\cal Q} = \pm 1$. The particle
charge is then given by $Q_\pi = {\cal Q}\, e$, where $e$ denotes the proton
charge.

The equation that defines the meson propagator is
\begin{eqnarray}
\left(\dcmu\, \dcmud + \mpi^2 \right) \piprop(x,y) = - \,\delta^{(4)}(x-y)\ .
\label{mesoprop1}
\end{eqnarray}
If we perform now a gauge transformation, using Eq.~(\ref{gaugecovdev}) we
get
\begin{equation}
\left(\tilde \dcmu\, \tilde \dcmud + \mpi^2 \right) e^{-iQ_\pi \Lambda(x)}\, \piprop(x,y)
= -\, e^{-iQ_\pi \Lambda(x)}\, \delta^{(4)}(x-y)\ ;
\end{equation}
hence, the propagator has to transform according to
\begin{equation}
\piprop(x,y) \ \rightarrow \ \pipropt(x,y) = e^{-iQ_\pi \Lambda(x)}\ \piprop(x,y)\ e^{iQ_\pi
\Lambda(y)}\ ,
\label{gauprop1}
\end{equation}
which is the natural extension of a gauge covariant transformation for
the case of a bilocal object. It is seen that the phase difference
appearing in the transformed propagator is just the same quantity that
appears in Eq.~(\ref {spgau}) for the gauge-transformed SP. Therefore, one
can always write the meson propagator as
\begin{equation}
\piprop(x,y) =  e^{i\SPpiQ(x,y)}\ \pipropb(x, y)\ ,
\label{sfx1}
\end{equation}
where $\pipropb(x, y)$ is a gauge invariant function; the gauge dependence
of the propagator is carried by the SP, which has a well defined expression.

Since we are dealing with a system subject to a static and uniform magnetic
field, the invariance under translations in time and space, under rotations
around any axis parallel to the magnetic field, and under boosts in
directions parallel to the magnetic field, is expected to be preserved.
Translations in time, as well as translations and boosts in the direction of
$\vec B$, can be treated in the same way as in the case of a free particle,
since they do not involve the axes 1 or 2. Thus, let us focus on the
translations in the plane perpendicular to $\vec B$ and in the rotations
around the $\vec B$ direction. Noticeably, the expected invariance seems to
be at odds with the fact that the charged pion propagator is known to be not
invariant under these transformations. The aim of the following discussion
is to clarify this point and see how the invariance implies further
constraints on the form of the propagator.

Let us first consider space translations in the perpendicular plane, i.e.\ a
general transformation of the form $x^{\mu} \rightarrow
x^{\prime\mu}=x^{\mu} +b^{\mu}_{\perp}$. From Eq.~(\ref{gengauge}), under
this transformation one has
\begin{equation}
\mathcal{A}^{\mu}(x) \ \rightarrow \ \mathcal{A}_{t}^{\mu}(x) \equiv
\mathcal{A}^{\mu}(x^{\prime}) =
\mathcal{A}^{\mu}(x)-\frac{1}{2}\,F^{\mu\nu}\,b_{\perp\nu}+
\partial^{\mu}\Psi\left(x^\prime\right)-\partial^{\mu}\Psi\left(x\right)\ .
\end{equation}
It is rather easy to see that this is fully equivalent to a gauge
transformation
\begin{equation}
\mathcal{A}_{\mu}(x) \rightarrow \tilde{\mathcal{A}}^{\mu}(x)=\mathcal{A}^{\mu}(x)+\partial^{\mu}\Lambda_{t}(x;
b_{\perp})\ ,
\label{gaugetran}
\end{equation}
with
\begin{equation}
\Lambda_{t}(x;b_{\perp})=\Psi(x^\prime)-\Psi(x)-\frac{1}{2}\
x_{\mu}\,F^{\mu\nu}\,b_{\perp\,\nu}\ .
\label{gengaugeT}
\end{equation}
{}From Eq.~(\ref{gengaugeT}) we can readily get the expressions of
$\Lambda_{t}(x;b_{\perp})$ in the particular gauges introduced in the
previous subsections. We have
\begin{align}
& \mbox{SG:} & & \hspace{-1.5cm} \Lambda_{t}(x;b_{\perp})= - \frac{B}{2} \left(x^{1}b^{2}-x^{2}b^{1}\right)\ ; \hspace{1.5cm} \\
& \mbox{LG1:} & & \hspace{-1.5cm} \Lambda_{t}(x;b_{\perp}) = \frac{B}{2}\, b^{2}\left(2 x^{1}+b^{1}\right)\ ; \hspace{1.5cm} \\
& \mbox{LG2:} & & \hspace{-1.5cm} \Lambda_{t}(x;b_{\perp}) = -\frac{B}{2}\, b^{1}\left(2 x^{2}+b^{2}\right)\ . \hspace{1.5cm}
\end{align}
A similar relation between the translation $x^{\mu} \rightarrow
x^{\prime\mu}=x^{\mu} +b^{\mu}_{\perp}$ and the gauge transformation in
Eqs.~(\ref{gaugetran}) and (\ref{gengaugeT}) can be obtained for the
Schwinger phase. Under the translation, the SP transforms as
\begin{equation}
\SPpiQ(x,y) \ \rightarrow \
\Phi_{\pi^{\scriptscriptstyle{\cal{Q}}},t}(x,y)\equiv \SPpiQ(x',y') = \frac{Q_\pi}{2}\ x'_{\mu}\, F^{\mu\nu}\, y'_{\nu} -
Q_\pi\left[\Psi(x')-\Psi(y')\right] \ ,
\end{equation}
whereas performing the corresponding gauge transformation one gets
\begin{equation}
\SPpiQ(x,y) \ \rightarrow \ \SPpiQtilde(x,y) = \SPpiQ(x,y)-Q_\pi\left[\Lambda_{t}(x; b_{\perp})-\Lambda_{t}(y;
b_{\perp})\right]\ .
\end{equation}
Taking into account the form of $\Lambda_{t}(x; b_{\perp})$ in
Eq.~(\ref{gengaugeT}) we observe that
$\Phi_{\pi^{\scriptscriptstyle{\cal{Q}}},t}=\SPpiQtilde(x,y)$. Now we can
turn back to Eq.~(\ref{mesoprop1}), writing the propagator as in
Eq.~(\ref{sfx1}). From the above equations it is seen that under the
considered translation the operator $\left(\dcmu\,\dcmud+\mpi^{2}\right)$
and the factor $\exp[ i\SPpiQ(x,y)]$ transform in the same way as under the
gauge transformation $\Lambda_{t}(x,b_{\perp})$. Together with the
requirement that Eq.~(\ref{mesoprop1}) be translational invariant, this
implies that the gauge invariant factor $\pipropb(x,y)$ has to be also
translational invariant. Thus, we can write
\begin{equation}
\pipropb(x, y)= \pipropb(x-y)\ ,
\label{piprogen}
\end{equation}
and it is possible to obtain a Fourier transform
$\pipropb(v_\parallel,v_\perp)$ that satisfies
\begin{eqnarray}
\pipropb(x-y) = \,\int \frac{d^4 v}{(2 \pi)^4} \ e^{-i\,
v\, (x-y)}\, \pipropb(v_\parallel,v_\perp) \ .
\label{pipropTinv}
\end{eqnarray}
Notice that in this expression we have made it explicit that one gets in
general different dependences on parallel and perpendicular momenta.

We consider now a rotation of arbitrary angle $\alpha$ around an axis
parallel to the magnetic field $\vec B$. The effect of such a rotation on an
arbitrary vector $v^{\mu}$ acts only on the perpendicular component
$v^{\mu}_{\perp}$.
Choosing the 3-axis in the direction of $\vec B$, for a rotation matrix
$R_{\hat{3}}(\alpha)$ we have $x^{\prime\mu}={\cal R}_{\,\,\nu}^{\mu}\
x^{\nu}$, with
\begin{equation}
{\cal R}_{\,\,\nu}^{\mu}\equiv R_{\hat{3}}(\alpha)_{\,\,\nu}^{\mu}=\left(\begin{array}{cccc}
1 & 0 & 0 & \ 0 \ \ \\
\ 0 \ \ & \cos\alpha & -\sin\alpha & \ 0 \ \ \\
0 & \sin\alpha & \cos\alpha & \ 0 \ \ \\
0 & 0 & 0 & \ 1 \ \
\end{array}\right)\ .
\end{equation}
At this point it is important to specify if we adopt an active or a passive
point of view for the rotation; here we adopt the passive point of view and
define $\bar x^{\prime \mu}={\bar {\cal R}}_{\,\,\nu}^{\mu}\, x^{\nu}$, with
${\bar{\cal R}} \equiv R_{\hat{3}}(-\alpha)$. From Eq.~(\ref{gengauge}),
under the rotation $R_{\hat 3}(\alpha)$ the electromagnetic field transforms
as
\begin{equation}
\mathcal{A}^{\mu}(x) \ \rightarrow \ \mathcal{A}_{r}^{\mu}(x) =
{\cal R}_{\,\,\nu}^{\mu}\
\mathcal{A}^\nu\left(\bar x^{\prime}\right) =
-\frac{1}{2}\ {\cal R}_{\,\,\tau}^{\mu}\,F^{\tau\delta}\,{\bar {\cal R}}_{\delta}^{\,\,\,\nu}\,x_{\nu}
+{\cal R}_{\,\,\tau}^{\mu}\,
\frac{\partial \Psi(z)}{\partial z_\tau}\Big|_{z\,=\,\bar x^{\prime}}\ .
\label{amurot}
\end{equation}
Noting that ${\cal R}_{\,\,\tau}^{\mu}\,F^{\tau\delta}\,{\bar {\cal
R}}_{\delta}^{\,\,\,\nu}=F^{\mu\nu}$, the result in Eq.~(\ref{amurot}) can
be reinterpreted as a gauge transformation
\begin{equation}
\mathcal{A}^{\mu}(x) \ \rightarrow \
\tilde{\mathcal{A}}^{\mu}(x) = \mathcal{A}^{\mu}(x)+\partial^{\mu}\Lambda_{r}(x;\alpha)\ ,
\end{equation}
where
\begin{eqnarray}
\Lambda_{r}(x,\alpha)=\Psi(\bar x^{\prime})-\Psi(x)\ .
\label{gengaugeR}
\end{eqnarray}
As in the case of the translations, the equivalence between the rotation
$R_{\hat{3}}(\alpha)$ and the gauge transformation $\Lambda_{r}(x;\alpha)$
is also obtained for the SP, i.e., one gets
$\Phi_{\pi^{\scriptscriptstyle{\cal{Q}}},r}(x,y)=\SPpiQtilde(x,y)$. The
explicit expressions for the function $ \Lambda_{r}(x;\alpha)$ in the
standard gauges read
\begin{align}
& \mbox{SG:} & & \Lambda_{r}(x;\alpha)= 0\ ; \\
& \mbox{LG1:} & & \Lambda_{r}(x;\alpha) = - \frac{B}{2}\,
\sin\alpha\,\Big[2 x^{1}x^{2}\sin\alpha + \big( (x^{1})^{2}-(x^{2})^{2}\big)\cos\alpha\Big]\ ; \\
& \mbox{LG2:} & & \Lambda_{r}(x;\alpha) =\frac{B}{2}\, \sin\alpha\,\Big[2 x^{1}x^{2}\sin\alpha +
\big( (x^{1})^{2}-(x^{2})^{2}\big)\cos\alpha\Big]\ .
\end{align}

Turning back once again to Eq.~(\ref{mesoprop1}), and writing the propagator
as in Eq.~(\ref{sfx1}), we observe that under a rotation
$R_{\hat{3}}(\alpha)$ the operator $\left(\dcmu\,\dcmud+\mpi^{2}\right)$ and
the factor $e^{i\SPpiQ(x,y)}$ transform in the same way as under the gauge
transformation  $\Lambda_{r}(x;\alpha)$.  This equivalence, together with
the requirement that  Eq.~(\ref{mesoprop1}) be invariant under
$R_{\hat{3}}(\alpha)$ rotations, implies that $\pipropb(x- y)$ has to be
invariant under these transformations. Since, in addition, the propagator
has to be invariant under boosts along the 3-axis, one can conclude that the
Fourier transform $\pipropb(v_\parallel,v_\perp)$ defined in
Eq.~(\ref{pipropTinv}) can depend only on the quantities $v_\parallel^2$ and
$v_\perp^2$.

An entirely similar analysis can be performed for the case of spin 1/2 and
spin 1 particles. Thus, the spin 1/2 fermion propagator $\qprop(x,y)$ can be
written as
\begin{equation}
\qprop(x,y) =  e^{i\SPq(x,y)}\, \qpropb(x-y) \ ,
\label{quarkpropTinv}
\end{equation}
with
\begin{equation}
\qpropb(x-y) = \,\int \frac{d^4 v}{(2 \pi)^4} \ e^{-i\, v (x-y)}\,
\qpropb(v_\parallel,v_\perp) \ .
\label{FermionPropTinv}
\end{equation}
Here the propagator is a matrix in Dirac space that involves products of the
$\gamma^\mu$ Dirac matrices. Due to the invariance under rotations around
the 3-axis (i.e., the $\vec B$ axis) and under boosts in that direction, it
is easy to see that $\qpropb(v_\parallel,v_\perp)$ has to be a function of
$v_\parallel^2$, $v_\perp^2$, $\gamma_\parallel \cdot v_\parallel$ and
$\vec{\gamma}_\perp \cdot \vec{v}_\perp$.

In the case of a charged vector meson propagator, for instance, a $\rho$
meson propagator $\rhopropnula(x,y)$, we can write
\begin{equation}
\rhopropnula(x,y) =  e^{i \SPrhoQ(x,y)} \rhopropnulab(x-y)\ ,
\label{rhopropTinv}
\end{equation}
with
\begin{eqnarray}
\rhopropnulab(x-y) = \,\int \frac{d^4 v}{(2 \pi)^4} \ e^{-i\,
v(x-y)}\, \rhopropnulab(v_\parallel,v_\perp) \ .
\label{rhopropTinvtra}
\end{eqnarray}
Similarly to the previous cases, invariance under rotations around the
3-axis and under boosts in that direction implies that
$\rhopropnulab(v_\parallel,v_\perp)$ will be given by a linear
combination of tensors of order two built from the tensors
$g_{\parallel}^{\mu\nu}$, $g_{\perp}^{\mu\nu}$, $F^{\mu\nu}$ and the vectors
$v_{\parallel}^{\mu}$, $v_{\perp}^{\mu}$, with coefficients given by
functions that depend only on $v_{\parallel}^{2}$ and $v_{\perp}^{2}$.

Obviously, the above statements can be corroborated by carrying out detailed
calculations of the propagators. This is sketched in Sec.~\ref{PiPropa},
where the explicit forms of $ \pipropb(v_\parallel,v_\perp)$,
$\qpropb(v_\parallel,v_\perp)$ and $\rhopropnulab(v_\parallel,v_\perp)$ are
given.

In conclusion, we have seen that the Schwinger phase carries all gauge,
translation and rotation non-invariance that is present in particle
propagators. In fact, this should not be surprising, since the breakdown of
translational and rotational symmetry is precisely produced by the gauge
choice. When calculating a physical quantity we specify a particular gauge
and use propagators that, in general, break both translational and
rotational invariance; however, all these symmetries are simultaneously
recovered in the final result.

\section{Quantum fields of charged particles in a magnetic field}
\setcounter{equation}{0}
\label{section:fields}

\subsection{A set of basic functions}
\label{Ffunctions}

Let us consider a charged scalar particle in a static and homogeneous
magnetic field. We introduce the scalar functions ${\calF}(x,\bar q)$,
solutions of the eigenvalue equation
\begin{equation}
\dcmu\, \dcmud \, {\calF}(x,\bar q) \, = \, f_{\bar q} \, {\calF}(x,\bar q)\ ,
\label{ecautovbox}
\end{equation}
where $Q$ is the particle electric charge, $\dcmu$ is the covariant
derivative defined in Eq.~(\ref{covdev}), and the corresponding
electromagnetic field ${\cal A}^\mu(x)$ is of the form given by
Eq.~(\ref{gengauge}). In Eq.~(\ref{ecautovbox}), $\bar q$ stands for a set
of four labels that are needed to completely specify each eigenfunction. One
can be more explicit and write the eigenvalue equation in the form
\begin{eqnarray}
\left[ \partial^\mu\, \partial_\mu \, - \, Q \, \vec B \cdot \vec L\; +\,
\frac{Q^2}{4} \Big( \vec x \times \vec B\Big)^2 \right] e^{iQ \Psi(x)}\,
{\calF}(x,\bar q) \ = \ f_{\bar q} \, e^{iQ \Psi(x)} \, {\calF}(x,\bar q) \ ,
\end{eqnarray}
where $L^k = i\epsilon_{klm}\, x_l\,\partial_m$. From this equation it is
seen that while the eigenfunctions ${\cal F}_Q(x,\bar q)$ are gauge
dependent, the eigenvalues $f_{\bar q}$ are not. As discussed in the
previous section, the magnetic field can always be taken to lie along the
3-axis, and then $\Psi(x)$ can be assumed to depend only on the two spatial
coordinates perpendicular to $\vec B$, $x^1$ and $x^2$. Consequently, as in
the case of a free particle, the eigenvalues of the components of the
four-momentum along the time direction, $q^0$, and the magnetic field
direction, $q^3$, can be taken as two of the labels required to specify
${\cal F}_Q(x,\bar q)$. On the other hand, as is well known, the eigenvalues
of Eq.~(\ref{ecautovbox}) are given by
\begin{eqnarray}
f_{\bar q} = -  \left[ (q^0)^2 - (2 \ellef +1) B_Q - (q^3)^2 \right]\ ,
\label{fdef}
\end{eqnarray}
where $B_Q \equiv |Q B|$, and $\ellef$ is a non-negative integer, to be
related with the so-called Landau level. This means that the eigenvalues
depend only on three of the labels included in $\bar q$. There is a
degeneracy, which arises, of course, as a consequence of gauge invariance;
to fully specify the eigenfunctions, a fourth quantum number $\chi$ is
required, i.e.\ one has $\bar q = (q^0,\ellef,\chi,q^3)$.

Although it is not strictly necessary, the quantum number $\chi$ can be
conveniently chosen according to the gauge in which the eigenvalue problem
is analyzed~\cite{Wakamatsu:2022pqo}. In particular, since for the standard
gauges SG, LG1 and LG2 one has unbroken continuous symmetries, in those
cases it is natural to consider quantum numbers $\chi$ associated with the
corresponding group generators. Usual choices are
\begin{align}
& \mbox{SG:} & & \chi = \imath\ , & & \mbox{nonnegative integer, associated to $L^3$} \\
& & & & & \mbox{(eigenvalue of $L^3$: $m = {\rm sign}(QB)(\imath -k)$)}\ ; \nonumber \\
& \mbox{LG1:} & & \chi = q^1\ , & & \mbox{real number, eigenvalue of}\ -i\frac{\partial\ }{\partial x^1}\ ; \\
& \mbox{LG2:} & & \chi = q^2\ , & & \mbox{real number, eigenvalue of}\ -i\frac{\partial\ }{\partial x^2}\ .
\end{align}

The explicit forms of the functions ${\calF}(x,\bar q)$ for the above
standard gauges and quantum numbers $\chi$ are given in
App.~\ref{functionF}. They are shown to satisfy the completeness and
orthogonality relations
\begin{eqnarray}
\sumint_{\bar q} \ {\calF}(x,\bar q)^\ast \, {\calF}(y,\bar q) & = &
\delta^{(4)}{(x- y)} \ ,
\label{compF} \\
\int d^4 x \ {\calF}(x,\bar q')^\ast \, {\calF}(x,\bar q) & = & \hat \delta_{\bar q \bar q'} \ .
\label{orthF}
\end{eqnarray}
Here we have introduced some shorthand notation whose explicit
form depends on the chosen gauge. For SG we have
\begin{equation}
\sumint_{\bar q} \equiv \dfrac{1}{(2\pi)^2}\sum_{\ellef,\imath=0}^\infty
\int \frac{dq^0}{2\pi}\frac{dq^3}{2\pi}\ , \quad \hat
\delta_{\bar q \bar q'} \equiv (2\pi)^4\, \delta_{\ellef\ellef'}\,
\delta_{\imath\imath'}\, \delta(q^0-q^{\prime 0})\,
\delta(q^3-q^{\prime 3})\ ,
\end{equation}
while for LG1 (LG2) we have
\begin{equation}
\sumint_{\bar q}\equiv \dfrac{1}{2\pi}\sum_{\ellef=0}^\infty \int
\frac{dq^0}{2\pi}\frac{dq^i}{2\pi}\frac{dq^3}{2\pi} \ ,
\quad \hat \delta_{\bar q \bar q'} \equiv (2\pi)^4\, \delta_{\ellef\ellef'}\,
\delta(q^0-q^{\prime 0})\, \delta(q^i-q^{\prime i})\,
\delta(q^3-q^{\prime 3})\ ,
\end{equation}
where $i=1$ ($i=2$). For later use we also find it convenient to define $\breve q
= (\ellef,\chi,q^3)$, $s=\mbox{sign}(Q B)$ and
\begin{equation}
\sumint_{\{\bar q_E\}} = \sumint_{\bar q} 2\pi \delta(q^0- E)\ .
\end{equation}

It can be shown that the functions ${\calF}(x,\bar q)$ satisfy the useful
relations~\cite{Coppola:2018ygv}
\begin{eqnarray}
{\cal D}^0\, {\calF}(x,\bar q) &=& -i q^0\, {\calF}(x,\bar q)\ ,  \nonumber \\
\left( {\cal D}^1 \pm i {\cal D}^2 \right) {\calF}(x,\bar q) &=& (\mp s)\,
\big[( 2 k+1 \mp s)B_Q\big]^{1/2}\, {\calF}(x,\bar q_{k\mp s})\ , \nonumber \\
{\cal D}^3\, {\calF}(x,\bar q) &=& -i q^3\, {\calF}(x,\bar q)\ ,
\label{fprop}
\end{eqnarray}
where $\bar q_{k\pm s} = (q^0,k\pm s,\chi,q^3)$. We also notice that
under a gauge transformation $\Lambda(x)$ the functions ${\calF}(x,\bar q)$
[with $\bar q = (q^0,\ellef,\chi,q^3)$] transform as
\begin{eqnarray}
{\calF}(x,\bar q) \ \rightarrow \ \tilde {\calF}(x,\bar q) = e^{-iQ \Lambda(x)}
\,{\calF}(x,\bar q)\ .
\label{gaugeF}
\end{eqnarray}

\subsection{Spin 0 charged particles: the charged pions}
\label{Spin-0-charged}


Let us start by considering the gauged Klein-Gordon action for a point-like
charged pion in the presence of a static and homogeneous magnetic field. We
have
\begin{equation}
{\cal S}_{KG} = -\,\int d^{4}x\
\picx^\ast\left(\dcmu\,\dcmud+\mpi^{2}\right)\picx\ ,
\label{KGaction}
\end{equation}
where, as in Sect.~\ref{section:chargedprop}, we have denoted the pion
charge by $Q_{\pi}={\cal Q}\, e$, with ${\cal Q}=\pm 1$. From the action in
Eq.~(\ref{KGaction}) one gets the associated gauged Klein-Gordon equation,
namely
\begin{eqnarray}
\left(\dcmu\, \dcmud + \mpi^2 \right) \picx = 0\ .
\label{KGeq0}
\end{eqnarray}
We notice that, taking into account Eq.~(\ref{gaugecovdev}), the gauge
invariance of the gauged Klein-Gordon action requires that under a gauge
transformation $\Lambda(x)$ the $\picx$ field transform as
\begin{equation}
\picx \ \rightarrow \ \pitcx = e^{-i Q_{\pi} \Lambda(x)}\, \picx \ .
\label{gaugepi}
\end{equation}

Using the notation introduced in the previous subsection, the quantized
charged pion field can be written as
\begin{equation}
\picx \ = \ \pimcx^\dagger \ = \
\,\sumint_{\{\bar q_{\Epi}\}} \frac{1}{2 \Epi}
\Big\{ \ach(\breve q)  \; \ffQxbarq + \amch(\breve q)^\dagger  \; \ffmQxbarq^\ast \Big\}\ .
\label{chargepionexp}
\end{equation}
Here the pion energy is given by $\Epi = \sqrt{\mpi^2+ (2\elle+1) \Bpi +
(q^3)^2}$, with $\elle \geq 0$, while the functions $\ffQxbarq$ are given by
\begin{equation}
\ffQxbarq = {\cal F}_{Q_{\pi}}(x,\bar q)\ .
\label{pionfunc}
\end{equation}
According to Eqs.~(\ref{compF}) and (\ref{orthF}), they satisfy the
relations
\begin{eqnarray}
\sumint_{\bar q} \, \ffQxbarq^\ast \, \ffQybarq & = & \delta^{(4)}{(x-y)} \ ,
\label{completfpion} \\
\int d^4 x \ \ffQxbarq^\ast \, \ffQxbarqprima & = &
\hat \delta_{\bar q \bar q'} \ .
\label{orthfpion}
\end{eqnarray}
On the other hand, the creation and annihilation operators in
Eq.~(\ref{chargepionexp}) satisfy the commutation relations
\begin{align}
\left[\, \ach(\breve q)\, , \, \apmch (\breve q\,')\,\right] &
\ = \ \left[\, \ach(\breve q)^\dagger\, , \, \apmch(\breve q\,')^\dagger\,\right]
\ = \ \left[\, \ach(\breve q)\, , \, \amch (\breve q\,')^\dagger\,\right]  \ = \ 0\ ,
\nonumber \\
\left[\, \ach(\breve q)\, , \, \ach(\breve q\,')^\dagger\,\right] & \ = \
2 \Epi  \,(2\pi)^3  \, \delta_{\elle\elle'}\, \delta_{\chi\chi'}\,
\delta(q^3-q^{\,\prime\, 3})\ .
\label{conbos}
\end{align}
Note that according to the above definitions the operators $\ach(\breve q)$
and $\amch(\breve q)$ turn out to have different dimensions from the
creation and annihilation operators that are usually defined in absence of
the external magnetic field.

\subsection{Spin 1/2 charged particles: the quarks}
\label{Spin-1/2-charged}

Let us consider the gauged Dirac action for a point-like quark of flavor $f$
in the presence of a static and homogeneous magnetic field. We express the
quark charge as $Q_{f}= {\cal Q}_f\,e$, with ${\cal Q}_u = 2/3$, ${\cal
Q}_d=- 1/3$ for $f=u,d$. The gauged action is given by
\begin{equation}
{\cal S}_{D} = \int d^{4}x\
\bar{\psif}(x)\left(i\,\rlap/\!{\cal D}-m_{f}\right)\psif(x)\ ,
\end{equation}
where, as usual, $\bar \psif = \psif^\dagger\, \gamma^0$ and $\rlap/\!{\cal
D} = \gamma_\mu \dcmu$; the associated gauged Dirac equation reads
\begin{eqnarray}
\left(i \, \rlap/\!{\cal D} - m_f \right) \psif(x) = 0\ .
\end{eqnarray}
In a similar way as in the case of the charged pion, gauge invariance of the
gauged Dirac action requires that under a gauge transformation $\Lambda(x)$
the field $\psi_f(x)$ transform as
\begin{equation}
\psif(x) \ \rightarrow \ \tilde{\psi}_f(x) = e^{-i Q_{f} \Lambda(x)}\, \psif(x) \ .
\label{gaugepsi}
\end{equation}

The quantized quark fields are given by
\begin{equation}
\psif (x) \ = \
\sumint_{\{\bar q_{E_f}\}} \ \sum_{a=1,2} \, \frac{1}{2 E_{f}}\,
\Big\{\,b_f(\breve {q},a) \, U_{f}(x,\bar q,a)
+ d_f(\breve q,a)^\dagger\,  V_{f}(x,\bar q,a)
\Big\}\ ,
\label{fermionfieldBpart}
\end{equation}
where the quark energy is given by $\Eq = \sqrt{\mq^2 + 2\elle \Bq +
(q^3)^2}$, with $\elle \geq 0$; for $k=0$, only the value $a=1$ in the sum
over $a$ is allowed. Once again we use here the definitions $\bar q =
(q^0,\elle,\chi,q^3)$, $\breve q =( \elle, \chi, q^3)$, $\Bq= |Q_f B|$ and
$s = \mbox{sign}(Q_f B)$. The spinors $U_{f}$ and $V_{f}$ in
Eq.~(\ref{fermionfieldBpart}) can be written as
\begin{align}
U_{f}\left(x,\bar q,a\right)  \ &= \
\mathbb{E}^{\scriptscriptstyle{{\cal Q}_f}}(x,\bar q) \ u_{\scriptscriptstyle{{\cal Q}_f}}(\elle,q^3,a)\ ,
\nonumber\\
V_{f}\left(x,\bar q,a\right) \ &= \ \mathbb{\tilde{E}}^{\scriptscriptstyle{-{\cal Q}_f}}(x,\bar q) \
v_{-\scriptscriptstyle{{\cal Q}_f}}(\elle,q^3,a)\ ,
\label{UVlept}
\end{align}
where $\mathbb{E}^{\scriptscriptstyle{{\cal Q}_f}}(x,\bar q)$ and
$\mathbb{\tilde{E}}^{\scriptscriptstyle{-{\cal Q}_f}}(x,\bar q)$ are Ritus
functions~\cite{Ritus:1978cj}. Their explicit forms are
\begin{eqnarray}
\mathbb{E}^{\scriptscriptstyle{{\cal Q}}}(x,\bar q) & = &
\sum_{\lambda=\pm} \ \Gamma^\lambda\, {\cal F}_{Q}(x,\bar q_\lambda) \ ,
\nonumber \\
\mathbb{\tilde{E}}^{\scriptscriptstyle{-{\cal Q}}}(x,\bar q) & = &
\sum_{\lambda=\pm} \ \Gamma^\lambda\, {\cal F}_{-Q}
(x,\bar q_{-\lambda})^\ast \ ,
\label{ep}
\end{eqnarray}
where $\Gamma^{\pm} = (1\pm S^3)/2$, $S^3 = i\gamma^1\gamma^2$ being the
3-component of the spin operator in the spin one-half representation. We
have also used the definition $\bar q_\lambda = (q^0, \elle_{s
\lambda},\chi,q^3)$, with
 $\elle_{s\pm}  \ = \ \elle - (1\mp s)/2$. The explicit form of the spinors
$u_{\scriptscriptstyle{{\cal Q}_f}}\left(\elle,q^3,a\right)$ and
$v_{-\scriptscriptstyle{{\cal Q}_f}}\left(\elle,q^3,a\right)$ in
Eq.~(\ref{UVlept}), as well as the anticommutation relations between the
fermion creation and annihilation operators and some properties of the
functions $\mathbb{E}^{\scriptscriptstyle{{\cal Q}}}(x,\bar q)$ are given in
App.~\ref{Spin-1/2-SpinorsPropa}. Using these properties it is easy to show
that the spinors $U_f$ and $V_f$ satisfy orthogonality and completeness
relations, namely~\cite{Coppola:2018ygv}
\begin{align}
\int d^{4}x\,\,\overline{U}_f(x,\bar{q}, a)\, U_f(x,\bar{q}', a') &
= 2 m_f \, \hat \delta_{\bar{q}\bar{q}'}\,\delta_{aa'}\ ,\nonumber \\
\int d^{4}x\,\,\overline{V}_f(x,\bar{q}, a)\, V_f(x,\bar{q}', a') &
= - 2 m_f \, \hat \delta_{\bar{q}\bar{q}'}\,\delta_{aa'} \ ,
\nonumber \\
\int d^{4}x\,\,\overline{V}_f(-x,\bar{q}, a)\, U_f(x,\bar{q}', a') &
=\int d^{4}x\,\,\overline{U}_f(x,\bar q, a)\, V_f(-x,\bar{q}',a') = 0
\label{D_L2_D_Norm1}
\end{align}
and
\begin{equation}
\frac{1}{2 m_f} \sumint_{\bar q} \sum_{a} \left[U_f(x,\bar{q}, a)\,
\overline{U}(x',\bar{q}, a) -
V_f(-x,\bar{q}, a)\,\overline{V}_f(-x',\,\bar{q}, a)\right]
= \delta^{\left(4\right)}(x - x')\ .
\label{103}
\end{equation}
On the right hand side of this last equation, an identity in Dirac space is
understood.

\subsection{Spin 1 charged particles: the charged rho mesons}
\label{Spin-1-charged}

We consider here the gauged Proca action for a charged point-like rho meson
in the presence of a static and homogeneous magnetic field. Expressing the
rho charge as $Q_{\rho}={\cal Q}e$, with ${\cal Q}=\pm 1$, we have
\begin{eqnarray}
{\cal S}_{P} & = &
\int d^{4}x\ \bigg\{\!-\frac{1}{2}\,\rhocmunu(x)^{\dagger}\,\rhocmunud(x) +
\mrho^{2}\, \rhocmu(x)^{\dagger} \,\rhocmud(x)\nonumber \\
 & &  + \, \frac{i}{2}\,Q_{\rho}\,F^{\mu\nu}
\Big[ \rhocmud(x)^{\dagger}\,\rhocnud(x) -
\rhocnud(x)^{\dagger}\, \rhocmud(x)\Big]\bigg\} \ ,
\end{eqnarray}
where $\rhocmunud = \dcmud \rhocnud - \dcnud \rhocmud$. The associated
gauged Proca equation reads
\begin{equation}
\dcmu \dcmud \, \rhocnud(x) + \mrho^2 \,\rhocnud(x) - 2 i\, Q_\rho\,F_\nu^{\ \alpha}\,
\rho^{\scriptscriptstyle \cal{Q}}_\alpha(x) = 0\ ,
\end{equation}
with
\begin{equation}
\dcmu \rhocmud(x) = 0\ .
\label{orthvec}
\end{equation}
In the same way as in the previous cases, gauge invariance of the gauged
Proca action requires that under a gauge transformation $\Lambda(x)$ the rho
field transform as
\begin{equation}
\rhocmud(x) \ \rightarrow \ {\tilde\rho}^{\,\scriptscriptstyle{{\cal Q}}}_\mu(x)
= e^{-i Q_\rho \Lambda(x)}\, \rhocmud(x)\ .
\end{equation}

The quantized charged rho field can be written as
\begin{equation}
\rhocmu (x) \ = \ \sumint_{\{\bar q_{\Erho}\}} \sum_{c} \
\frac{1}{2 \Erho} \,\Big[\, \arhoch (\breve {q}, c) \,
\wqchmu(x,\bar q,c) + \arhomch (\breve q,c)^\dagger \,
\wqmchmu(x,\bar q, c)^\ast\, \Big]\ ,
\label{rhofieldBpart}
\end{equation}
where the rho energy is given by $\Erho = \sqrt{\mrho^2+ (2\elle+1) \Brho +
(q^3)^2}$, and we have used the definitions of $\bar q$ and $\breve q$
introduced in the previous subsections, together with $\Brho= |Q_\rho B|$
and $s = \mbox{sign}(Q_\rho B)$. It is important to point out that in this
case the sum over the integer index $k$ starts at $k=-1$, instead of zero.

The functions $\wqchmu\left(x,\bar q,c\right)$ in Eq.~(\ref{rhofieldBpart})
are given by
\begin{eqnarray}
\wqchmu(x,\bar q,c) = \mathbb{R}^{{\scriptscriptstyle{{\cal Q}}},\mu\nu}
(x,{\bar q}) \; \epsilon_{{\scriptscriptstyle{\cal{Q}}},\nu}(\elle,q^3,c)\ ,
\label{funrho}
\end{eqnarray}
where, as in the case of spin 1/2 fields [see Eqs.~(\ref{UVlept})], we
have separated the wavefunction into a function
$\mathbb{R}^{{\scriptscriptstyle{{\cal Q}}},\mu\nu}$ that depends on $x$ and
$\bar q$ and a polarization vector
$\epsilon_{{\scriptscriptstyle{\cal{Q}}},\nu}(k,q^3,c)$, the index $c$
denoting the polarization state. Explicit expressions for these vectors
---dictated by the orthogonality relation Eq.~(\ref{orthvec})--- are given
in App.~\ref{Spin-1-Pola_Propa}. In fact, we note that for $\elle=-1$ there
is only one possible polarization vector; this means that the index $c$ can
only take the value $c=1$ in this case. For $\elle=0$ two polarization
vectors can be constructed, thus in that case $c$ can take values 1 and 2,
while for $\elle\geq 1$ the sum over $c$ in Eq.~(\ref{rhofieldBpart}) runs
over the full set of values $c=1,2,3$.

The functions $\mathbb{R}^{{\scriptscriptstyle{{\cal Q}}},\mu\nu}$ are given by
\begin{eqnarray}
\mathbb{R}^{{\scriptscriptstyle{{\cal Q}}},\mu\nu}(x,{\bar q}) =
\sum_{\lambda=-1,0,1} {\cal F}_{Q_{\rho}}(x,\bar q_\lambda) \ \Upsilon^{\mu\nu}_\lambda
\ ,
\label{rhodef}
\end{eqnarray}
where $\bar q_\lambda= (q^0, k-s\lambda, \chi,q^3)$ (notice that this
definition of $\bar q_\lambda$ is different from the one used in the case of
charged fermions). There are various possible choices for the tensors
$\Upsilon^{\mu\nu}_\lambda$; here we use
\begin{equation}
\Upsilon^{\mu\nu}_{0} = g^{\mu\nu}_{\parallel}\ ,
\qquad \qquad \Upsilon^{\mu\nu}_{\pm 1} =
\frac{1}{2}(g^{\mu\nu}_{\perp} \mp S_3^{\mu\nu})\ ,
\label{upsidef}
\end{equation}
where $S_3^{\mu\nu}=i (\delta^{\mu}_{\ 1}\delta^{\nu}_{\ 2} -
\delta^{\mu}_{\ 2}\delta^{\nu}_{\ 1})$ is the 3-component of the spin
operator in the spin one representation. Orthogonality and completeness
relations for the functions $\mathbb{R}^{{\scriptscriptstyle{{\cal
Q}}},\mu\nu}$, as well as other useful relations involving these functions
and the $\Upsilon^{\mu\nu}_\lambda$ tensors, are given in
App.~\ref{Spin-1-Pola_Propa}. In that appendix we also quote the commutation
relations between the creation and annihilation operators for the charged
rho fields.

As discussed in App.~\ref{Spin-1-Pola_Propa}, for $k \ge 0$ an
additional vector, orthogonal to the physical polarization vectors, can be
introduced [see Eq.~(\ref{long})]. We keep for this new vector the notation
$\epsilon^{\mu}_{\scrcalQ}(\elle, q^3 ,c)$, taking for the polarization
index the value $c=0$, and we refer to the associated polarization as
``longitudinal''. If we extend the set of charged rho meson wavefunctions
$\wqchmu(x,\bar q,c)$ by including the corresponding ``longitudinal''
wavefunction $\wqchmu(x,\bar q,0)\equiv
\mathbb{R}^{{\scriptscriptstyle{{\cal Q}}},\mu\nu}(x,{\bar
q})\,\epsilon_{{\scriptscriptstyle{\cal{Q}}},\nu}(\elle,q^3,0)$, we get for
these functions orthogonality and completeness relations, namely
\begin{equation}
\int d^{4}x\;W_{\scrcalQ}^{\mu}(x,\bar{q}',c')^\ast\,W_{{\scrcalQ},\mu}(x,\bar{q},c)
 = -\,\zeta_{c}\ \hat \delta_{\bar{q}\bar{q}'}\,\delta_{cc'}
\label{orthfrho}
\end{equation}
and
\begin{equation}
\sumint_{\bar q}\sum_{c=c_{\rm min}}^{c_{\rm max}}\;\zeta_c
\,W_{\scrcalQ}^{\mu}(x,\bar{q},c)^\ast\,
W_{\scrcalQ}^{\nu}(x^{\prime},\bar{q},c) =
-\,g^{\mu\nu}\,\delta^{(4)}(x-x')\ .
\label{compfrho}
\end{equation}
In these equations the coefficients $\zeta_c$ are defined as $\zeta_0=-1$,
$\zeta_{1}=\zeta_{2}=\zeta_{3}=1$, while $c_{\rm min}$ and $c_{\rm max}$ are
given by
\begin{equation}
c_{\rm min}=\left\{ \begin{array}{ccc}
1 & \quad\text{if}\quad & k=-1\\
0 & \quad\text{if}\quad & k\ge0
\end{array}\right.\ ,
\qquad\qquad c_{\rm max}=\left\{ \begin{array}{ccc}
1 & \quad\text{if}\quad & k=-1\\
2 & \quad\text{if}\quad & k=0\\
3 & \quad\text{if}\quad & k\ge1
\end{array}\right. \ .
\label{ces1}
\end{equation}

\section{Explicit form of the charged particle propagators}
\label{PiPropa}
\setcounter{equation}{0}

\subsection{The spin 0 charged particle propagator}

As discussed above, the charged pion propagator $\piprop(x,y)$ satisfies
Eq.~(\ref{mesoprop1}), and its behavior under a gauge transformation is
given by Eq.~(\ref{gauprop1}). Using the functions $\ffQxbarq$ defined in
Eq.~(\ref{pionfunc}) and the properties of the functions ${\cal
F}_{Q}(x,\bar q)$ discussed in Sec.~\ref{Ffunctions}, it can be easily seen
that $\piprop(x,y)$ can be expressed as
\begin{eqnarray}
\piprop(x,y) & = & \sumint_{\bar q} \ \ffQxbarq \,
\pipropq \, \ffQybarq^\ast\ ,
\label{propPi}
\end{eqnarray}
with
\begin{eqnarray}
\pipropq  = \frac{1}{q_\parallel^2 - \mpi^2 -(2\elle+1)\Bpi + i\epsilon}\ .
\label{pipropq}
\end{eqnarray}
{}In fact, from this expression of the propagator, taking into account the
gauge transformation properties of the functions ${\cal F}_{Q}(x,\bar q)$ in
Eq.~(\ref{gaugeF}), it is immediately seen that it transforms in the
covariant way given by Eq.~(\ref{gauprop1}).

In addition, as is well known, an alternative form for the charged pion
propagator can be obtained using Schwinger proper time
method. If $\piprop(x,y)$ is written as in Eqs.~(\ref{sfx1})
and (\ref{pipropTinv}), i.e.
\begin{equation}
\piprop(x,y) =  e^{i\SPpiQ (x,y)}\int \frac{d^4 v}{(2 \pi)^4} \ e^{-i\, v (x-y)}\,
\pipropb(v_\parallel,v_\perp) \ ,
\end{equation}
one gets
\begin{equation}
\pipropb(v_{\parallel},v_{\perp}) =
-\,i\int_{0}^{\infty}d\sigma\,\frac{1}{\cos(\sigma\Bpi)}\,
\exp\!\bigg[\!-i\sigma\Big(\mpi^{2}-v_{\parallel}^{2}+
\vec{v}_{\perp}^{\,2}\,\dfrac{\tan(\sigma\Bpi)}{\sigma\Bpi}-
i\epsilon\Big)\bigg]\ .
\label{sfp_schw1}
\end{equation}
This expression can be also obtained starting from Eq.~(\ref{propPi}), as
shown e.g.\ in App.~D of Ref.~\cite{Andersen:2014xxa}. Notice that, as
expected from the discussion in Sec.~\ref{section:chargedprop},
$\pipropb(v_{\parallel},v_{\perp})$ depends only on $v_\parallel^2$ and
$v_\perp^2$.

\subsection{The spin 1/2 charged particle propagator}

The spin 1/2 charged particle propagator $\qprop(x,y)$ satisfies the
equation
\begin{eqnarray}
\left(i \, \rlap/\!{\cal D} - m_f \right) \qprop(x,y) =
\delta^{(4)}(x-y)\ .
\label{quarkprop}
\end{eqnarray}
In terms of the Ritus functions in Eq.~(\ref{ep}), it can be expressed as
\begin{eqnarray}
\qprop(x,y) & = & \sumint_{\bar q} \ \mathbb{E}^{\scriptscriptstyle{{\cal Q}_f}}(x,\bar q)
\; \qpropq \; \bar{\mathbb{E}}^{\scriptscriptstyle{{\cal Q}_f}}(y,\bar q)\ ,
\label{propF}
\end{eqnarray}
where
\begin{eqnarray}
\qpropq  =
\frac{\rlap/\!\hat{\Pi}_s + m_f}
{q_\parallel^2 - \mq^2 - 2\elle B_f +i\epsilon}\ ,
\end{eqnarray}
with the definitions $\hat{\Pi}_s^\mu = (q^0,\,0,\,-s\sqrt{2kB_f},\,q^3)$
and $\bar{\mathbb{E}}^{\scriptscriptstyle{{\cal Q}_f}}(y,\bar q)=\gamma^0\,
\mathbb{E}^{\scriptscriptstyle{{\cal Q}_f}}(y,\bar q)^\dagger\,\gamma^0$.
The above expression can be obtained using the relations in
App.~\ref{Spin-1/2-SpinorsPropa}. Moreover, taking into account the gauge
transformation properties of these functions [see Eq.~(\ref{gaugeF})], from
Eq.~(\ref{propF}) it is easy to see that under a gauge transformation
$\Lambda(x)$ the propagator transforms, as it should, in the covariant way
\begin{equation}
\qprop(x,y) \ \rightarrow \ \tilde S_f(x,y) = e^{-iQ_f \Lambda(x)}\,
\qprop(x,y)\, e^{iQ_f \Lambda(y)}\ .
\end{equation}

As in the case of spin 0 particles, an alternative form of this propagator
can be obtained using Schwinger proper time method. If $S_f(x,y)$ is written
as in Eqs.~(\ref{quarkpropTinv}) and (\ref{FermionPropTinv}), i.e.\
\begin{equation}
\qprop(x,y) =  e^{i\SPq(x,y)}\int \frac{d^4 v}{(2 \pi)^4} \ e^{-i\, v (x-y)}\,
\qpropb(v_\parallel,v_\perp) \ ,
\end{equation}
one gets
\begin{eqnarray}
\qpropb(v_\parallel,v_\perp) & = & -\, i \int_0^{\infty} d\sigma\
\exp\!\bigg[\!-i\sigma\Big(\mq^2-v_\parallel^2+\vec{v}_{\perp}^{\,2}\,\dfrac{\tan(\sigma
\Bq)}{\sigma \Bq} - i\epsilon\Big) \bigg]
\nonumber \\
&& \times \left[ \left(v_\parallel \cdot \gamma_\parallel + \mq
\right)(1 - s\, \gamma^1 \gamma^2 \tan(\sigma \Bq) ) -
\frac{\vec{v}_\perp \cdot \vec{\gamma}_\perp}{\cos^2(\sigma \Bq)}
\right] \ .
\label{sfp_schw_a}
\end{eqnarray}
The derivation of this expression starting from Eq.~(\ref{propF}) can be
found e.g.\ in Ref.~\cite{Watson:2013ghq}. We note that
$\qpropb(p_\parallel,p_\perp)$ satisfies the constraints imposed by the
invariance under rotations around the 3-axis (i.e.\ the $\vec B$ axis) and
under boosts in that direction discussed in Sec.~\ref{section:chargedprop}.

\subsection{The spin 1 charged particle propagator}

The spin 1 charged particle propagator $\rhopropnula(x,y)$ satisfies the
equation
\begin{eqnarray}
\Big[\left(\dcga\dcgad+\mrho^{2}\right)\,g_{\mu\nu}-\dcmud\dcnud+2i\,Q_{\rho}\,F_{\mu\nu}\Big]\rhopropnula(x,y)=
\delta_{\mu}^{\ \gamma}\, \delta^{(4)}(x-y)\ ;
\label{rhoprop}
\end{eqnarray}
it can be expressed as
\begin{eqnarray}
\rhopropnula(x,y) & = &  \sumint_{\bar q} \ \mathbb{R}^{{\scriptscriptstyle{{\cal Q}}},\nu\alpha}(x,{\bar q})
\; \hat D_{\rho^{\scriptscriptstyle{\cal{Q}}},\alpha\beta}(k,q_\parallel)\;
\mathbb{R}^{{\scriptscriptstyle{{\cal Q}}},\gamma\beta}(y,\bar q)^\ast\ ,
\label{rhopropagDGD}
\end{eqnarray}
with
\begin{eqnarray}
\hat D_{\rho^{\scriptscriptstyle{\cal{Q}}},\alpha\beta}(k,q_\parallel))
 = \frac{\displaystyle -\, g_{\alpha\beta} + (1-\delta_{k,-1})\
\Pi_\alpha(k,q_\parallel) \, \Pi_\beta(k,q_\parallel)^\ast /
m_\rho^2}{q_\parallel^2 - m^2 -(2\elle+1) \Brho + i \epsilon}\ .
\label{rhopropagDGD1correcta}
\end{eqnarray}
Here we have introduced the four vector $\Pi^\mu(k,q_\parallel)$, given by
\begin{equation}
\Pi^\mu(k,q_\parallel) = \left(q^0\, , \, i\sqrt{B_\rho/2}\, \Big(\sqrt{k+1}
- \sqrt{k}\Big)\, , \, -s \sqrt{B_\rho/2}\, \Big(\sqrt{k+1}+ \sqrt{k}\Big)\,
,\, q^3\right) .
\label{pimu}
\end{equation}
This vector, which is defined only for $k\geq 0$, plays in some cases a role
equivalent as the one played by the four-momentum vector for $B=0$. It is
easy to see that
\begin{eqnarray}
\Pi_\mu(k,q_\parallel)^\ast\; \Pi^\mu(k,q_\parallel) =
(q^0)^2 - (2 \ellef +1) B_Q - (q^3)^2 \ ,
\end{eqnarray}
which is equal to $m_\rho^2$ for $q^0 = E_\rho$.

Taking into account the properties of $\mathbb{R}^{{\scriptscriptstyle{{\cal
Q}}},\mu\nu}$ functions quoted it App.~\ref{Spin-1-Pola_Propa}, it can be
shown that $\rhopropnula(x,y)$, expressed as in Eq.~(\ref{rhopropagDGD}),
satisfies Eq.~(\ref{rhoprop}). Moreover, using the gauge transformation
properties of the functions ${\cal F}_{Q}(x,\bar q)$ [see
Eq.~(\ref{gaugeF})] it is easy to see that, as in the case of spin 0 and
spin 1/2 particles, the propagator transforms in a covariant way under a
gauge transformation.

As in the previous cases, an alternative form for the charged $\rho$ meson
propagator can be obtained using Schwinger proper time method. If
$\rhopropmunu(x,y)$ is written as in Eqs.~(\ref{rhopropTinv}) and
(\ref{rhopropTinvtra}), i.e.\
\begin{equation}
\rhopropmunu(x,y) =  e^{i \SPrhoQ(x,y)}\int \frac{d^4 v}{(2 \pi)^4} \ e^{-i\, v (x-y)}\,
\rhopropbarmunu(v_\parallel,v_\perp) \ ,
\end{equation}
one gets
\begin{eqnarray}
\hspace{-1.5cm} \rhopropbarmunu\left(v_{\parallel},v_{\perp}\right) & = &
i\int_{0}^{\infty}\frac{d\sigma}{\cos(\sigma\Brho)}\,
\exp\left[-i\sigma\left(\mrho^{2}-v_{\parallel}^{2}+\vec{v}_{\perp}^{\,2}\,\frac{\tan(\sigma\Brho)}{\sigma\Brho}
-i\epsilon\right)\right]\nonumber \\
&& \hspace{-1.7cm} \times\, \Bigg\{\mathbb{O}^{\mu\nu}_{1}(v)
 -\left[2\,\sin^{2}(\sigma\Brho)-1+
 \frac{\vec{v}_{\perp}^{\,2}}{\mrho^{2}}\,\tan^{2}(\sigma\Brho)+
 i\,\frac{\Brho}{2\mrho^{2}}\,\tan(\sigma\Brho)\right]\mathbb{O}^{\mu\nu}_{2}(v)
 \nonumber \\
&& \hspace{-1cm} -\,\frac{1}{m_\rho^2}\, \bigg[\mathbb{O}^{\mu\nu}_{3}(v) +
 \left[ 1+ \tan^{2}(\sigma\Brho)\right] \, \mathbb{O}^{\mu\nu}_{4}(v) +
 \, \mathbb{O}^{\mu\nu}_{5}(v) \bigg] \nonumber \\
&& \hspace{-1cm} +\, i \left[\sin(2\sigma\Brho)+ \,\frac{\vec{v}_{\perp}^{\,2}}{\mrho^{2}}\,
\tan(\sigma\Brho)+\frac{i\Brho}{2\mrho^{2}}\right]\mathbb{O}^{\mu\nu}_{6}(v)
+ \frac{i \tan(\sigma\Brho)}{m_\rho^2} \  \mathbb{O}^{\mu\nu}_{7}(v) \Bigg\}\ ,
\label{RhoPropa}
\end{eqnarray}
where
\begin{align}
& \mathbb{O}^{\mu\nu}_{1}(v) = g^{\mu\nu}_\parallel \ , & &
\mathbb{O}^{\mu\nu}_{2}(v) = g^{\mu\nu}_\perp \ , & &
\mathbb{O}^{\mu\nu}_{3}(v) = v^\mu_\parallel \, {v^\nu_\parallel}^\ast \ ,
\nonumber \\
& \mathbb{O}^{\mu\nu}_{4}(v) = v^\mu_\perp \, {v^\nu_\perp}^\ast \ ,
& & \mathbb{O}^{\mu\nu}_{5}(v) =  v^\mu_\perp \, {v^\nu_\parallel}^\ast +
v^\mu_\parallel \, {v^\nu_\perp}^\ast\ , & &
\mathbb{O}^{\mu\nu}_{6}(v) = -i Q_\rho F^{\mu\nu}/{B_\rho}\ ,
\nonumber \\
& \mathbb{O}^{\mu\nu}_{7}(v) = i Q_\rho \big[ F^{\mu\alpha} \,
v_{\perp\alpha} \, {v^\nu_\parallel}^\ast + v^\mu_\parallel \, v_{\perp\alpha}^{\ \,\ast} \,
F^{\alpha\nu}\big]/B_\rho \ .
\hspace{-5cm}
\label{oprho}
\end{align}
This expression can be obtained from Eq.~(\ref{rhoprop}) using the same
methods as in the previous cases. In fact, an equivalent result has been
obtained for the $W$ boson propagator in Ref.~\cite{Erdas:1990gy}. Once
again, it is found that $\rhopropbarmunu(p_{\parallel},p_{\perp})$ satisfies
the constraints imposed by the invariance under rotations around the 3-axis
(i.e.\ the $\vec B$ axis) and under boosts in that direction discussed in
Sec.~\ref{section:chargedprop}.

\section{Meson-quark interactions and one-loop corrections to charged meson two-point correlators}
\label{loc_correlators}

In the previous section we have considered charged non-interacting boson and
fermion fields in the presence of an external magnetic field; let us now
analyze the situation in which these particles interact with each other. The
type of interactions to be considered here are quite generic. In fact, they
can be found in several effective approaches for low energy strong
interactions, such as e.g.\ meson-quark models in hadronic physics and
meson-nucleon models in nuclear physics. As simple but relevant issues, in
this section we discuss the one-loop corrections to the charged pion and rho
meson two point correlators. It is worth mentioning that in the presence of
an external magnetic field these mesons turn out to get mixed. Since in this
article we are mainly concerned on the role played by the Schwinger phase in
this type of calculation, these mixing terms will be neglected; i.e., the
corrections to spin 0 and spin 1 meson self-energies will be treated
separately.

\subsection{Pion-quark interactions and one-loop correction to the
charged pion two-point correlator}

Let us consider the quark-pion interaction Lagrangian
\begin{equation}
{\cal L}_{\rm int}^{(\pi q)} =  g_s \, \bar \psi(x)\, i
\gamma_5 \, \vec \tau \, \psi(x)\, \vec \pi(x)\ .
\label{intlag}
\end{equation}
Here, $\psi(x)$ stands for a fermion field doublet; for definiteness, we
take it to be
\begin{eqnarray}
\psi(x) = \left(
\begin{array}{c}
  \psi_u(x) \\
  \psi_d(x) \\
\end{array}
\right)\ ,
\label{q_doublet}
\end{eqnarray}
where the fields $\psi_f(x)$, with $f=u,d$, are associated to $u$ and $d$
quarks. Using the same notation as in previous sections, we have $Q_u
=2e/3$, $Q_d = -e/3$, $e$ being the proton charge. As usual, $\tau_i$ are
the Pauli matrices, and pion charge and isospin states are related by
$\pi^\pm = (\pi_1 \mp i \pi_2)/\sqrt 2$, $\pi^0 = \pi_3$. The gauge
transformation properties for charged fields given in the previous section
guarantee that the interaction Lagrangian in Eq.~(\ref{intlag}) is gauge
invariant.

We analyze now the leading order correction (LOC) to the two-point $\pi^+$
correlator. One has
\begin{equation}
i \Delta^{\rm (LOC)}_{\pi^+}(y,y') =
\frac{i^2}{2} \int d^4x\; d^4x'\;
\langle 0|\,T\big[ \pi^+(y)\, \pi^+(y')^\dagger \,
{\cal L}_{\rm int}^{(\pi q)}(x)\, {\cal L}_{\rm int}^{(\pi q)}(x')\big] | 0\rangle\ ,
\end{equation}
where the contributions that lead to vacuum-vacuum subdiagrams have been
omitted~\cite{Itzykson:1980rh}. Considering the relevant terms in ${\cal
L}_{\rm int}^{(\pi q)}$ we have
\begin{equation}
\Delta^{\rm (LOC)}_{\pi^+}(y,y') = - i\, g_s^2
\int d^4x\; d^4x' \; \Delta_{\pi^+}(y,x) \;  J_{\pi^+}(x,x')\;
\Delta_{\pi^+}(x',y')\ ,
\label{deltaLOC}
\end{equation}
where $J_{\pi^+}(x,x')$ is the polarization function in coordinate space,
\begin{eqnarray}
J_{\pi^+}(x,x') = - 2 N_c \ \mbox{tr}_D \big[ i S_u(x,x')\, i\gamma_5 \, i S_d(x',x)
\, i\gamma_5 \big]\ ,
\label{jxxpPi}
\end{eqnarray}
tr$_D$ denoting trace in Dirac space.

We also introduce the $\pi^+$ polarization function in $\bar q$-space (or
Ritus space), $J_{\pi^+}(\bar q, \bar q')$, defined by
\begin{eqnarray}
J_{\pi^+}(\bar q, \bar q') = \int d^4x\, d^4x' \ \ffpxbarq^\ast \, J_{\pi^+}(x,x') \,
\ffpxprimabarqprima \ .
\label{sigmapi}
\end{eqnarray}
This equation can be inverted using the completeness relation for the
functions ${\cal F}_Q(x,\bar q)$, Eq.~(\ref{completfpion}), obtaining
\begin{eqnarray}
J_{\pi^+}(x,x') = \sumint_{\bar q,\,\bar q'} \ \ffpxbarq \,
J_{\pi^+}(\bar q, \bar q') \, \ffpxprimabarqprima^\ast\ .
\label{jpolqspace}
\end{eqnarray}
{}From this last relation, together with Eq.~(\ref{propPi}) and the
orthogonality relation Eq.~(\ref{orthfpion}), the leading order correction
to the $\pi^+$ propagator can be written as
\begin{equation}
\Delta^{\rm (LOC)}_{\pi^+}(y,y') = -\, i\, g_s^2 \,\sumint_{\bar q,\, \bar q'}
\, \ffpybarq \; \pipropqplus \; J_{\pi^+}(\bar q, \bar q') \;
\pipropqprimaplus \; \ffpyprimabarqprima^\ast \ ,
\end{equation}
where $\pipropqplus$ is given by Eq.~(\ref{pipropq}).

Note that none of the functions appearing in the rhs of Eq.~(\ref{sigmapi})
is by itself an invariant quantity. However, being ${\cal L}_{\rm int}^{(\pi
q)}(x) $ gauge invariant, so must be $J_{\pi^+}(\bar q, \bar q')$. In fact,
on the basis of gauge, translational and rotational symmetries, we expect
$J_{\pi^+}(\bar q, \bar q')$ to be of the form $J_{\pi^+}(\bar q, \bar q') =
\hat \delta_{\bar q \bar q'}  \, \hat J_{\pi^+}(\elle,q_\parallel)$. In the
following we will see how this comes out by explicit calculation.

We start by considering Eq.~(\ref{jxxpPi}), writing the quark propagators in
the form given by Eqs.~(\ref{quarkpropTinv}) and (\ref{FermionPropTinv}). In
this way we get
\begin{eqnarray}
J_{\pi^+}(x,x') =  e^{i \SPpiplus(x,x')}\, \bar J_{\pi^+}(x-x')\ ,
\label{jpixxdos}
\end{eqnarray}
where
\begin{eqnarray}
\bar J_{\pi^+}(x-x') = \int \frac{d^4v}{(2\pi)^4}\, e^{-i v (x- x')}\, \bar
J_{\pi^+}(v_\parallel,v_\perp)\ ,
\label{jinvPidos}
\end{eqnarray}
with
\begin{eqnarray}
 \bar J_{\pi^+}(v_\parallel,v_\perp) =  - 2 N_c \int \frac{d^4p}{(2\pi)^4}\,
\mbox{tr}_D \Big[ i \bar S^u({p_\parallel^+},{p_\perp^+})\,
i\gamma_5 \, i \bar S^d({p_\parallel^-},{p_\perp^-}) \, i\gamma_5 \Big]\ .
\label{jPiquarkprop}
\end{eqnarray}
Here we have used the definition $p_\mu^\pm = p_\mu \pm v_\mu/2$. In
addition, in Eq.~(\ref{jpixxdos}) we have made use of the relation
\begin{eqnarray}
\Phi^{Q_u}(x,x') + \Phi^{Q_d}(x',x) = \Phi^{Q_u-Q_d}(x,x') =
\Phi^{Q_{\pi^+}}(x,x')\ .
\end{eqnarray}
We see from the above equations that $J_{\pi^+}(x,x')$ can be written as the
product of a SP and a function $\bar J_{\pi^+}(x-x')$, which is both gauge
and translational invariant. Thus, under a gauge transformation the
polarization function transforms in the same way as the SP. On the other
hand, as in the case of bare charged pion propagator, invariance under
rotations around the 3-axis (i.e. the $\vec B$ axis) and under boosts in
that direction imply that $\bar J_{\pi^+}(v_\parallel,v_\perp)$ has to be a
function of $v_\parallel^2$ and $v_\perp^2$; this is indeed corroborated by
the explicit form given below.

Replacing Eq.~(\ref{jpixxdos}) in Eq.~(\ref{sigmapi}), and taking into
account Eq.~(\ref{pionfunc}), we get
\begin{equation}
J_{\pi^+}(\bar q,\bar q') =  \int \frac{d^4v}{(2\pi)^4} \, h_{\pi^+}(\bar q, \bar q', v_\parallel,v_\perp)
\, \bar J_{\pi^+}(v_\parallel,v_\perp)\ ,
\label{jpibarqq}
\end{equation}
where
\begin{equation}
h_{{\mbox{{\scriptsize P}}}}(\bar q, \bar q', v_\parallel,v_\perp) = \int d^4x\, d^4x'
\,  {\cal F}_{Q_{\mbox{{\scriptsize P}}}}(x,\bar q)^\ast\,  {\cal F}_{Q_{{\mbox{{\scriptsize P}}}}}(x',\bar q')
 \, e^{i \Phi_{{\mbox{{\scriptsize P}}}}(x,x')} \, e^{-i v (x- x')}\ .
\label{hPiInt}
\end{equation}
It is easy to see that $h_{{\mbox{{\scriptsize P}}}}(\bar q, \bar q',
v_\parallel,v_\perp)$ is gauge invariant, given the gauge transformation
properties of the SP and the functions ${\cal F}_{Q_{\mbox{{\scriptsize
P}}}}(x,\bar q)$. One can carry out its explicit calculation in any of the
standard gauges SG, LG1 and LG2, obtaining
\begin{eqnarray}
h_{{\mbox{{\scriptsize P}}}}(\bar q, \bar q', v_\parallel,v_\perp) =
\delta_{\chi\chi'}  \, \check{h}_{{\mbox{{\scriptsize P}}}}
(\elle,q_\parallel, \elle',q_\parallel', v_\parallel,v_\perp)\ .
\label{hPiGauge0}
\end{eqnarray}
Here $\delta_{\chi\chi'}$ stands for $\delta_{\imath\imath'}$,
$\delta(q^1-q^{\prime 1})$ and $\delta(q^2-q^{\prime 2})$ for SG, LG1 and
LG2, respectively, while the function $\check{h}_{{\mbox{{\scriptsize P}}}}
(\elle,q_\parallel, \elle',q_\parallel', v_\parallel,v_\perp)$ is given by
\begin{equation}
\check{h}_{{\mbox{{\scriptsize P}}}}(\elle,q_\parallel, \elle',q_\parallel',v_\parallel,v_\perp) =
\left(2\pi\right)^{4}\,\delta^{\left(2\right)}(q_\parallel-q^\prime_\parallel)\,
\left(2\pi\right)^{2}\,\delta^{\left(2\right)}(q_{\parallel}-v_{\parallel}) \,
f_{\elle\elle'}(v_\perp) \ ,
\label{hPiGauge}
\end{equation}
with
\begin{equation}
f_{\elle\elle'}(v_\perp) = \frac{4\pi(-i)^{\elle+\elle'}}{B_P}\,
\sqrt{\frac{\elle!}{\elle^{\prime}!}} \, \left(\frac{2 \vec
v^{\,2}_\perp}{B_P}\right)^{\frac{\elle'-\elle}{2}}
L_{\elle}^{\elle^{\prime}-\elle}\Big(\frac{2 \vec v^{\,2}_\perp}{B_P}\Big)\;
e^{-\vec v^{\,2}_\perp/B_P}\; e^{i s (\elle-\elle')\phi_\perp}\ .
\label{fkkp}
\end{equation}
We have used here the definition $B_P = |B\QP|$ and introduced the angle
$\phi_\perp$, given by $\vec v_\perp = |\vec v_\perp|\,(\cos \phi_\perp,\sin
\phi_\perp)$. Note that in the present case $B_P= B_\pi = e|B|$ and
$s=\mbox{sign}(B)$.

As stated, $\bar J_{\pi^+}(v_\parallel,v_\perp)$ is found to be a function
of $\vec v_{\perp}^{\,2}$ [see Eq.~(\ref{jpi}) below]. Performing the
integral over $\phi_\perp$, one arrives to the expected form
\begin{equation}
J_{\pi^+}(\bar q, \bar q') = \hat \delta_{\bar q \bar q'}  \,
\hat J_{\pi^+}(\elle,q_\parallel) \ ,
\label{JpiDiagonal0}
\end{equation}
where
\begin{equation}
\hat J_{\pi^+}(\elle,q_\parallel) =
\int_0^\infty d|\vec v_{\perp}|^2\, \bar J_{\pi^{+}}
\left(q_{\parallel},v_{\perp}\right)\,
\rho_{\elle}(\vec v_\perp^{\, 2})\ ,
\label{sigmapibar}
\end{equation}
with
\begin{eqnarray}
\rho_{\elle}(\vec v_\perp^{\, 2}) =
 \frac{(-1)^{\elle}}{B_{\pi}}\,e^{-\vec v^{\,2}_\perp/B_{\pi}}\,
L_{\elle}\left(\frac{2 \vec v^{\,2}_\perp}{B_{\pi}}\right)\ .
\label{vPerpDis}
\end{eqnarray}
It is worth recalling that, due to the presence of the nonvanishing
Schwinger phase $\Phi^{Q_{\pi^+}}(x,x')$, the polarization function
$J_{\pi^+}(x,x')$ is not just a function of $x-x'$, and therefore
it cannot be diagonalized by a Fourier transform into momentum space.
Instead, one can obtain a diagonal, gauge invariant function $J_{\pi^+}(\bar
q, \bar q')=\hat \delta_{\bar q \bar q'}\,\hat J_{\pi^+}(\elle,q_\parallel)$
through the above described transformation into $\bar q$-space. As shown by
Eq.~(\ref{sigmapibar}), the Fourier transform $\bar
J_{\pi^{+}}(q_{\parallel},v_{\perp})$ of the translational invariant part of
$J_{\pi^+}(x,x')$ does not coincide with $\hat
J_{\pi^+}(\elle,q_\parallel)$; in fact, the latter can be obtained from the
integration of $\bar J_{\pi^{+}}(q_{\parallel},v_{\perp})$ over the
perpendicular momentum $v_\perp$, weighted by the function $\rho_k(\vec
v^{\, 2}_\perp)$ given by Eq.~(\ref{vPerpDis}). On the other hand, in
absence of the SP in Eq.~(\ref{hPiInt}) one could replace the functions
${\cal F}_{Q_{\mbox{{\scriptsize P}}}}(x,\bar q)$ by plane waves, and the
function $h_{\pi^+}$ in Eq.~(\ref{jpibarqq}) would be simply given by
$(2\pi)^8\,\delta^{(4)}(q-q')\delta^{(4)}(q-v)$; this is, indeed, what is
done in the case of neutral mesons.

Given that $J_{\pi^+}(\bar q, \bar q')$ is diagonal in $\bar q$-space, from
Eq.~(\ref{deltaLOC}) the LOC to the propagator can be written as
\begin{eqnarray}
\Delta^{\rm (LOC)}_{\pi^+}(y,y') = \sumint_{\bar q} \, \ffpybarq \, \pipropqplusLOC \,
\ffpyprimabarq^\ast\ ,
\label{Deltapiqq}
\end{eqnarray}
where
\begin{eqnarray}
\pipropqplusLOC= \pipropqplus\, \hat \Sigma_{\pi^+}(\elle, q_\parallel) \,
\pipropqplus \ ,
\label{piprophat}
\end{eqnarray}
with
\begin{eqnarray}
\hat \Sigma_{\pi^+}(\elle, q_\parallel) = - i\, g_s^2\, \hat J_{\pi^+}(\elle,
q_\parallel)\ .
\end{eqnarray}

To get the final form of $\hat J_{\pi^+}(\elle,q_\parallel)$ from
Eq.~(\ref{sigmapibar}) we need the explicit expression of $\bar
J_{\pi^+}(v_\parallel,v_\perp)$. The latter can be readily obtained from
Eq.~(\ref{jPiquarkprop}), taking into account the invariant part of the quark
propagators given in Eq.~(\ref{sfp_schw_a}). One has
\begin{eqnarray}
\bar J_{\pi^+}(v_\parallel,v_\perp) & = & - \frac{i N_c}{4\pi^2}
\int_{-1}^1 dx \int_0^\infty \frac{dz}{ \ t_+} \,  e^{-z\,
\phi(x,v_\parallel^2)} \,  e^{- (t^2_+ - t^2_-)\, \vec v_\perp^{\, 2} /(4
t_+)}
\nonumber \\
& & \times \, \bigg\{ \bigg[ m_u m_d +\frac{1}{z} + (1-x^2) \frac{v_\parallel^2}{4} \bigg] (1- t_{u}\, t_{d})
\nonumber \\
& & \ \ \ \ \ + \bigg[\frac{1}{t_+} - \bigg( 1 - \frac{t^2_-}{t^2_+} \bigg)
\frac{\vec v_\perp^{\, 2}}{4}\bigg] (1-t_u^2) (1-t_d^2)\bigg\} \ ,
\label{jpi}
\end{eqnarray}
where we have used the definition
\begin{equation}
\phi(x,v_\parallel^2) = (m_u^2 + m_d^2)/2 - x (m_u^2 - m_d^2)/2 - (1-x^2) \,
v_\parallel^2/4\ ,
\label{phi}
\end{equation}
as well as
\begin{equation}
t_u = \tanh\left[(1-x) z B_u/2\right]\ , \qquad t_d = \tanh\left[(1+x) z
B_d/2\right]\ , \qquad t_\pm = t_u/B_u \pm t_d/B_d\ .
\label{tf}
\end{equation}
Replacing this expression in Eq.~(\ref{sigmapibar}) and performing the
integral over $v_\perp$ we finally obtain
\begin{eqnarray}
\hat J_{\pi^+}(\elle,q_\parallel) & = &
 - \frac{i N_c}{4\pi^2}  \int_{-1}^1 dx \int_0^\infty dz
\ e^{-z \phi(x,q_\parallel^2)} \, \frac{1}{\alpha_+} \left(\frac{\alpha_-}{\alpha_+}\right)^\elle
\nonumber \\
& & \times \, \bigg\{ \bigg[ m_u m_d +\frac{1}{z} + (1-x^2) \frac{q_\parallel^2}{4}\bigg] (1- t_{u}\, t_{d})
\nonumber \\
& & \ \ \ \ \ \ + \frac{\alpha_- + \elle (\alpha_- - \alpha_+)}{\alpha_+ \alpha_-} \, (1-t_u^2) (1-t_d^2)\bigg\}\ ,
\label{jpiint}
\end{eqnarray}
where we have defined $\alpha_\pm$ as
\begin{eqnarray}
\alpha_\pm = \frac{t_u}{B_u} +  \frac{t_d}{B_d} \pm B_{\pi}  \frac{t_u}{B_u}
\frac{t_d}{B_d} =t_+ \pm B_{\pi}  \frac{t_u}{B_u}  \frac{t_d}{B_d} \ .
\label{alpha}
\end{eqnarray}

The integral on the rhs of Eq.~(\ref{jpiint}) is divergent, so it has to be
regularized. This can be done e.g.\ by subtracting the $B=0$ contribution,
leaving a finite $B$-dependent piece. In addition, an analytical extension
of the function $\hat J_{\pi^+}(\elle,q_\parallel)$ can be performed for
large positive values of $q_\parallel^2$.

We end this subsection by noting that $\Delta^{\rm (LOC)}_{\pi^+}(y,y')$ can
be expressed in an alternative way. By looking at Eqs.~(\ref{sfx1}) and
(\ref{piprogen}) for the bare propagator together with Eq.~(\ref{jpixxdos})
for the polarization function, one can foresee that the translational
non-invariance of the dressed propagator will be carried by Schwinger phases
at any order of the calculation. On this basis, we explicitly separate the
corresponding SP in the LOC to the propagator, writing
\begin{equation}
\Delta^{\rm (LOC)}_{\pi^+}(y,y') =  e^{i\SPpiplus(y,y')} \,
\bar{\Delta}^{\rm (LOC)}_{\pi^+}(y,y')\ .
\label{deltaLOC11}
\end{equation}
Then we can use Eqs.~(\ref{sfx1}), (\ref{piprogen}), (\ref{deltaLOC}) and
(\ref{jpixxdos}) to obtain
\begin{equation}
\bar{\Delta}^{\rm (LOC)}_{\pi^+}(y,y')
= -ig_s^2 \int d^4x\, d^4x' \,  e^{i \varphi(y-x',x-y')}
\, \bar{\Delta}_{\pi^+}(y-x) \, \bar{J}_{\pi^+}(x-x')\, \bar{\Delta}_{\pi^+}(x'-y') \ ,
\end{equation}
where
\begin{align}
\varphi\left(y-x',\,x-y^{\prime}\right) & =
\Phi_{\pi^{+}}(y,x)+\Phi_{\pi^{+}}(x,x')+\Phi_{\pi^{+}}(x',y')+\Phi_{\pi^{+}}(y',y)
\nonumber \\
& = \frac{Q_{\pi}}{2}\,(y_{\mu}-x_{\mu}^{\prime})\,F^{\mu\nu}\,(x_{\nu}-y_{\nu}^{\prime})\ .
\label{SPsum}
\end{align}
It is worth noticing that the phase
$\varphi\left(y-x',\,x-y^{\prime}\right)$ is in general non-vanishing for
nonzero $B$. Moreover, it is invariant under gauge transformations,
translations, rotations around the direction of $\vec B$ and boosts in that
direction. This implies that once the SP has been extracted, the remaining
factor $\bar{\Delta}^{\rm (LOC)}_{\pi^+}(y,y')$ should have all the
associated invariance properties; in particular, one should be able to write
$\bar{\Delta}^{\rm (LOC)}_{\pi^+}(y,y')=\bar{\Delta}^{\rm
(LOC)}_{\pi^+}(y-y')$. Indeed, using the Fourier transforms defined in
Eqs.~(\ref{pipropTinv}) and (\ref{jinvPidos}) changing variables $x$ and
$x'$ to $z=y-x'$ and $z'=x-y'$ we obtain
\begin{align}
\bar{\Delta}_{\pi^{+}}^{\rm (LOC)}\left(y,y'\right) = &
 -ig_{s}^{2}\,\,\int
\frac{d^{4}r}{\left(2\pi\right)^{4}}\,\frac{d^{4}s}{\left(2\pi\right)^{4}}\,\frac{d^{4}t}{\left(2\pi\right)^{4}}
\;\bar{\Delta}_{\pi^{+}}(r_\parallel,r_\perp)\,\bar{J}_{\pi^{+}}(s_\parallel,s_\perp)\,
\bar{\Delta}_{\pi^{+}}(t_\parallel,t_\perp) \nonumber \\
& \times \,e^{-i(r-s+t)\,(y-y')}\, \int d^{4}z\,d^{4}z' \,e^{i\varphi(z,\,z^{\prime})}
  \,e^{i(r-s)\,z'}\,e^{i(t-s)\,z}\ ;
\end{align}
thus, we can write
\begin{equation}
\bar{\Delta}^{\rm (LOC)}_{\pi^+}(y,y') = \bar{\Delta}^{\rm (LOC)}_{\pi^+}(y-y') =
\int \frac{d^4v}{(2\pi)^4}\; e^{-i v (y- y')} \, \bar{\Delta}^{\rm
(LOC)}_{\pi^+}(v_\parallel,v_\perp)\ .
\end{equation}
Now, noting that $\varphi(z,z')$ depends only on the perpendicular components of
$z$ and $z'$, we get
\begin{align}
\bar{\Delta}^{\rm (LOC)}_{\pi^+}(v_\parallel,v_\perp) = & -ig_{s}^{2}\,\int
\frac{d^{2}r_{\perp}}{\left(2\pi\right)^{2}}\,\frac{d^{2}t_{\perp}}{\left(2\pi\right)^{2}}
\; \bar{\Delta}_{\pi^{+}}(v_{\parallel},r_\perp)\,\bar{J}_{\pi^{+}}(v_{\parallel},r_{\perp}+ t_{\perp}- v_{\perp})
\,\bar{\Delta}_{\pi^{+}}(v_{\parallel},t_{\perp})
\nonumber \\
 & \times \, \int d^{2}z_{\perp}\,d^{2}z_{\perp}^{\prime}\;
 e^{i\varphi(z_\perp,\,z^{\prime}_\perp)}\,
 e^{-i\,(\vec{v}_{\perp}-\vec{t}_{\perp})\,\cdot\vec{z}_{\perp}^{\,\prime}}\,
 e^{-i\,(\vec{v}_{\perp}-\vec{r}_{\perp})\,\cdot\vec{z}_{\perp}}\ ,
\label{PropLOCAltern0}
\end{align}
and finally we can perform the integrals over $z_\perp$ and $z_\perp'$ and make
the change of variables
\begin{equation}
r_{\perp}=v_{\perp}-\sqrt{\frac{B_{\pi}}{2}}\,u_{\perp}\ ,\qquad\qquad
t_{\perp}=v_{\perp}-\sqrt{\frac{B_{\pi}}{2}}\,u_{\perp}^{\,\prime}\ ,
\end{equation}
obtaining
\begin{eqnarray}
\bar{\Delta}_{\pi^{+}}^{\rm (LOC)}
(v_\parallel,v_\perp) & =&
-i\frac{g_{s}^{2}}{\left(2\pi\right)^{2}}\,\int d^{2}u_{\perp}\,\,d^{2}u_{\perp}^{\prime}
\,e^{i\varphi\left(\sqrt{\frac{2}{B_{\pi}}}\,u_\perp^{\prime},\,\sqrt{\frac{2}{B_{\pi}}}\,u_\perp\right)}\,\nonumber \\
 & & \!\!\!\! \!\!\!\! \!\!\!\! \!\!\!\! \!\!\!\! \!\!\!\! \!\!\!\!
 \!\!\!\!\!\!\!\!\!\! \times \
 \bar{\Delta}_{\pi^{+}}\Big(v_{\parallel},v_{\perp}-\sqrt{\frac{B_{\pi}}{2}}\,u_{\perp}\Big)\,
 \bar{J}_{\pi^{+}}
 \Big(v_{\parallel},v_{\perp}-\sqrt{\frac{B_{\pi}}{2}}\,(u_{\perp}^{\,\prime}+u_{\perp})\Big)
 \,\bar{\Delta}_{\pi^{+}}\Big(v_{\parallel},v_{\perp}-\sqrt{\frac{B_{\pi}}{2}}\,u_{\perp}^{\,\prime}\Big)\ .
 \nonumber \\
&&
\label{PropLOCAltern1}
\end{eqnarray}
At this point we can remark the important role played by the Schwinger
phase, which is responsible for the existence of the phase
$\varphi\left(\sqrt{\frac{2}{B_{\pi}}}\,u_\perp^{\prime},
\,\sqrt{\frac{2}{B_{\pi}}}\,u_\perp\right)$ [see  Eq.~(\ref{SPsum})]. As
shown in Eq.~(\ref{PropLOCAltern1}), the latter drives the fluctuation of
transverse momenta suffered by the charged particles when they propagate in
the presence of the magnetic field. Were the phase
$\varphi(z_\perp,\,z^{\prime}_\perp)$ omitted in Eq.~(\ref{PropLOCAltern0}),
one would directly obtain
\begin{equation}
\bar{\Delta}_{\pi^{+}}^{\rm (LOC)}(v_\parallel,v_\perp) = -ig_{s}^{2}\,
\bar{\Delta}_{\pi^{+}}(v_{\parallel},v_{\perp})\,\bar{J}_{\pi^{+}}(v_{\parallel},v_{\perp})
\,\bar{\Delta}_{\pi^{+}}(v_{\parallel},v_{\perp})\ ,
\label{DlocPWA}
\end{equation}
loosing any transverse momentum fluctuation.

A final comment on the $B\rightarrow0$ limit of Eq. (\ref{PropLOCAltern1})
is pertinent. It is easy to see that in this limit the integral over
$u_\perp$ and $u_\perp'$ only affects the phase
$\varphi\left(\sqrt{\frac{2}{B_{\pi}}}\,u_\perp^{\prime},\,
\sqrt{\frac{2}{B_{\pi}}}\,u_\perp\right)$. To regulate the oscillatory
integrals one can introduce factors $e^{-\epsilon\left|u^{i}\right|}$,
$e^{-\epsilon\left|u^{\prime i}\right|}$, with $i=1,2$, and then take
$\epsilon\rightarrow0^+$, obtaining
\begin{equation}
\lim_{B\rightarrow0}\bar{\Delta}_{\pi^{+}}^{\rm (LOC)}\left(v_{\parallel},v_{\perp}\right) = -ig_{s}^{2}
\left[\lim_{B\rightarrow0}\,\bar{\Delta}_{\pi^{+}}(v_{\parallel},v_{\perp})\right]
\left[\lim_{B\rightarrow0}\,\bar{J}_{\pi^{+}}(v_{\parallel},v_{\perp})\right]
\left[\lim_{B\rightarrow0}\,\bar{\Delta}_{\pi^{+}}(v_{\parallel},v_{\perp})\right]\ .
\end{equation}
This is the expected result. If there is no magnetic field, there is no
fluctuation of transverse momenta; this is a consequence of translation
invariance, which implies the conservation of the four components of the
momentum. On the contrary, in the presence of a static and uniform
magnetic field the situation is different. As we have seen in
Sect.~\ref{section:SP}, in that case translation invariance in the
transverse direction is realized in a nontrivial way, being related to gauge
transformations. In addition, in Sect.~\ref{section:fields} we have seen
that the wavefunctions associated to charged particles depend on the
chosen gauge, and cannot be written in terms of definite four-momentum
states. In fact, this leads to a fluctuation in the transverse spatial
directions that is translated into a fluctuation in the transverse
directions of the momentum. Our result in Eq.~(\ref{PropLOCAltern1}) shows
how these fluctuations affect the evaluation of the LOC to the pion
propagator, in particular, the part of the propagator that is invariant
under gauge transformations, translations, rotations around the $\vec B$
axis and boosts in the spatial direction parallel to the magnetic field.

On the basis of charge conservation, it is not difficult to realize that the
appearance of a SP as in Eq.~(\ref{deltaLOC11}) will be valid at any order
of correction, and, therefore, it also applies to the full propagator.

\subsection{Rho meson-quark interactions and one loop correction to the
charged rho meson two-point correlator}
\label{onelooprho}

Let us consider the rho meson-quark interaction lagrangian
\begin{eqnarray}
{\cal L}_{\rm int}^{(\rho q)} =  g_v\, \vec \rho_\mu(x) \, \bar \psi(x)\,
\gamma^\mu \vec \tau\, \psi(x)\ .
\end{eqnarray}
As usual, $\rho$ charge states are related to isospin states by
$\rho_\mu^\pm = (\rho_{1,\mu} \mp i \rho_{2,\mu})/\sqrt 2$ and $\rho_\mu^0 =
\rho_{3,\mu}$.

The leading order correction to the two-point $\rho^+$ correlator is given
by
\begin{eqnarray}
i D^{\rm (LOC)}_{\rho^+,\mu\nu}(y,y') =\frac{ i^2 }{2} \int d^4x\, d^4x'
\, \langle 0 \big| T\big[\rho^+_\mu(y)\, {\rho^+_\nu(y')}^\dagger\,
{\cal L}_{\rm int}^{(\rho q)}(x)\, {\cal L}_{\rm int}^{(\rho q)}(x')\big] \big|0
\rangle \ .
\end{eqnarray}
Considering the relevant terms in ${\cal L}_{\rm int}^{(\rho q)}$ we have
\begin{eqnarray}
D^{\rm (LOC)}_{\rho^+,\mu\nu}(y,y') = - i g_v^2 \int d^4x\, d^4x' \, D_{\rho^+,\mu\alpha}(y,x)
\, J_{\rho^+}^{\alpha\beta}(x,x')\, D_{\rho^+,\beta\nu}(x',y')\ ,
\label{dLOC}
\end{eqnarray}
where $J_{\rho^+}^{\alpha\beta}(x,x')$ is the polarization function in coordinate
space,
\begin{eqnarray}
J_{\rho^+}^{\alpha\beta}(x,x') = - 2 N_c \, \mbox{tr}_D \Big[ i S_u(x,x')\,
\gamma^\beta \, i S_d(x',x) \, \gamma^\alpha \Big]\ .
\label{jxxpRho}
\end{eqnarray}
As in the charged pion case, we introduce the polarization function in $\bar
q$-space (or Ritus space), $J^{\alpha\alpha'}_{\rho^+}(\bar q,\bar
q')$, given by
\begin{eqnarray}
J^{\alpha\alpha'}_{\rho^+}(\bar q,\bar q') = \int d^4x\, d^4x' \
{\mathbb{R}^{+,\mu\alpha}(x,\bar q)}^\ast  \,
J_{\rho^+,\mu\nu}(x,x') \, \mathbb{R}^{+,\nu\alpha'}(x',\bar q')\ .
\label{sigmarho}
\end{eqnarray}
Using the completeness relation, Eq.~(\ref{compR}), one gets
\begin{equation}
J_{\rho^{+}}^{\mu\nu}(x,x')= \sumint_{\bar{q},\,\bar q'} \, \mathbb{R}^{+,\mu\alpha}(x,\bar{q})\,
J_{\rho^{+},\alpha\alpha'}(\bar{q},\bar{q}')\,{\mathbb{R}^{+,\nu\alpha'}(x',\bar{q}')}^{\ast}\ .
\label{polrhodiagxx1}
\end{equation}
Then, from Eq.~(\ref{rhopropagDGD}) and the orthogonality relation
Eq.~(\ref{orthormunu}), the LOC to the propagator can be written as
\begin{equation}
D^{\rm (LOC)}_{\rho^+,\mu\nu}(y,y') = -i g_v^2 \sumint_{\bar q,\, \bar q'}\,
\mathbb{R}^+_{\mu\alpha}(y,\bar q) \, \hat D_{\rho^+}^{\alpha\alpha'}(k,q_\parallel)\,
J_{\rho^+,\alpha'\beta'}(\bar q,\bar q')\, \hat D_{\rho^+}^{\beta'\beta}(k',q_\parallel)
\, {\mathbb{R}^+_{\nu\beta}(y',\bar q')}^\ast\ .
\label{DLOC}
\end{equation}
One can also take into account the explicit form of the functions
$\mathbb{R}^+_{\mu\nu}$ in Eq.~(\ref{rhodef}) to write $J^{\alpha\alpha'}_{\rho^+}(\bar q,\bar q')$ as
\begin{equation}
J^{\alpha\alpha'}_{\rho^+}(\bar q, \bar q') = \sum_{\lambda,\lambda'=-1}^{+1} (\Upsilon^{\mu\alpha}_\lambda)^\ast
\, \Upsilon^{\nu\alpha'}_{\lambda'} \int d^4x\, d^4x'
\; {\cal F}_{Q_{\rho^+}}(x,{\bar q}_\lambda)^\ast \, {\cal F}_{{Q_{\rho^+}}}(x',{\bar q'}_{\lambda'})
  \, J_{\rho^+,\mu\nu}(x,x') \ ,
\label{calJa}
\end{equation}
where $\bar q_\lambda = (q^0,\elle_\lambda,\chi,q^3)$, $\elle_\lambda =
\elle-s\lambda$, $s = \mbox{sign}(Q_{\rho^+} B)=\mbox{sign}(B)$.

Proceeding as in the $\pi^+$ case, we go back to Eq.~(\ref{jxxpRho}) and
write the quark propagators in the form given by Eqs.~(\ref{quarkpropTinv})
and (\ref{FermionPropTinv}). This leads to
\begin{eqnarray}
J_{\rho^+}^{\mu\nu}(x,x') =  e^{i \SPrhoplus(x,x')}
  \int \frac{d^4v}{(2\pi)^4}\, e^{-i v (x- x')}\, \bar
  J_{\rho^+}^{\mu\nu}(v_\parallel,v_\perp)\ ,
\label{jpixx}
\end{eqnarray}
where
\begin{equation}
\bar J_{\rho^+}^{\mu\nu}(v_\parallel,v_\perp) = - 2 N_c \int \frac{d^4p}{(2\pi)^4}
\,\mbox{tr}_D \left[ i \bar S^u({p_\parallel^+},{p_\perp^+})\,
\gamma^\nu \, i \bar S^d({p_\parallel^-},{p_\perp^-}) \, \gamma^\mu \right]\ ,
\label{barjpimunu}
\end{equation}
with $p^\pm_\mu = p^\pm_\mu + v_\mu/2$. Replacing these equations into
Eq.~(\ref{calJa}) we get
\begin{equation}
J^{\alpha\alpha'}_{\rho^+}(\bar q, \bar q') = \int \frac{d^4v}{(2\pi)^4} \sum_{\lambda,\lambda'=-1}^{+1}
(\Upsilon^{\mu\alpha}_\lambda )^\ast
\, \Upsilon^{\nu\alpha'}_{\lambda'}\, \bar J_{\rho^+,\mu\nu}(v_\parallel,v_\perp)
\, h_{\rho^+}({\bar q}_\lambda,{\bar q'}_{\lambda'},v_\parallel,v_\perp)\ ,
\label{jaahp}
\end{equation}
where the function $h_{{\mbox{{\scriptsize P}}}}$ is given by
Eq.~(\ref{hPiInt}). As in the case of the charged pion,
in the standard gauges one can carry out explicit calculations that lead to
\begin{equation}
J^{\alpha\alpha'}_{\rho^+}(\bar q, \bar q') = \left(2\pi\right)^{4}
\delta^{(2)}(q_\parallel-q^\prime_\parallel)\, \delta_{\chi\chi'}
 \int \frac{d^2v_\perp}{(2\pi)^2} \sum_{\lambda,\,\lambda'}\;
 (\Upsilon^{\mu\alpha}_\lambda)^\ast
\, \Upsilon^{\nu\alpha'}_{\lambda'}\, \bar J_{\rho^+,\mu\nu}(q_\parallel,v_\perp)
\, f_{\elle_\lambda \elle'_{\lambda'}}(v_\perp) \ ,
\end{equation}
where $f_{kk'}(v_\perp)$ is given by Eq.~(\ref{fkkp}) and
$\delta_{\chi\chi'}$ stands for $\delta_{\imath\imath'}$,
$\delta(q^1-q^{\prime 1})$ and $\delta(q^2-q^{\prime 2})$ for SG, LG1 and
LG2, respectively.

To proceed further, one can carry out the calculation of $\bar
J_{\rho^+}^{\mu\nu}(v_\parallel,v_\perp)$ from Eq.~(\ref{barjpimunu}). As
expected form symmetry arguments, the explicit calculation shows that one
can write
\begin{eqnarray}
 \bar J_{\rho^+}^{\mu\nu}(v_\parallel,v_\perp) = \sum_{i=1}^7\, c_i(v_\parallel,v_\perp) \,
 \mathbb{O}^{\mu\nu}_{i}(v) \ ,
\label{barjpimunudos}
\end{eqnarray}
where $\mathbb{O}^{\mu\nu}_{i}(v)$ are the operators defined in
Eq.~(\ref{oprho}) and $c_i(v_\parallel,v_\perp)$ are scalar functions that
depend on $v_\parallel^2$ and $v_\perp^2$. Then one has
\begin{equation}
J^{\alpha\alpha'}_{\rho^+}(\bar q, \bar q') =  \left(2\pi\right)^{4}
\delta^{(2)}(q_\parallel-q^\prime_\parallel)\, \delta_{\chi\chi'}\,
\sum_{i=1}^7 \int \frac{d |\vec v_\perp|^2}{8\pi^2}\, c_i(q_\parallel,v_\perp) \,
Z^{\alpha\alpha'}_i(\elle,\elle',v_\perp^2)\ ,
\label{calJinterm}
\end{equation}
where
\begin{equation}
Z^{\alpha\alpha'}_i(\elle,\elle',v_\perp^2) =
\int_0^{2\pi} d\phi_\perp\,
\sum_{\lambda,\,\lambda'}\, (\Upsilon^{\mu\alpha'}_\lambda)^\ast\,
\Upsilon^{\nu\alpha}_{\lambda'} \ \mathbb{O}_{i,\mu\nu}(v) \,
f_{\elle_\lambda \elle'_{\lambda'}}(v_\perp)\ .
\end{equation}
By performing the above integral for each one of the operators
$\mathbb{O}_{i,\mu\nu}(v)$, it is seen that
$Z^{\alpha\alpha'}_i(\elle,\elle',v_\perp^2) \propto \delta_{\elle\elle'}$,
and consequently $J^{\alpha\alpha'}_{\rho^+}(\bar q,\bar q')$ can be
written as
\begin{equation}
J^{\alpha\alpha'}_{\rho^+}(\bar q,\bar q') =
\hat \delta_{\bar q \bar q'}\ {\hat J}^{\alpha\alpha'}_{\rho^+}(\elle,
q_\parallel)\ .
\label{polrhodiag}
\end{equation}
The expression for ${\hat J}^{\alpha\alpha'}_{\rho^+}(\elle, q_\parallel)$
can be obtained taking into account the Schwinger form of the translational
invariant part of quark propagators, see Eq.~(\ref{sfp_schw_a}). In this
way, for $k\geq 0$ we obtain
\begin{equation}
{\hat J}^{\alpha\alpha'}_{\rho^+}(\elle,q_\parallel) =
\sum_{i=1}^7 \ d_i(\elle,q_\parallel) \, \mathbb{O}^{\alpha\alpha'}_{i}(\Pi)\ ,
\label{boldj}
\end{equation}
where $\Pi^\mu(k,q_\parallel)$ is the four-vector defined in
Eq.~(\ref{pimu}). The explicit expressions of the functions
$d_i(\elle,q_\parallel)$ are given in App.~\ref{dis}. As one can see from
Eqs.~(\ref{efes}), for the particular case $k=-1$ (where $\Pi^\mu$ is not
defined), we get $d_2(-1,q_\parallel)=d_6(-1,q_\parallel)$, while the
remaining coefficients are zero. In this case one has ${\hat
J}^{\alpha\alpha'}_{\rho^+}(-1,q_\parallel) \propto
\mathbb{O}^{\alpha\alpha'}_{2}\! + \mathbb{O}^{\alpha\alpha'}_{6} = 2
\Upsilon_{-s}^{\alpha\alpha'}$.

{}From Eq.~(\ref{DLOC}), we see now that the one-loop correction to the
charged rho meson two-point correlator can be expressed as
\begin{equation}
D^{\rm (LOC)}_{\rho^+,\mu\nu}(y,y') = \sumint_{\bar q}\, \mathbb{R}_{\mu\alpha}^+(y,\bar q)\,
\hat D^{\rm (LOC)\,\alpha\alpha'}_{\rho^+}(k,q_\parallel)\,
{\mathbb{R}^+_{\nu\alpha'}(y',\bar q')}^\ast\ ,
\end{equation}
with
\begin{eqnarray}
\hat D^{\rm (LOC)\,\alpha\alpha'}_{\rho^+}(k,q_\parallel) =
\hat D_{\rho^+}^{\,\alpha\beta}(k,q_\parallel) \, \hat \Sigma_{{\rho^+},\beta\beta'}(\elle,q_\parallel)
\,\hat D^{\,\beta'\alpha}_{\rho^+}(k,q_\parallel)\ ,
\end{eqnarray}
$\hat \Sigma_{\rho^+}(\elle,q_\parallel)$ being the one-loop $\rho^+$ meson
self energy, related to the polarization function $\hat
J_{\rho^+}(\elle,q_\parallel)$ by
\begin{equation}
\hat \Sigma^{\alpha\alpha'}_{\rho^+}(\elle,q_\parallel) = -i g_v^2 \,
{\hat J}_{\rho^+}^{\alpha\alpha'}(\elle, q_\parallel)\ .
\label{sigmarho}
\end{equation}

For the description of physical $\rho^+$ meson states, it is also useful to
project ${\hat J}_{\rho^+}(\elle,q_\parallel)$ on the polarization
state basis. In this way, one can define a matrix ${\bf
J}^{cc'}_{\rho^+}(\elle, q_\parallel)$ given by
\begin{equation}
{\bf J}^{cc'}_{\rho^+}(\elle, q_\parallel) =
\epsilon_{+,\alpha}(\elle,q^3,c)^\ast \,  {\hat
J}_{\rho^+}^{\alpha\alpha'}(\elle, q_\parallel) \,
\epsilon_{+,\alpha'}(\elle,q^3,c')\ ,
\label{jrhoccprima}
\end{equation}
where $\epsilon^+_\alpha(\elle,q^3,c)$ are the polarization vectors
introduced in Eq.~(\ref{funrho}). In the case $\elle=-1$, i.e., the Lowest
Landau Level (LLL), only $c=1$ is allowed. One has
\begin{eqnarray}
{\hat {\bf J}}_{\rho^+}^{11}(-1,q_\parallel) & = &
- i\, \frac{N_c}{4\pi^2}\,\int_{-1}^1 dx \int_0^\infty dz
\ e^{-z \phi(x,q_\parallel^2)} \nonumber \\
&& \times \, \frac{(1+t_u)\,(1+t_d)}{\alpha_+}
\,\Big[ m_u m_d + \frac{1}{z} + \frac{1-x^2}{4}\,q_\parallel^2\Big]\ ,
\label{rhorhom1}
\end{eqnarray}
where $\phi(x,q_\parallel^2)$, $t_f$ and $\alpha^+$ have been defined in
Eqs.~(\ref{phi}), (\ref{tf}) and (\ref{alpha}), respectively. As in the case
of the charged pion [see Eq.~(\ref{jpiint}) for $\hat
J_{\pi^+}(\elle,q_\parallel)$], this expression is divergent and has to be
regularized. Again, this can be done by subtracting the $B=0$ contribution,
leaving a well-defined $B$-dependent piece. In addition,  the function $\hat
J_{\rho^+}(\elle,q_\parallel)$ can be analytically extended for large
positive values of $q_\parallel^2$.

\section{Magnetized charged pion and rho masses in the Nambu-Jona-Lasinio model}
\setcounter{equation}{0}

In this section we consider an extended NJL model in the presence of
external magnetic field. The corresponding lagrangian reads
\begin{equation}
\mathcal{L} = \bar{\psi}\left(i\ \rlap/\!{\cal D}-m_{0}\right)\psi + G_s
\left[\left(\bar{\psi}\psi\right)^{2}+\left(\bar{\psi}i\gamma_{5}\vec{\tau}\,\psi\right)^{2}\right]
 -G_{v}\left(\bar{\psi}\gamma_{\mu}\,\vec{\tau}\,\psi\right)^{2}\ ,
\label{NJL_Lagrangian}
\end{equation}
where $\psi (x)$ is the $u$-$d$ quark doublet defined in
Eq.~(\ref{q_doublet}) and ${\cal D}^\mu$ is the covariant derivative in
Eq.~(\ref{covdev}). Models like the one described by
Eq.~(\ref{NJL_Lagrangian}) have often been used to study the influence of an
external magnetic field on meson masses. In fact, the NJL model was
introduced more that 60 years ago for the description of spontaneous chiral
symmetry breaking and dynamical mass
generation~\cite{Nambu:1961tp,Nambu:1961fr}; then, during the late 80's and
earlier 90's, the approach was re-interpreted as an effective model for low
energy QCD~\cite{Vogl:1991qt,Klevansky:1992qe,Hatsuda:1994pi}. For a large
enough value of the coupling constant $G_s$, it is seen that the model
describes adequately the breakdown of chiral symmetry, and leads to a
phenomenologically reasonable value for the chiral quark-antiquark
condensate at the mean field level. In turn, this implies that the quarks
acquire an effective dynamical mass $M_f \approx 300-400\;{\rm MeV}\,\gg
m_0$. In the simple model given by Eq.~(\ref{NJL_Lagrangian}), it turns out
that $M_u=M_d$ even in the presence of an external magnetic field; however,
the magnetic field can break this degeneracy if more general flavor mixing
interactions are included (for details see e.g.\
Ref.~\cite{Carlomagno:2022inu}).

In the above framework, mesons can be described as quantum fluctuations in
the large $N_c$ approximation (which, in this context, is equivalent to the
well known random phase approximation); i.e., they can be introduced via the
summation of an infinite number of quark loops. Here we are particularly
interested in the masses of the charged pion (lightest charged meson in the
absence of the external magnetic field) and the charged rho meson.
Concerning the latter, we recall that there has been some discussion about
the possibility that the presence of a strong magnetic field may induce
$\rho^\pm$ condensation. Our interest here is not to perform a detailed
analysis of meson masses in the presence of the magnetic field but to study
the effect of Schwinger phases, showing how the results get modified if SPs
are neglected. Therefore, as done in Sec.~\ref{loc_correlators}, we study
here $\pi^\pm$ and $\rho^\pm$ masses separately. A full analysis, in which
$\pi^+-\rho^+$ mixing is explicitly considered, can be found in
Ref.~\cite{Carlomagno:2022arc}.

Let us first take $G_v=0$ in Eq.~(\ref{NJL_Lagrangian}) and concentrate just
on the charged pion mass. Following Ref.~\cite{Bijnens:1995ww} we introduce
the charged pseudoscalar currents
\begin{eqnarray}
j_+(x) = \sqrt 2\ \bar \psi_u(x)\, i\gamma_5\, \psi_d(x) \ , \qquad \qquad
j_-(x) = \sqrt 2\ \bar \psi_d(x)\, i\gamma_5\, \psi_u(x)\ .
\end{eqnarray}
Next, we define the two-point function $\Pi_{P^+}(x,x')$ as the two-point
correlator between these two currents. To zeroth order in $G_s$ we have
\begin{eqnarray}
\Pi^{(0)}_{P^+}(x,x') = \langle 0 | T[j_-(x) j_+(x')]  |0\rangle =
J_{\pi^+}(x,x')\ ,
\end{eqnarray}
where $J_{\pi^+}(x,x')$ is given by Eq.~(\ref{jxxpPi}). The full two-point
functions in the large $N_c$ approximation is obtained as
\begin{eqnarray}
\Pi_{P^+}(x,x') &=&  J_{\pi^+}(x,x') \, + \,  2 i G_s \int d^4 z \ J_{\pi^+}(x,z)\, J_{\pi^+}(z,x') \nonumber \\
            & & + \,  (2 i G_s)^2  \int d^4 z\, d^4 z' \ J_{\pi^+}(x,z)\, J_{\pi^+}(z,z')\, J_{\pi^+}(z',x')\, + \,
            \dots
\end{eqnarray}
Then from Eqs.~(\ref{jpolqspace}) and (\ref{orthfpion}) one readily gets
\begin{eqnarray}
\Pi_{P^+}(x,x') &=& \sumint_{\bar q} \ \ffpxbarq \ \hat
J_{\pi^+}(k,q_\parallel)\nonumber \\
& & \times \; \bigg\{ 1 + 2 i G \hat J_{\pi^+}(k,q_\parallel) +
\Big[ 2 i G \hat J_{\pi^+}(k,q_\parallel)\Big]^2 + \dots\bigg\}\;
\ffpxprimabarq^\ast
\nonumber \\
&=& \sumint_{\bar q} \ \ffpxbarq \ \hat J_{\pi^+}(k,q_\parallel)\,
K_{\pi^+}^{\; -1}\ \ffpxprimabarq^\ast \ ,
\label{pionSDeq}
\end{eqnarray}
where we have defined
\begin{equation}
K_{\pi^+}(k,q_\parallel) = 1 - 2 i G \hat J_{\pi^+}(k,q_\parallel)
\end{equation}
and $\hat J_{\pi^+}(k,q_\parallel)$ is the function given by
Eq.~(\ref{jpiint}), in which we have replaced the quark masses $m_f$ by the
dynamical masses $M_f$. Thus, one can obtain the $\pi^+$ pole mass for each
Landau level $k$ by solving the equation
\begin{equation}
K_{\pi^+}|_{q_\parallel^2=m^2_{\pi^+}} = 0\ .
\label{keq0pi}
\end{equation}
Charged pion masses have been determined for the first time in this way in
Refs.~\cite{Coppola:2018vkw, Coppola:2019uyr}. For the lowest Landau level
$k=0$ one obtains
\begin{eqnarray}
\hat J_{\pi^+}(0,q_\parallel) & = & -\frac{iN_{c}}{4\pi^{2}}\int_{-1}^{1}dx\int_{0}^{\infty}dz\;\frac{1}{\alpha_{+}}
\;e^{-z\phi(x,q_{\parallel}^2)}\,\nonumber \\
& & \times\; \bigg\{
\bigg[M_{u}\,M_{d}+\frac{1}{z}+(1-x^{2})\frac{q_{\parallel}^{2}}{4}\bigg]
(1-t_{u}\,t_{d})+\frac{1}{\alpha_{+}}(1-t_{u}^{2})(1-t_{d}^{2})\bigg\}\ . \ \
\label{jpik0}
\end{eqnarray}

In the derivation of Eq.~(\ref{pionSDeq}), it is worth paying attention to
Eqs.~(\ref{jpolqspace}) and (\ref{JpiDiagonal0}), which show that
$J_{\pi^{+}}\left(x,x'\right)$ is diagonal in the basis of eigenstates of
the Klein-Gordon operator in Eq.~(\ref{KGeq0}). In usual quantum field
theory, particle states are given at zero order in perturbation theory by
plane waves (i.e., they have definite four-momentum). In contrast, in our
case there is an external static and uniform magnetic field that plays the
role of a background; consequently, as discussed in Sec.~III, the zero order
charged particle states correspond to wavefunctions expressed in terms of
the functions $\mathcal{F}_{Q}(x,\bar{q})$.

Tracing back the derivation of Eq.~(\ref{pionSDeq}), we can see what happens
if the SP is neglected. The diagonal condition in Eq.~(\ref{JpiDiagonal0})
arises in fact from Eqs.~(\ref{jpibarqq}) and (\ref{hPiInt}); if one intends
to make an approximation in which the SP in Eq.~(\ref{hPiInt}) is removed,
one should also replace the wavefunctions by plane waves, ${\cal
F}_{Q_{\mbox{{\scriptsize P}}}}(x,\bar q)\to \exp(-iqx)$, in order to
guarantee translational invariance. Thus, we denote this procedure by
``plane wave approximation'' (PWA). Within this approximation, the
two-interacting quark state
---or the pion, in the context discussed in Sec.~\ref{loc_correlators}--- is
no longer specified by the set of quantum numbers
$\bar{q}=(q^{0},\,k,\,\chi,\,q^{3})$ but by the four-momentum
$q^\mu=(q^{0},\,q^{1},\,q^{2},\,q^{3})$. In this way one obtains
\begin{equation}
h_{{\mbox{{\scriptsize P}}}}^{\text{PWA}}(q,q',v) =
(2\pi)^{8}\,\delta^{(4)}(q-q')\;\delta^{(4)}(q-v)\ ,
\label{hppwa}
\end{equation}
losing the effect of the magnetic field in this part of the calculation.
After a trivial integration over $v$, according to Eq.~(\ref{jpibarqq}) one
gets
\begin{equation}
J_{\pi^{+}}^{\text{PWA}}(q,q') =
(2\pi)^{4}\,\delta^{(4)}(q-q')\,\hat{J}_{\pi^{+}}^{\text{PWA}}(q_\parallel,q_\perp)
\ ,
\label{jpipwa}
\end{equation}
where
\begin{equation}
\hat{J}_{\pi^{+}}^{\text{PWA}}(q_\parallel,q_\perp) =
\bar{J}_{\pi^{+}}(q_{\parallel},q_{\perp})\ ,
\end{equation}
with $\bar{J}_{\pi^{+}}(q_{\parallel},q_{\perp})$ given by Eq.~(\ref{jpi})
(notice that the calculation of
$\bar{J}_{\pi^{+}}(q_{\parallel},q_{\perp})$ only involves the translational
invariant part of quark propagators, hence it is not affected by Schwinger
phases).

We notice that, in some sense, the result in Eq.~(\ref{jpipwa}) can be
misleading. Given that the magnetic field is assumed to be uniform, one
would expect the system to be invariant under translations in space-time,
and this seems to be confirmed by the conservation of four-momentum arising
from Eq.~(\ref{jpipwa}). Nevertheless, in the presence of the magnetic field
it is found that translational invariance (in the plane perpendicular to
$\vec B$) is realized in a nontrivial way, related to gauge transformations.
We refer here to Sec.~\ref{section:SP}, in which this issue has been
discussed in detail.

{}From the above results it is easy to see that within the PWA one can
define a $\pi^+$ pole mass (taking $q_\perp=0$) as the solution of the
equation
\begin{equation}
K_{\pi^+}^{\text{PWA}}|_{q_\parallel^2=m^2_{\pi^+}} = 1 - 2 i\,  G
\hat{J}_{\pi^{+}}^{\text{PWA}}(q_\parallel,0)|_{q_\parallel^2=m^2_{\pi^+}} =
0 \ ,
\label{keq0pipwa}
\end{equation}
where, according to Eq.~(\ref{jpi}),
\begin{eqnarray}
\hat{J}_{\pi^{+}}^{\text{PWA}}(q_\parallel,0) & = &
-\frac{iN_{c}}{4\pi^{2}}\int_{-1}^{1}dx\int_{0}^{\infty}dz\;\frac{1}{t_{+}}
\;e^{-z\phi(x,q_{\parallel}^2)}\,\nonumber \\
& & \times\; \bigg\{
\bigg[M_{u}\,M_{d}+\frac{1}{z}+(1-x^{2})\frac{q_{\parallel}^{2}}{4}\bigg]
(1-t_{u}\,t_{d})+\frac{1}{t_{+}}(1-t_{u}^{2})(1-t_{d}^{2})\bigg\}\ .\ \ \ \
\ \
\label{jpiPWA}
\end{eqnarray}
Comparing Eq.~(\ref{jpiPWA}) with our result in Eq.~(\ref{jpik0}), it is
seen that the PWA expression can be obtained from the full result by the
replacement $\alpha_+ \to t_+$ in the integrand. We also note that
Eq.~(\ref{jpiPWA}) is consistent with Eq.~(80) of Ref.~\cite{Li:2020hlp},
where an alternative method has been used to evaluate the effects of the
magnetic field on charged pion masses. The difference between
Eqs.~(\ref{jpiPWA}) and (\ref{jpik0}) shows that the approach in
Ref.~\cite{Li:2020hlp} does not fully take into account the effects arising
from the presence of the Schwinger phase.

An important point to be stressed is the fact that within the PWA the
two-quark state has zero total transverse momentum. One can see, however,
that this cannot be possible: the two-quark state, as a whole, has to behave
as a charged bound system immerse in a magnetic field, whose quantum ground
state ---which must have some nonvanishing zero-point energy--- cannot be
described by a particle at rest. In fact, the charged meson state cannot
have any definite momentum in the plane perpendicular to the magnetic field.
The situation can be better understood by looking at Eq.~(\ref{sigmapibar}),
which shows that our result for $\hat J_{\pi^+}(k,q_\parallel)$ arises from
the convolution of $\bar{J}_{\pi^{+}}(q_{\parallel},v_{\perp})$ with the
function $\rho_{k}(\vec v_{\perp}^{\,2})$ given in Eq.~(\ref{vPerpDis}). In
fact, this function describes the total transverse momentum distribution due
to the vibration of the two-quark quantum state in the presence of the
external magnetic field. Notice that, expressed in this way, the plane wave
approximation would correspond to a distribution $\rho_{k}^{\text{PWA}}(\vec
v_{\perp}^{\,2})= \delta(|\vec v_{\perp}|^{2} - |\vec q_\perp|^2)$.

Let us consider now the rho meson sector. As mentioned above, for simplicity
we analyze the situation in which the $\rho^+-\pi^+$ mixing is neglected.
The study of the $\rho^+$ meson in this simplified scenario can be performed
by eliminating the pseudoscalar-pseudoscalar coupling
$(\bar{\psi}i\gamma_{5}\vec{\tau}\,\psi)^{2}$ in
Eq.~(\ref{NJL_Lagrangian}). To proceed we introduce the charged vector
currents
\begin{eqnarray}
j^\mu_+(x) = \sqrt 2\, \bar \psi_u(x) \gamma^\mu \psi_d(x)\ , \qquad\qquad
j^\mu_-(x) = \sqrt 2\, \bar \psi_d(x) \gamma^\mu \psi_u(x)\ ,
\end{eqnarray}
and define the two-point function $\Pi^{\mu\nu}_{V^+}(x,x')$ as the
two-point correlator between both currents. To zero order in $G_v$ we have
\begin{eqnarray}
\Pi^{(0)\,\mu\nu}_{V^+}(x,x') = \langle 0 |\, T\,[j^\mu_-(x) j^\nu_+(x')] \, |0\rangle
= J^{\mu\nu}_{\rho^+}(x,x')\ ,
\end{eqnarray}
where $J^{\mu\nu}_{\rho^+}(x,x')$ is given by Eq.~(\ref{jxxpRho}). Now, as
in the case of the $\pi^+$, we can evaluate the full vector two-point
function in the large $N_{c}$ approximation,
\begin{align}
\Pi_{V^{+}}^{\mu\nu}(x,x')  = & J_{\rho^{+}}^{\mu\nu}(x,x')+ (-2iG_{V})
\int d^{4}z\ J_{\rho^{+}}^{\mu\alpha}(x,z)\,g_{\alpha\beta}\,J_{\rho^{+}}^{\beta\nu}(z,x')
\nonumber \\
 & +(-2iG_{V})^{2} \int d^{4}z\;d^{4}z'\ J_{\rho^{+}}^{\mu\alpha}(x,z)\,g_{\alpha\alpha'}\,
 J_{\rho^{+}}^{\alpha'\beta'}(z,z')\,g_{\beta'\beta}\,J_{\rho^{+}}^{\beta\nu}(z',x')\, +\, \dots
\end{align}
Using Eq.~(\ref{polrhodiagxx1}) together with Eqs.~(\ref{orthormunu})
and (\ref{polrhodiag}), and resumming the loop contributions, we obtain
\begin{equation}
\Pi_{V^{+}}^{\mu\nu}(x,x')  = \sumint_{\bar{q},\,\bar q'} \, \mathbb{R}^{+,\mu\alpha}(x,\bar{q})\,
{\hat J}_{\rho^+,\alpha\alpha'}(\elle,q_\parallel)\,
(K_{\rho^+}^{-1})^{\alpha'}_{\ \beta}\,{\mathbb{R}^{+,\nu\beta}(x',\bar{q})}^{\ast}\ ,
\end{equation}
where
\begin{equation}
K_{\rho^+}^{\alpha\beta}(\elle,q_\parallel) = g^{\alpha\beta} + 2iG_V\,
{\hat J}_{\rho^+}^{\alpha\beta}(\elle,q_\parallel)\ .
\end{equation}
In this way, taking $q^3=0$, the charged rho pole masses $m_{\rho}$ can be
obtained for each Landau level by solving the equation
\begin{equation}
\det\, K_{\rho^+} = 0
\label{rhoMassNJL}
\end{equation}
for $q_\parallel^\mu = (E_\rho,0,0,0)$, $E_\rho^2 =
m_{\rho}^2+(2k+1)B_\rho$.

In the limit $B\to 0$, it can be seen from the coefficients
$d_i(k,q_\parallel)$ in App.~\ref{dis} that ${\hat
J}_{\rho^+}^{\alpha\beta}(\elle,q_\parallel)$ can be written in terms of
$\mathbb{O}_{1}^{\alpha\beta}+\mathbb{O}_{2}^{\alpha\beta} =
g^{\alpha\beta}$ and
$\mathbb{O}_{3}^{\alpha\beta}(\Pi)+\mathbb{O}_{4}^{\alpha\beta}(\Pi) +
\mathbb{O}_{5}^{\alpha\beta}(\Pi) = \Pi^\alpha{\Pi^\beta}^\ast$, where
$\Pi^{\alpha} \to (q^0,0,0,q^3)$. In this limit the $\rho^+$ meson can be
taken to be at rest, and one gets three degenerate masses that correspond to
the $\rho^+$ polarization states. On the other hand, for nonzero $B$ the
mass states depend on the value of $k$. For the lowest Landau level $k=-1$,
which corresponds to the lightest charged $\rho$ state, from the results in
App.~\ref{dis} it is seen that $d_2(-1,q_\parallel) =d_6(-1,q_\parallel)$,
while the remaining coefficients $d_i(-1,q_\parallel)$ are zero. In
addition, according to the definitions in App.~\ref{Spin-1-Pola_Propa}, one
has
\begin{eqnarray}
-\,\frac{1}{2}\,\Big(
\mathbb{O}_{2}^{\alpha\beta}+\mathbb{O}_{6}^{\alpha\beta} \Big) & = &
-\,\Upsilon^{\alpha\beta}_{-s} \nonumber \\
& = & \epsilon^{\alpha}_{+}(-1, q^3 ,1)\,
\epsilon^{\beta}_{+}(-1, q^3 ,1)^\ast\ ,
\end{eqnarray}
so we can write
\begin{equation}
{\hat J}_{\rho^+}^{\alpha\beta}(-1,q_\parallel) =
-2\,d_2(-1,q_\parallel) \; \epsilon^{\alpha}_{+}(-1, q^3 ,1)\,
\epsilon^{\beta}_{+}(-1, q^3 ,1)^\ast\ .
\label{coefd2}
\end{equation}
Hence, it is found that for the lowest Landau level there is only one mass
eigenstate, which corresponds to $c=1$ in the polarization state basis. From
the expressions in App.~\ref{dis} the coefficient in the right hand side of
Eq.~(\ref{coefd2}) is given by
\begin{eqnarray}
-2\,d_2(-1,q_\parallel) & = &
-i\, \frac{N_c}{4\pi^2}\,\int_{-1}^1 dx \int_0^\infty dz
\ e^{-z \phi(x,q_\parallel^2)}\, \frac{(1+t_u)\,(1+t_d)}{\alpha_+}\nonumber \\
& & \times\,\Big[ M_u M_d + \frac{1}{z} +
\frac{1-x^2}{4}\,q_\parallel^2\Big] \nonumber \\
& = & {\hat {\bf J}}_{\rho^+}^{11}(-1,q_\parallel)\ ,
\label{j11full}
\end{eqnarray}
consistently with the result found in Sec.~\ref{onelooprho}, see
Eq.~(\ref{rhorhom1}) (here the quark masses have been replaced by the
effective masses $M_u$ and $M_d$). We see that the $\rho^+$ state is in this
case also an eigenstate of the spin operator $S_3$, with eigenvalue $s_3 = s
= {\rm sg}(Q_\rho B)$. From Eq.~(\ref{rhoMassNJL}), the mass of this state
can be obtained as the solution of
\begin{equation}
1-2iG_V\,{\hat {\bf J}}_{\rho^+}^{11}(-1,q_\parallel) = 0\ .
\label{j11eq0rho}
\end{equation}

As in the charged pion case, it is interesting to see how the results get
modified if the plane wave approximation is used. As stated, in the PWA the
Schwinger phase $\Phi_{\rho^+}(x,x')$ is neglected and one should replace
$\mathcal{F}_{Q_{P}}(x,\bar{q})\to \exp(-iqx)$, which leads to
$\mathbb{R}^{{\scriptscriptstyle{{\cal Q}}},\mu\nu} = \exp(-i q x)
\sum_\lambda \Upsilon^{\mu\nu}_\lambda = \exp(-i q x)\, g^{\mu\nu}$. In this
way, from Eqs.~(\ref{polrhodiagxx1}) and (\ref{jpixx}) one gets
\begin{eqnarray}
J_{\rho^{+}}^{\text{PWA},\mu\nu}(x,x') & = &
\int \frac{d^4q}{(2\pi)^4}\,\frac{d^4q'}{(2\pi)^4}\, e^{-iqx}\,
g^{\mu\alpha}\, J^{\text{PWA}}_{\rho^{+},\alpha\alpha'}(q,q')\,
e^{iq'x'}\, g^{\nu\alpha'}
\nonumber \\
& = & \int \frac{d^4v}{(2\pi)^4}\, e^{-i v (x- x')}\, \bar
  J_{\rho^+}^{\mu\nu}(v_\parallel,v_\perp)\ ,
\end{eqnarray}
and consequently
\begin{equation}
J^{\text{PWA},\alpha\alpha'}_{\rho^{+}}(q,q') =
(2\pi)^4 \delta^{(4)}(q-q')\,
\bar J_{\rho^+}^{\alpha\alpha'}(q_\parallel,q_\perp)\ ,
\label{jqqpwa}
\end{equation}
where $\bar J_{\rho^+}^{\alpha\alpha'}(q_\parallel,q_\perp)$ is the quark
loop function given by Eq.~(\ref{barjpimunu}). Notice that
Eq.~(\ref{jqqpwa}) can also be obtained from Eq.~(\ref{jaahp}), taking into
account the PWA result in Eq.~(\ref{hppwa}).

Within the PWA approximation the lowest energy $\rho^+$ states correspond to
the situation in which the meson is at rest. It is easy to see that the pole
masses for the different polarization states can be obtained as the
solutions of
\begin{equation}
\det\, K_{\rho^+}^{\text{PWA}} = 0 \ ,
\end{equation}
where
\begin{equation}
K_{\rho^+}^{\text{PWA},\alpha\beta} = g^{\alpha\beta} + 2iG_V\,
{\bar J}_{\rho^+}^{\alpha\beta}(q_\parallel,0)\ ,
\end{equation}
with $q^3 = 0$. We can compare the PWA result with the full result for the
lowest Landau level $k=-1$ by taking the projection of ${\bar
J}_{\rho^+}^{\alpha\beta}(q_\parallel,0)$ onto the polarization state
$s_3=s$, i.e., taking the piece of ${\bar
J}_{\rho^+}^{\alpha\beta}(q_\parallel,0)$ proportional to
$-\Upsilon^{\alpha\beta}_{-s}$. The explicit calculation of the quark loop
leads to the equation
\begin{equation}
1-2iG_V\,{\hat {\bf J}}_{\rho^+}^{11,\text{PWA}}(q_\parallel,0) = 0\ ,
\label{j11eq0pwa}
\end{equation}
where
\begin{eqnarray}
{\hat {\bf J}}_{\rho^+}^{11,\text{PWA}}(q_\parallel,0) & = &
 -i\, \frac{N_c}{4\pi^2}\,\int_{-1}^1 dx \int_0^\infty dz
\ e^{-z \phi(x,q_\parallel^2)}\, \frac{(1+t_u)\,(1+t_d)}{t_+}\nonumber \\
& & \times\,\Big[ M_u M_d + \frac{1}{z} +
\frac{1-x^2}{4}\,q_\parallel^2\Big]\ .
\label{J11pwa}
\end{eqnarray}
Hence, in the same way as in the case of the charged pion, the mass of the
lowest $\rho^+$ state within the PWA can be obtained from the full result in
Eq.~(\ref{j11full}) by replacing the factor $1/\alpha_+$ by $1/t_+$ in the
integrand. It can be seen that the expressions in
Eqs.~(\ref{j11eq0pwa}-\ref{J11pwa}) are consistent with Eq.~(24) of
Ref.~\cite{Cao:2019res}, which shows that the method used in that reference
turns out to be equivalent to the PWA.

To complete this section, we find it worth to estimate the importance of
taking into account Schwinger phases in the calculation of charged meson
properties as functions of the magnetic field. Therefore, in what follows we
analyze the $B$ dependence of $\pi^+$ and $\rho^+$ masses, comparing the
results obtained from Eqs.~(\ref{keq0pi}) and (\ref{j11eq0rho}) with those
found within the plane wave approximation, i.e.\ those obtained from
Eqs.~(\ref{keq0pipwa}) and (\ref{j11eq0pwa}).

We recall that the above expressions for the quark loop integrals are
divergent and have to be regularized. Here, as done e.g.\ in
Refs.~\cite{Cao:2019res,Li:2020hlp,Carlomagno:2022arc}, we use the so-called
magnetic field independent regularization (MFIR), in which we subtract from
the integrals the corresponding expressions in the $B \rightarrow 0$ limit,
and then we add them in a regularized form. In fact, as noticed in
Ref.~\cite{Cao:2019res}, to properly regularize the function ${\hat {\bf
J}}_{\rho^+}^{11,\text{PWA}}(q_\parallel,0)$ in Eq.~(\ref{J11pwa}) it is
necessary to introduce a modification of the method, considering not only
the $B\to 0$ limit but also a linear term in $B$. To perform the numerical
calculations, for definiteness we choose here the same set of model
parameters as in Ref.~\cite{Carlomagno:2022arc}, viz.\ $m_0 = 5.833$~MeV,
$\Lambda = 587.9$~MeV and $G_s\Lambda^2 = 2.44$, where $\Lambda$ is a 3D
cutoff parameter that is introduced to regularize the ultraviolet divergent
quark loops in the $B=0$ limit. For vanishing external field, this
parametrization leads to an effective quark mass $M_0 = 400\ \mbox{MeV}$ and
a quark-antiquark condensate $\langle\psi_f \bar \psi_f\rangle_0 = -(241 \
\mbox{MeV})^3$; in addition, one obtains the empirical values of the pion
mass and decay constant in vacuum, namely $m_{\pi,0} = 138$~MeV and
$f_{\pi,0} = 92.4$~MeV. Regarding the vector couplings, we take $G_v
\Lambda^2 = 2.651$, which leads to $m_{\rho,0} = 770$~MeV for $B = 0$. It is
worth mentioning that we have checked that our results remain basically
unchanged if one uses other standard parameters, like e.g.~those considered
in Refs.~\cite{Cao:2019res} and~\cite{Li:2020hlp}.

\begin{figure}[hbt]
\includegraphics[width=0.85\textwidth]{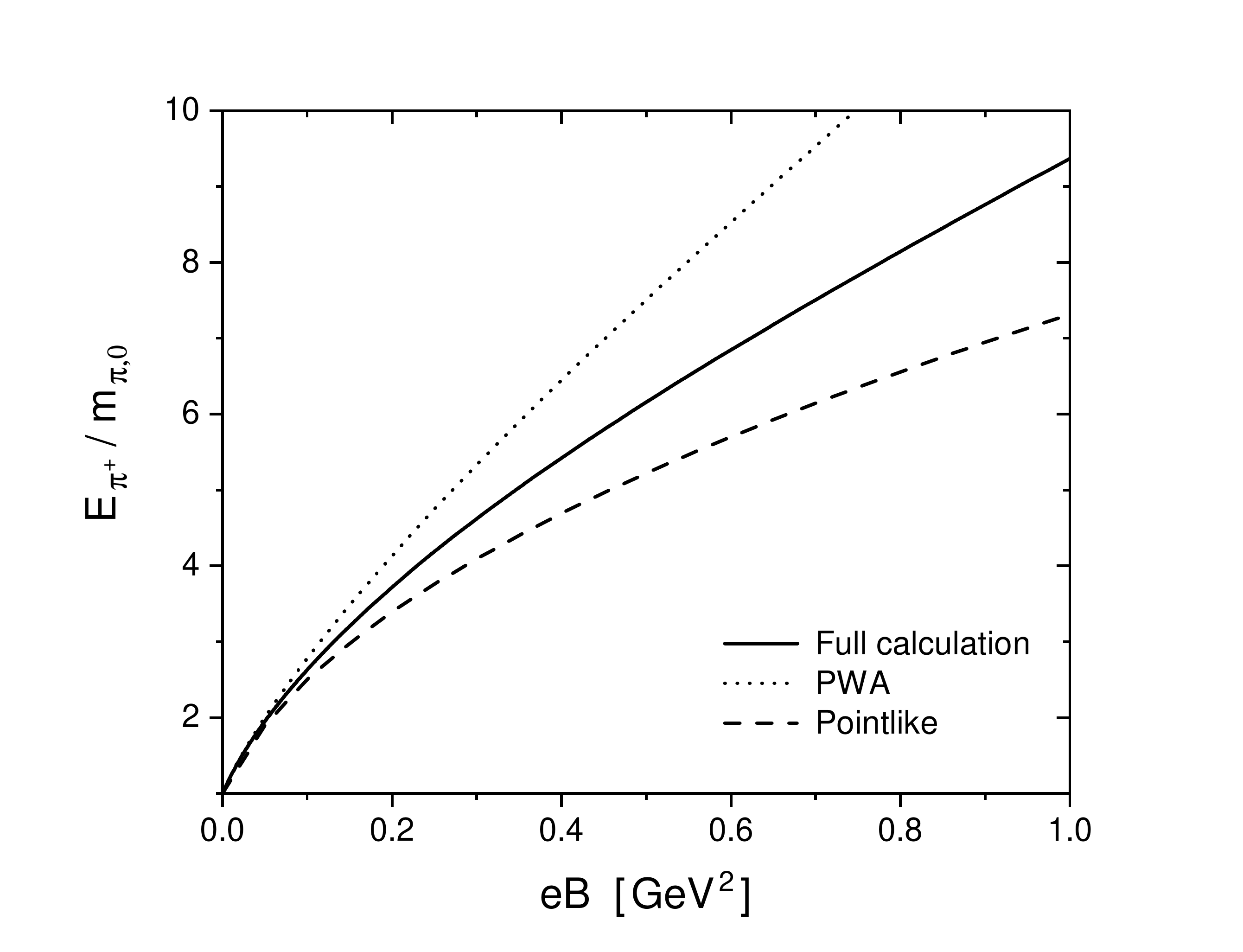}
\caption{Ratio $E_{\pi^+}/m_{\pi,0}$ as a function of $eB$. Here
$E_{\pi^+}$ stands for the energy of the lowest $\pi^+$ state (corresponding
to the Landau level $k=0$), while $m_{\pi,0}$ is the charged pion mass at
vanishing external magnetic field.}
\label{fig1}
\end{figure}

\begin{figure}[hbt]
\includegraphics[width=0.85\textwidth]{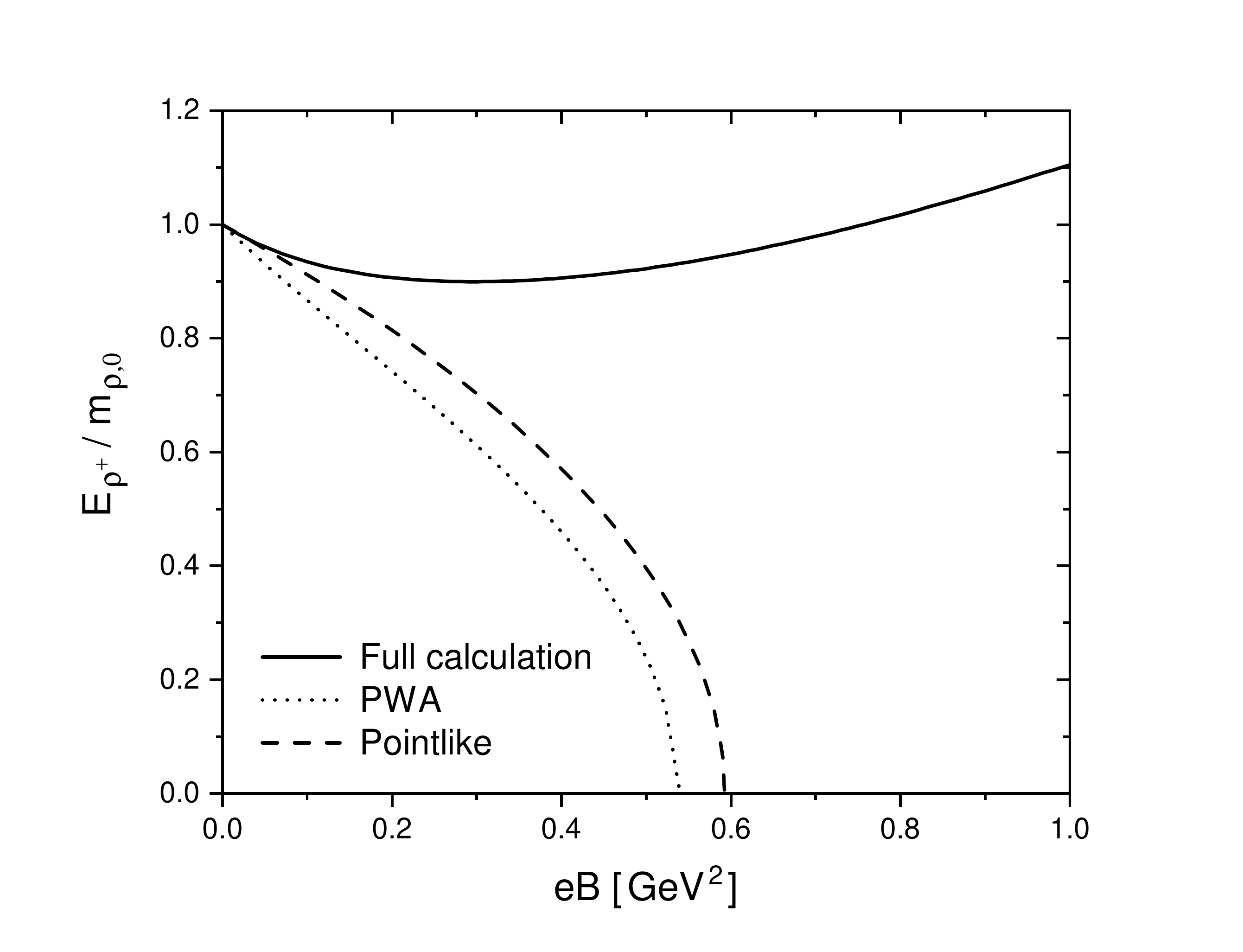}
\caption{Ratio $E_{\rho^+}/m_{\rho,0}$ as a function of $eB$. Here
$E_{\rho^+}$ stands for the energy of the lowest $\rho^+$ state
(corresponding to the Landau level $k=-1$), while $m_{\rho,0}$ is the
charged rho mass at vanishing external magnetic field.}
\label{fig2}
\end{figure}

In Fig.~\ref{fig1} and Fig.~\ref{fig2} we display our numerical results for
the charged pion and charged rho mesons, respectively. The curves show the
values of the ratio $E_P/m_{P,0}$, where
$P=\pi^+,\rho^+$, as functions of $eB$. Here $m_{P,0}$ is the
particle mass at $B=0$, while $E_P$ stands for the energy of
the $P$ meson in its lowest state, i.e.\ $E_{\pi^+} = \sqrt{m_{\pi^+}^2 +
B_{\pi}}$ and $E_{\rho^+} = \sqrt{m_{\rho^+}^2 - B_{\rho}}$, where
$m_P$ is the meson mass for nonzero $B$. We stress that to
determine the mass of the lowest energy state from our full calculation one
has to take $q^3=0$ and $k=0$ ($k=-1$) for the pion (rho meson), whereas
within the PWA one has to take $\vec q=0$.

{}From Fig.~\ref{fig1} it is seen that, for the whole considered range of
values of $eB$, the PWA leads to values of the ratio $E_{\pi^+}/m_{\pi,0}$
that are larger than those obtained from the full calculation, in which the
SP is properly taken into account. In turn, the latter are larger than those
obtained within the ``pointlike approximation'' (PLA), in which the meson is
considered as a particle with no internal structure (in this pointlike
limit, one has $E^{\rm PLA}_{\pi^+} = \sqrt{m_{\pi,0}^2 + e B}$). On the
other hand, the results can be compared with the values arising from LQCD
calculations~\cite{Bali:2017ian,Ding:2020hxw}. These are found to be close
or even lower than those corresponding to the PLA, which implies that the
proper treatment of Schwinger phases improves the agreement between LQCD
results and NJL model predictions for the dependence of $E_{\pi^+}$ with the
magnetic field. It is also worth mentioning that, as shown in
Ref.~\cite{Carlomagno:2022arc}, $\rho^+$$-$$\pi^+$ mixing effects (which
have been neglected in the calculations shown in Fig.~\ref{fig1}) tend to
bring NJL results even closer to LQCD values.

Now, as can be seen from Fig.~\ref{fig2}, the effect of taking into account
the SP is even more striking in the case of the charged rho meson energy.
Indeed, the results from PWA and PLA approximations (dotted and dashed lines
in the figure) seem to indicate that $E_{\rho^+}$ vanishes at some critical
magnetic field
---driving in this way a possible $\rho^+$ meson condensation---, while this
is not what comes out from the full calculation, in which the SP is properly
included (full line in Fig.~\ref{fig2}). Regarding LQCD calculations, in
this case it is found~\cite{Bali:2017ian,Hidaka:2012mz,Andreichikov:2016ayj}
that the ratio $E_{\rho^+}/m_{\rho,0}$ shows some decrease for low values of
$eB$, while for $eB \agt 0.7$~GeV$^2$ it tends to stabilize at a value of
about 0.7; hence, no $\rho^+$ meson condensation is expected from these
results. In fact, values consistent with the behavior predicted by LQCD can
be obtained within the NJL model ---taking into account the effect of the
SP--- by considering $B$-dependent couplings~\cite{Carlomagno:2022arc}. We
recall that these results correspond to the Landau level $k=-1$, for which
there is no mixing between the $\rho^+$ and the charged pion.

\section{Conclusions}

In this paper we have studied the role of the Schwinger phases appearing in
the propagators of charged particles in the presence of a static and uniform
magnetic field $\vec{B}$. These propagators are not gauge invariant objects;
if one performs a gauge transformation, they transform in a well defined
covariant way. In fact, it is seen that the non-invariance can be isolated
in a Schwinger phase $\Phi_P$, in such a way that the propagator can be
written as $\exp(i\Phi_P)$ times a gauge invariant function. As a first
result we have shown that the SP cannot be removed by a gauge
transformation; far from this, we have seen that it plays an important role
in the restoration of the symmetries of the system.

The presence of a static and uniform magnetic field does not alter the
homogeneity of space-time, although it does break space isotropy. Still,
isotropy is preserved in transverse directions, taking the direction of
$\vec{B}$ as a symmetry axis. Therefore, the studied system has to be
invariant under translations, under rotations around the direction of
$\vec{B}$ and under boost transformations in this direction. As a
consequence of the existence of this set of symmetries, for any Lorentz
tensor one should distinguish between ``longitudinal'' components (time
components and spatial components in the direction of $\vec{B}$) and
``perpendicular components'' (spatial components in the direction
perpendicular to $\vec{B}$), which in general will show different behaviors.

The equations that describe the dynamics of a charged particle in a static
and uniform magnetic field involve the electromagnetic field $A^{\mu}$. Even
if one assumes that the physical system has the above stated symmetries, it
has to be taken into account that both the homogeneity and the transverse
isotropy of space become broken when one chooses a specific gauge to set
$A^{\mu}$. Looking at the propagators of charged particles, we have shown
that this breakdown manifests itself in the SP, whereas the part of the
propagators that is invariant under gauge transformations is found to be
also invariant under translations and rotations around the direction of
$\vec{B}$. Additionally, we have seen that a translation in a direction
perpendicular to $\vec{B}$, as well as a rotation around the direction of
$\vec{B}$, are equivalent to gauge transformations. Explicit expressions
have been given for some gauges that are usually considered in the
literature, namely the symmetric gauge and the Landau gauges 1 and 2.

As an application to particular physical quantities, we have analyzed the
effect of the SP in the one-loop corrections to charged pion and rho meson
selfenergies. To carry out this analysis we have firstly considered standard
meson-quark interactions, and then we have studied the $\pi^+$ and $\rho^+$
propagators within the Nambu-Jona-Lasinio model, performing a numerical
analysis of the $B$ dependence of meson lowest energy states. For both
$\pi^+$ and $\rho^+$ mesons (for simplicity, $\pi^+$$-$$\rho^+$ mixing has
not been considered), we have compared the numerical results arising from
the full calculation
---in which SPs are included in the propagators, and meson wavefunctions
correspond to states of definite Landau quantum number--- and those obtained
within a Plane Wave Approximation (PWA) ---in which SPs are neglected (or
simply eliminated) and meson states are described by plane waves of definite
four-momentum.

In the case of the pion, from our analysis it is seen that
the polarization function is diagonal in the basis of $\pi^+$ eigenfunctions
${\cal F}_+(x,\bar q)$, and can be written as a convolution of a gauge
invariant function $\bar{J}_{\pi^{+}}(v_{\parallel},v_{\perp})$
---calculated from the gauge invariant part of the polarization function,
after a transformation to momentum space--- with a function
$h_{\pi^+}(\bar{q},\bar{q}',v_{\parallel},v_{\perp})$ given by a projection
of these eigenfunctions onto plane waves, modulated by the SP [see
Eqs.~(\ref{jpibarqq}-\ref{hPiInt})]. Moreover, after some integration we
have found that the polarization function can be written as an integral of
the function $\bar{J}_{\pi^{+}}(q_{\parallel},v_{\perp})$ over the
perpendicular momentum $v_\perp$, weighted by a given function
$\rho_{k}(\vec v_{\perp}^{\;2})$ [see Eq.~(\ref{sigmapibar})]; i.e., the
polarization function depends on definite values of the energy $q^0$ and the
parallel momentum $q^3$ of the two-quark system as a whole, while the
perpendicular momentum has no definite value but some distribution. This is
what one would expect for a charged particle, which must have some
zero-point energy when it is submerged in a magnetic field. In contrast,
within the PWA the polarization function can be transformed to four-momentum
space as $\bar{J}_{\pi^{+}}(q_{\parallel},q_{\perp})$, where $q_{\perp}$
would be the perpendicular momentum of the two-quark system (the pion, in
the case of the NJL model). Formally this would correspond to take
$\rho_{k}(\vec v_{\perp}^{\;2})=\delta(|\vec v_{\perp}|^{2}-|\vec
q_{\perp}|^{2})$, although the perpendicular momentum $\vec q_\perp$ is not
a well defined quantity for a charged particle in a magnetic field.
Alternatively, the effect of the SP on the one-loop correction to the
$\pi^+$ propagator can be seen from Eq.~(\ref{PropLOCAltern1}), where once
again our result shows the fluctuations of the perpendicular momenta of the
coupled two-quark system. These fluctuations are due to the presence of a
gauge invariant phase $\varphi$, which arises from a combination of
Schwinger phases along a closed path [see Eq.~(\ref{SPsum})]; if this phase
is eliminated, the effect of the fluctuations gets lost, as shown in
Eq.~(\ref{DlocPWA}).

It is worth to emphasize that even though the function
$h_{\pi^+}(\bar{q},\bar{q}',v_{\parallel},v_{\perp})$ involves several gauge
dependent quantities, its explicit expression, given by
Eqs.~(\ref{hPiGauge0}-\ref{fkkp}), is itself gauge independent. This has
been checked by performing the corresponding calculations in the three
standard gauges mentioned above. In this way, it has been shown that the
inclusion of the SP allows us to carry out a full calculation of the
polarization function, using the proper wavefunctions of charged particles
and preserving both the invariance under gauge transformations and the
symmetries under translations and rotations around the direction of
$\vec{B}$.

The above qualitative discussion applies also to the case of the $\rho^+$
meson propagator, although the explicit expressions are more involved due to
the more complex Lorentz structure. It is worth mentioning that we have
introduced in this case a description of spin one charged fields in the
presence of the magnetic field. As proposed by Ritus for the case of spin
1/2 fermion fields~\cite{Ritus:1978cj}, in our formalism we have separated
the meson wavefunctions as a product of a tensor that carries the spatial
coordinates and a polarization vector. Then the explicit expression for the
meson propagator and the corresponding one-loop correction have been
obtained.

Finally, as mentioned above, we have carried out a numerical analysis of the
$B$ dependence of $\pi^+$ and $\rho^+$ masses (and lowest state energies)
within the NJL model. Using a three-momentum cutoff and a so-called magnetic
field independent
regularization~\cite{Cao:2019res,Li:2020hlp,Carlomagno:2022arc}, we have
found that our full calculation leads to a result for the $B$ dependence of
the charged pion mass that clearly improves the agreement with LQCD results,
in comparison with the one obtained using the PWA. Moreover, there is still
room for further improvement, e.g.\ by considering $\rho^{+}$$-$$\pi^{+}$
mixing as done in Ref.~\cite{Carlomagno:2022arc}. Concerning the charged rho
meson, we have found a qualitative difference between our results and those
obtained within the PWA. Indeed, it has been shown that if the presence of
the SP is properly taken into account, the numerical results are not
compatible with the phenomenon of $\rho^{+}$ condensation for any value of
the magnetic field. This is in the same line of LQCD
results~\cite{Bali:2017ian,Hidaka:2012mz,Andreichikov:2016ayj}, which
indicate that the value of the energy of the lowest $\rho^+$ state tends to
stabilize at $E_{\rho^{+}}/m_{\rho,0}\sim 0.7$ for $eB>0.8$~GeV$^2$. Let us
recall that this state corresponds to the Landau level $k=-1$, which does not
mix with the pion. We have also checked that our numerical results do not
suffer significant changes if one uses other standard model parameters, like
e.g.\ those considered in Refs.~\cite{Cao:2019res,Li:2020hlp}.

\section*{Acknowledments}

NNS would like to thank the Department of Theoretical Physics of the
University of Valencia, where part of this work has been carried out, for
their hospitality within the Visiting Professor program of the University of
Valencia. This work has been partially funded by CONICET (Argentina) under
Grant No.\ PIP17-700, by ANPCyT (Argentina) under Grants No.\ PICT17-03-0571
and PICT20-01847, by the National University of La Plata (Argentina),
Project No.\ X824, by Ministerio de Ciencia e Innovaci\'on and Agencia
Estatal de Investigaci\'on (Spain), and European Regional Development Fund
Grant PID2019-105439 GB-C21, by EU Horizon 2020 Grant No.\ 824093
(STRONG-2020), and by Conselleria de Innovaci\'on, Universidades, Ciencia y
Sociedad Digital, Generalitat Valenciana GVA PROMETEO/2021/083.

\section*{\large Appendices}

\setcounter{section}{0}
\renewcommand{\thesection}{\Alph{section}}
\global\long\def\theequation{\thesection.\arabic{equation}}
\setcounter{equation}{0}
\global\long\def\thesubsection{\thesection.\arabic{subsection}}

\section{The functions ${\cal F}_Q(x,\bar q)$ in the standard gauges}
\label{functionF}

We give here the expressions for the functions ${\calF}(x,\bar q)$,
eigenfunctions of the operator $\dcmu\,\dcmud$ [see Eq.~(\ref{ecautovbox})],
in the standard gauges SG, LG1 and LG2. As in the main text, we choose the
3-axis in the direction of the magnetic field, and use the notation $B_Q =
|Q\, B|$ and $s = \mbox{sign}(Q B)$.

It is worth pointing out that the functions ${\calF}(x,\bar q)$ can be
determined up to a global phase, which in general can depend on $k$. In the
following expressions for SG, LG1 and LG2 the corresponding phases have been
fixed by requiring ${\calF}(x,\bar q)$ to satisfy Eqs.~(\ref{hPiInt}) to
(\ref{hPiGauge}), with $f_{kk'}(v_\perp)$ given by Eq.~(\ref{fkkp}).

\subsection{Symmetric gauge}

In the SG we take $\chi=\imath$, where $\imath$ is a nonnegative integer.
Thus, the set of quantum numbers used to characterize a given particle state
is $\bar q =(q^0, \ellef, \imath, q^3)$. In addition, we introduce polar
coordinates $r,\phi$ to denote the vector $\vec x_\perp = (x^1,x^2)$ that
lies in the plane perpendicular to the magnetic field. The functions ${\cal
F}_Q(x,\bar q)$ in this gauge are given by
\begin{equation}
\calF(x,\bar q)^{\rm (SG)} \ = \
\sqrt{2\pi} \ e^{-i(q^0\, x^0 - q^3 x^3)}\, e^{-i s (\elle - \imath) \phi} \, R_{\ellef,\imath}(r)\ ,
\label{efes}
\end{equation}
where
\begin{equation}
R_{\ellef,\imath}(r) \ = \ N_{\ellef,\imath} \, \xi^{(\ellef - \imath)/2} \, e^{-\xi/2}
\, L_\imath^{\ellef-\imath}(\xi)\ ,
\end{equation}
with $\xi = B_Q \, r^2/2\,$. Here we have used the definition $N_{\ellef,
\imath} = (B_Q \ \imath! / \ellef!)^{1/2}$, while $L_j^m(x)$ are
generalized Laguerre polynomials.

\subsection{Landau gauges LG1 and LG2}

For the gauges LG1 and LG2 we take $\chi=q^j$ with $j=1$ and $j=2$,
respectively. Thus, we have $\bar q =(q^0, \ellef, q^j, q^3)$. The
corresponding functions $\calF(x,\bar q)$ are given by
\begin{eqnarray}
\calF(x,\bar q)^{\rm (LG1)} & = & (-is)^k\, N_\elle \, e^{-i( q^{0}\,x^{0} - q^{1}x^{1}
- q^{3}x^{3})}\,D_{\ellef}(\rho_s^{(1)})\ , \\
\calF(x,\bar q)^{\rm (LG2)} & = &  N_\elle \, e^{-i( q^{0}\,x^{0} - q^{2}x^{2}
- q^{3}x^{3})}\,D_{\ellef}(\rho_s^{(2)})\ ,
\end{eqnarray}
where $\rho_s^{(1)}= \sqrt{2B_{Q}}\, (x^{2} + s\, q^{1}/B_Q)$,
$\rho_s^{(2)}= \sqrt{2B_{Q}}\, (x^{1} - s\, q^{2}/B_Q)$ and $N_\ellef =
\left(4\pi B_{Q}\right)^{1/4}/\sqrt{\ellef!}$. The cylindrical parabolic
functions $D_\ellef(x)$ in the above equations are defined as
\begin{equation}
D_{\ellef}\left(x\right)=2^{-\elle/2}\,e^{-x^{2}/4}\,
H_{\ellef}\big(x/\sqrt 2\big)\ ,
\end{equation}
where $H_{\ellef}(x)$ are Hermite polynomials, with the standard convention
$H_{-1}(x) = 0$.

\section{Wavefunction properties and anticommutation relations for spin 1/2 charged particles
in a uniform magnetic field}
\label{Spin-1/2-SpinorsPropa}
\setcounter{equation}{0}

As stated in Eqs.~(\ref{UVlept}), the fermion wavefunctions can be written as
\begin{align}
U_{f}\left(x,\bar q,a\right)  \ &= \
\mathbb{E}^{\scriptscriptstyle{{\cal Q}_f}}(x,\bar q) \ u_{\scriptscriptstyle{{\cal Q}_f}}(\elle,q^3,a)\ ,
\nonumber\\
V_{f}\left(x,\bar q,a\right) \ &= \ \mathbb{\tilde{E}}^{\scriptscriptstyle{-{\cal Q}_f}}(x,\bar q) \
v_{-\scriptscriptstyle{{\cal Q}_f}}(\elle,q^3,a)\ ,
\label{spinors}
\end{align}
where the functions $\mathbb{E}^{\scriptscriptstyle{{\cal
Q}_f}}(x,\bar q)$ and $\mathbb{\tilde{E}}^{\scriptscriptstyle{-{\cal
Q}_f}}(x,\bar q)$ are defined by Eqs.~(\ref{ep}). It is easy to see that the
matrices $\Gamma^\pm$ appearing in these definitions satisfy
\begin{equation}
\Gamma^\lambda\, \Gamma^{\lambda} = \Gamma^\lambda\ ,
\qquad
\Gamma^\lambda\, \Gamma^{-\lambda} = 0\ ,
\qquad
\Gamma^\lambda\, \gamma_{\parallel}^\mu = \gamma_{\parallel}^\mu\, \Gamma^\lambda\ ,
\qquad
\Gamma^\lambda\, \gamma_{\perp}^\mu = \gamma_{\perp}^\mu \, \Gamma^{-\lambda}\ .
\label{progam}
\end{equation}
It can be shown that the functions $\mathbb{E}_{\bar p}(x)$ satisfy
orthogonality and completeness relations, namely
\begin{equation}
\int d^4x \; \bar{\mathbb{E}}^{\scriptscriptstyle{{\cal Q}_f}}(x,\bar q)\,
\mathbb{E}^{\scriptscriptstyle{{\cal Q}_f}}(x,\bar q') =
\hat \delta_{\bar q\bar q\,'}\, \big[\,
\mathcal{I} + \delta_{k0}\, ( \Gamma^s-\mathcal{I})\big]
\end{equation}
and
\begin{equation}
\sumint_{\bar q} \;
\mathbb{E}^{\scriptscriptstyle{{\cal Q}_f}}(x,\bar q)
\,\bar{\mathbb{E}}^{\scriptscriptstyle{{\cal Q}_f}}(x',\bar q)
= \delta^{(4)}(x-x')\,\mathcal{I}\ ,
\end{equation}
where $\mathcal{I}$ stands for the identity in Dirac space, and, as in the
main text, we have used the definitions ${\cal Q}_f=Q_f/e$, $s={\rm
sign}(Q_fB)$ and $\bar{\mathbb{E}}^{\scriptscriptstyle{{\cal Q}_f}}(x,\bar
q)=\gamma^0\, \mathbb{E}^{\scriptscriptstyle{{\cal Q}_f}}(x,\bar
q)^\dagger\,\gamma^0$. In addition, they satisfy the useful relation
\begin{equation}
i \, \rlap/\!{\cal D}\, \mathbb{E}^{\scriptscriptstyle{{\cal Q}_f}}(x,\bar q)
= \mathbb{E}^{\scriptscriptstyle{{\cal Q}_f}}(x,\bar q)\; \rlap/\! \hat{\Pi}_s(q^0,k,q^3) \ ,
\label{impid}
\end{equation}
where $\hat{\Pi}_s^\mu(q^0,k,q^3) = (q^0,\,0,\,-s\sqrt{2k|Q_fB|},\,q^3)$.

On the other hand, the spinors $u_{\scriptscriptstyle{{\cal
Q}_f}}(\elle,q^3,a)$ and $v_{-\scriptscriptstyle{{\cal Q}_f}}(\elle,q^3,a)$,
$a=1,2$, in Eqs.~(\ref{spinors}) are given by
\begin{align}
u_{\scriptscriptstyle{{\cal Q}_f}}(\elle,q^3,a)\ &=\
\frac{1}{\sqrt{2(\Eq + \mq)}}\,
\Big[\;\rlap/\! \hat{\Pi}_s(\Eq,k,q^3) + \mq\,{\cal I}\,\Big]\,
\left( \begin{array}{c}
  \phi^{(a)} \\
  \phi^{(a)}\\
\end{array}
\right)\ ,
\label{uspindos}
\\
v_{-\scriptscriptstyle{{\cal Q}_f}}(\elle,q^3,a)\ &=\
\frac{1}{\sqrt{2(\Eq + \mq)}}\,
\Big[\!-\rlap/\! \hat{\Pi}_{-s}(\Eq,k,q^3) + \mq\,{\cal I}\,\Big]\,
\left( \begin{array}{c}
\tilde \phi^{(a)} \\
- \tilde \phi^{(a)}\\
\end{array}
\right)\ ,
\label{vspindos}
\end{align}
where $\phi^{(1)}{}^\dagger = -\tilde \phi^{(2)}{}^\dagger = (1,0)$ and
$\phi^{(2)}{}^\dagger = \tilde \phi^{(1)}{}^\dagger = (0,1)$. We use here
the Weyl representation for Dirac matrices, namely
\begin{equation}
 \gamma_0 =
\begin{pmatrix} 0 & \mathcal{I}  \\ \mathcal{I} & 0 \end{pmatrix} \ ,
\qquad\qquad
\vec \gamma=\begin{pmatrix} 0 & \vec \sigma
\\ - \vec \sigma & 0 \end{pmatrix}\ ,
\end{equation}
where $\sigma_i$, with $i=1,2,3$, are the Pauli matrices. It can be
shown that the spinors satisfy the relations
\begin{eqnarray}
\sum\limits_{a=1,2} u_{\scriptscriptstyle{{\cal Q}_f}}(\elle,q^3,a)
\ \bar u_{\scriptscriptstyle{{\cal Q}_f}}(\elle,q^3,a) &=& \;\rlap/\! \hat{\Pi}_s(\Eq,k,q^3) + \mq\,{\cal I}\ ,
\nonumber\\
\sum\limits_{a=1,2} v_{-\scriptscriptstyle{{\cal Q}_f}}(\elle,q^3,a)\
\bar v_{-\scriptscriptstyle{{\cal Q}_f}}(\elle,q^3,a) &=& \;\rlap/\! \hat{\Pi}_{-s}(\Eq,k,q^3)
- \mq\,{\cal I}\ .
\label{sumspinl}
\end{eqnarray}

We finally quote the anticommutation relations between creation and
annihilation operators in Eqs.~(\ref{fermionfieldBpart}). They read
\begin{eqnarray}
\left\{b_f(\breve{q},a), b_f(\breve{q}\,',a')\right\} & = &
\left\{d_f(\breve{q},a), d_f(\breve{q}\,',a')\right\} = \, 0\ , \nonumber \\
\left\{b_f(\breve{q},a), d_f(\breve{q}\,',a')\right\} & = &
\left\{b_f(\breve{q},a), d_f(\breve{q}\,',a')^\dagger\right\} = \, 0\ , \nonumber \\
\left\{b_f(\breve{q}, a), b_f(\breve{q}\,',a')^\dagger\right\} & = &
\left\{d_f(\breve{q}, a), d_f(\breve{q}\,',a')^\dagger\right\}  =  2 E_f\,  \delta_{a a'}\,(2\pi)^3  \, \delta_{\elle\elle'}\,
\delta_{\chi\chi'}\, \delta(q^3-q^{\,\prime\, 3})\ . \qquad \
\label{conferB}
\end{eqnarray}

\section{Wavefunction properties and commutation relations for massive spin 1
charged particles in a uniform magnetic field}
\label{Spin-1-Pola_Propa}\setcounter{equation}{0}

According to Eq.~(\ref{funrho}), the wavefunctions $\wqchmu\left(x,\bar
q,c\right)$ are given by
\begin{equation}
\wqchmu(x,\bar q,c) = \mathbb{R}^{{\scriptscriptstyle{{\cal Q}}},\mu\nu}
(x,{\bar q}) \, \epsilon_{{\scriptscriptstyle{\cal{Q}}},\nu}(\elle,q^3,c)\ ,
\end{equation}
where $\mathbb{R}^{{\scriptscriptstyle{{\cal Q}}},\mu\nu}(x,{\bar q})$ is
given by Eqs.~(\ref{rhodef}) and (\ref{upsidef}), while
$\epsilon_{{\scriptscriptstyle{\cal{Q}}},\nu}(\elle,q^3,c)$ are the charged
rho meson polarization vectors. As in the main text, we define ${\cal
Q}={\rm sign}(Q_\rho)$.

The tensors $\mathbb{R}^{{\scriptscriptstyle{{\cal Q}}},\mu\nu}$ involve the
functions ${\cal F}_Q(x,\bar q)$ and the tensors
$\Upsilon_\lambda^{\mu\nu}$, defined by Eq.~(\ref{upsidef}). The latter obey
some useful relations, namely
\begin{equation}
\Upsilon_\lambda^{\mu\nu} = (\Upsilon_\lambda^{\nu\mu})^\ast =
\Upsilon_{-\lambda}^{\nu\mu}\ , \qquad
\Upsilon_\lambda^{\mu\alpha} \ \Upsilon_{\lambda',\alpha\nu} = \delta_{\lambda\lambda'} {\Upsilon_{\lambda}}^\mu_{\
\nu}\ , \qquad \sum_{\lambda=-1,0,1} \Upsilon_\lambda^{\mu\nu} = g^{\mu\nu}\ .
\label{defp}
\end{equation}
It is also useful to introduce the projector $({\cal P}_{k,s})^{\mu\nu}$,
defined by
\begin{eqnarray}
({\cal P}_{k,s})^{\mu\nu} &=& g^{\mu \nu}\ - \delta_{k,-1} \, \Upsilon_{0}^{\mu \nu}
- (\delta_{k,-1}+\delta_{k,0})  \, \Upsilon_{s}^{\mu \nu}\nonumber
\\
&=&  \Upsilon_{-s}^{\mu \nu} + (1- \delta_{k,-1})\, \Upsilon_{0}^{\mu \nu}
+ (1-\delta_{k,-1}-\delta_{k,0})\, \Upsilon_{s}^{\mu \nu}\ ,
\end{eqnarray}
which satisfies $({\cal P}_{k,s})^{\mu\alpha}\ ({\cal P}_{k,s})_{\alpha\nu} = ({\cal P}_{k,s})^\mu_{\
\nu}$. Here $s={\rm sign}(Q_\rho B)=\pm 1$.

The functions $\mathbb{R}^{{\scriptscriptstyle{{\cal Q}}},\mu\nu}$ are shown to satisfy
orthogonality and completeness relations, viz.
\begin{equation}
\int d^{4}x\,\mathbb{R}^{{\scriptscriptstyle{{\cal Q}}},\mu\alpha}(x,{\bar q})
\,{\mathbb{R}^{{\scriptscriptstyle{{\cal Q}}}}_{\nu\alpha}(x,{\bar q}')}^\ast
= \hat \delta_{\bar{q}\bar{q}'}\,({\cal P}_{k,s})^\mu_{\ \nu}
\label{orthormunu}
\end{equation}
and
\begin{equation}
\sumint_{\bar q}\; \mathbb{R}^{{\scriptscriptstyle{{\cal Q}}},\mu\alpha}(x,\bar q) \,
{\mathbb{R}^{\scriptscriptstyle{{\cal Q}}}_{\nu\alpha}(x',\bar q)}^\ast
= \delta^{(4)}(x-x') \, \delta^\mu_{\ \nu} \ .
\label{compR}
\end{equation}
It can also be seen that
\begin{equation}
({\cal P}_{k,s})^\mu_{\ \alpha}\, \mathbb{R}^{{\scrcalQ},\alpha\nu}(x,\bar q) =
\mathbb{R}^{{\scrcalQ},\mu\nu}(x,\bar q)\ .
\end{equation}

For $k\geq 0$, one can find some other useful relations that involve the
four-vector $\Pi^\mu(k,q_\parallel)$ defined in Eq.~(\ref{pimu}). These are
\begin{eqnarray}
 \dcmud\, \mathbb{R}^{{\scriptscriptstyle{{\cal Q}}},\mu\nu} (x,{\bar q}) & = & -i \, \calF(x,\bar q) \
 {\Pi^\nu(k,q_\parallel)}^\ast \ ,
\label{dmuR} \\
\dcmud\, \calF(x,\bar q) & = & -i\,\mathbb{R}^{\scriptscriptstyle{{\cal Q}}}_{\mu\nu}(x,{\bar q})
\,\Pi^\nu(k,q_\parallel)\ , \\
({\cal P}_{k,s})^\mu_{\ \alpha}\Pi^\alpha(k,q_\parallel)
& = & \Pi^\mu(k,q_\parallel)\ .
\end{eqnarray}

Let us consider now the polarization vectors $\epsilon_\nu(\elle,q^3,c)$.
Their form is dictated by the transversality condition $\dcmu \rhocmud(x)=0$
in Eq.~(\ref{orthvec}), which implies that for $q^0 = E_\rho$ one must have
\begin{eqnarray}
\dcmud \wqchmu\left(x,\bar q,c\right) = \dcmud\,
\mathbb{R}^{{\scriptscriptstyle{{\cal Q}}},\mu\nu} (x,{\bar q}) \,
\epsilon_{{\scriptscriptstyle{\cal{Q}}},\nu}(\elle,q^3,c) = 0 \ .
\label{transv}
\end{eqnarray}
Taking into account Eq.~(\ref{dmuR}), it is seen that the transversality is
trivially satisfied for $\elle=-1$, since in that case $\calF(x,\bar q)$ is
zero. For $k\geq 0$, according to Eq.~(\ref{dmuR}) the condition
(\ref{transv}) can be expressed as
\begin{equation}
{\Pi^\mu(\elle,q_\parallel)}^\ast\big|_{q^0 = E_\rho}\,
\epsilon_{{\scriptscriptstyle{\cal{Q}}},\mu}(\elle,q^3,c) = 0\ .
\label{transvcond}
\end{equation}

For $\elle \geq 1$ there are three linearly independent vectors that
satisfy Eq.~(\ref{transvcond}). A convenient choice is
\begin{eqnarray}
\epsilon_{\scrcalQ}^{\mu}(\elle, q^3,1) &=& \frac{1}{\sqrt2}\,
\frac{1}{m_\perp \, m_{2\perp} }\, \bigg[ \Pi_+ \Big(
  \Erho ,\, 0 , \,  0, \,  \, q^3 \Big) + m_\perp^2\, \Big(
  0, \, 1 , \,  is, \, 0 \Big) \bigg]\ ,
\nonumber \\[2.mm]
\epsilon_{\scrcalQ}^\mu(\elle, q^3,2) &=&
\frac{1}{m_\perp} \Big(  q^3 ,\,  0 ,\,  0,\,  \Erho \Big)\ ,
\nonumber  \\[2.mm]
\epsilon_{\scrcalQ}^{\mu}(\elle, q^3,3) &=&
\frac{1}{\sqrt2}\, \frac{1}{m_\rho \, m_{2\perp} } \bigg[
\Pi_- \Big( \Erho, \, \frac{\Pi^\ast_+}{2} ,
\, i s \frac{\Pi^\ast_+}{2},\, q^3 \Big) + m_{2\perp}^2\,
\Big(0, \, 1, \, -i s ,\, 0\Big)\bigg]\ ,\qquad
\end{eqnarray}
where we have used the definitions
\begin{eqnarray}
m_\perp & = & \sqrt{m_\rho^2 + ( 2\elle +1)\Brho}\ , \nonumber \\
m_{2\perp} & = & \sqrt{m_\rho^2 + \elle \Brho}\ , \nonumber \\
\Pi_+ & = & - \Pi^1(\elle,q_\parallel) + i s\, \Pi^2(\elle,q_\parallel) =
-i\sqrt{2(\elle+1)\Brho}\ , \nonumber \\
\Pi_- & = & - \Pi^1(\elle,q_\parallel) - i s\,
\Pi^2(\elle,q_\parallel) = i\sqrt{2\elle\Brho} \ ,
\end{eqnarray}
with $B_\rho = |Q_\rho B|$. Using these polarization vectors one recovers
the known expressions for a vector boson in a constant magnetic field, see e.g.\
Ref.~\cite{Nikishov:2001fd}.

For $\elle=0$ two independent nontrivial transverse polarization
vectors can be constructed. A suitable choice is
\begin{eqnarray}
\epsilon_{\scrcalQ}^{\mu}(0, q^3,1) &=&
\frac{1}{\sqrt2} \frac{1}{m_\perp \, m_{2\perp}  } \Big(
  \Erho \, \Pi_+ ,\,
  m_\perp^2 , \,
  is \, m_\perp^2  \, , \,  q^3 \, \Pi_+ \Big)\ ,
\nonumber \\[2.mm]
\epsilon_{\scrcalQ}^\mu(0, q^3,2) &=&
\frac{1}{m_\perp} \Big(
  q^3 ,\,
  0 ,\,
  0,\,
  \Erho
\Big)\ ,
\end{eqnarray}
where $m_\perp$, $m_{2\perp}$, $\Pi_+$ and $E_\rho$ are understood to be
evaluated at $\elle=0$. It can be seen that $\epsilon^{\mu}_{\scrcalQ}(0, q^3 ,2)$
satisfies
\begin{equation}
S_3^{\mu\nu} \epsilon_{{\scrcalQ},\nu}(0, q^3 ,2)  = 0\ ,
\end{equation}
while $\epsilon^{\mu}(0, q^3 ,1)$ is not an eigenvector of $S_3$.

For $\elle=-1$, one has $\mathbb{R}^{{\scriptscriptstyle{{\cal
Q}}},\mu\nu}(x,{\bar q})\propto \Upsilon_{-s}^{\mu\nu}$. This leaves only
one nontrivial polarization vector, which can be conveniently written as
\begin{equation}
\epsilon^{\mu}_{\scrcalQ}(-1, q^3 ,1) = \frac{1}{\sqrt 2}\Big(
 0 ,\,
1 ,\,
 is ,\,
0 \Big)\ .
\label{polm1}
\end{equation}
As in the case of $\epsilon_{{\scrcalQ},\nu}(0, q^3 ,2)$, it is easy to see
that this vector has a definite spin projection in the direction of the
magnetic field. Indeed, one has
\begin{equation}
S_3^{\mu\nu} \epsilon_{{\scrcalQ},\nu}(-1, q^3 ,1)  = s\,
\epsilon^{\mu}_{\scrcalQ}(-1, q^3 ,1)\ .
\end{equation}

Finally, for $\elle \ge 0$ one can also define an additional, ``longitudinal'',
polarization vector. We keep for this vector the notation
$\epsilon^{\mu}_{\scrcalQ}(\elle, q^3 ,c)$, taking for the polarization index
the value $c=0$. It is given by
\begin{equation}
\epsilon^{\mu}_{\scrcalQ}(\elle, q^3 ,0) = \frac{1}{m_\rho}\,
\Pi^\mu(\elle,q_\parallel)\big|_{q^0 = E_\rho} \ ,
\label{long}
\end{equation}
where, as stated, $k\geq 0$. For $\elle = -1$ no longitudinal vector is
introduced.

It is worth noticing that the full set of four polarization vectors satisfy
orthogonality and completeness relations, namely
\begin{eqnarray}
{\epsilon^{\mu}_{\scrcalQ}(\elle, q^3 ,c)}^\ast \,
\epsilon_{{\scrcalQ},\mu}(\elle, q^3 ,c') = -\, \zeta_{c}\,
\delta_{cc'}
\label{orthpol}
\end{eqnarray}
and
\begin{eqnarray}
\sum_{c=c_{\rm min}}^{c_{\rm max}} \zeta_{c}\ \epsilon^{\mu}_{\scrcalQ}(\elle, q^3 ,c)
\, {\epsilon^\nu_{\scrcalQ}(\elle, q^3 ,c)}^\ast =
-\, ({\cal P}_{k,s})^{\mu\nu}\ ,
\label{completeness}
\end{eqnarray}
where $\zeta_0=-1$ and $\zeta_{1}=\zeta_{2}=\zeta_{3}=1$, while
$c_{\rm min}$ and $c_{\rm max}$ are given by
\begin{equation}
c_{\rm min}=\left\{ \begin{array}{ccc}
1 & \quad\text{if}\quad & k=-1\\
0 & \quad\text{if}\quad & k\ge0
\end{array}\right.\ ,
\qquad\qquad c_{\rm max}=\left\{ \begin{array}{ccc}
1 & \quad\text{if}\quad & k=-1\\
2 & \quad\text{if}\quad & k=0\\
3 & \quad\text{if}\quad & k\ge1
\end{array}\right. \ .
\label{ces1}
\end{equation}
For $k \ge 1$, from Eqs.~(\ref{completeness}), (\ref{long}) and (\ref{defp})
it is seen that the sum over the physical polarizations $c=1,2,3$ satisfies
\begin{eqnarray}
\sum_{c=1}^3 \epsilon^{\mu}_{\scrcalQ}(\elle, q^3 ,c) \
{\epsilon^\nu_{\scrcalQ}(\elle, q^3 ,c)}^\ast = - \left[\, g^{\mu\nu}
- \frac{\Pi^\mu(\elle,q_\parallel)\,
{\Pi^\nu(\elle,q_\parallel)}^\ast}{m_\rho^2}\,
\right]\ ,
\label{orthpolphys}
\end{eqnarray}
where the vectors $\Pi^\mu(\elle,q_\parallel)$ are assumed to be ``on
shell'', i.e., one has to take $q^0 = E_\rho$.

As stated in the main text, one can also extend the set of charged rho meson
wavefunctions $\wqchmu(x,\bar q,c)$ including a ``longitudinal''
wavefunction $\wqchmu(x,\bar q,0)\equiv
\mathbb{R}^{{\scriptscriptstyle{{\cal Q}}},\mu\nu}(x,{\bar
q})\,\epsilon_{{\scriptscriptstyle{\cal{Q}}},\nu}(\elle,q^3,0)$. In this way
one gets for these functions the orthogonality and completeness relations in
Eqs.~(\ref{orthfrho}) and (\ref{compfrho}).

We conclude this appendix by quoting the commutation relations for the
creation and annihilation operators in Eq.~(\ref{rhofieldBpart}). One has
\begin{align}
\left[\, \arhoch(\breve q,c), \, \arhopmch (\breve q\,'\!,c')\right] &
\ = \ \left[\, \arhoch(\breve q,c)^\dagger\, , \, \arhopmch(\breve q\,'\!,c')^\dagger\right]
\ = \ \left[\, \arhoch(\breve q,c)\, , \, \arhomch (\breve q\,'\!,c')^\dagger\right]  \ = \ 0\ ,
\nonumber \\
\left[\arhoch(\breve q , c) , \arhoch(\breve q\,'\! ,
c')^\dagger\right] & \ = \ \left[\arhomch(\breve q , c) ,
\arhomch(\breve q\,'\! , c')^\dagger\right] \ = \ 2 \Erho
\,(2\pi)^3  \, \delta_{cc'} \, \delta_{\elle\elle'}\,
\delta_{\chi\chi'}\, \delta(q^3-q^{\,\prime\, 3})\ .
\end{align}

\section{Explicit form of the coefficients of the operators for the one-loop $\rho^+$
meson polarization function}
\label{dis} \setcounter{equation}{0}

As stated in Sec.~\ref{onelooprho}, the one-loop correction to the
$\rho^{+}$ propagator in $\bar{q}$-space can be written in terms of a set of
tensors $\mathbb{O}_{i}^{\alpha\alpha'}(\Pi)$, with $i=1,\dots7$. We give
here the explicit expressions for the corresponding coefficients
$d_{i}(k,q_{\parallel})$, introduced in Eq.~(\ref{boldj}). The latter have
been obtained taking into account the Schwinger form of quark propagators in
Eq.~(\ref{sfp_schw_a}). In general they can be written in the form
\begin{equation}
d_{i}(k,q_{\parallel})= -i\,\frac{N_{c}}{4\pi^{2}}\int_{-1}^{1}dx\int_{0}^{\infty}dz\;
\frac{e^{-z\phi(x,q_{\parallel}^{2})}}{\alpha_{+}}\;
\left(\frac{\alpha_{-}}{\alpha_{+}}\right)^{k}f_{k,q_{\parallel}}^{(i)}(x,z)\ ,
\end{equation}
where $\phi(x,q_{\parallel}^{2})$ and $\alpha^{\pm}$ are defined
in Eqs.~(\ref{phi}) and (\ref{alpha}), respectively. After some calculation, the functions
$f_{k,q_{\parallel}}^{(i)}(x,z)$ are found to be given by
\begin{eqnarray}
f_{k,q_{\parallel}}^{(1)}(x,z) & = &
-(1-t_{u}t_{d})\bigg[m_{u}m_{d}+(1-x^{2})\frac{q_{\parallel}^{2}}{4}\bigg]
-\frac{\alpha_{-}+\elle(\alpha_{-}-\alpha_{+})}{\alpha_{+}\alpha_{-}}\,(1-t_{u}^{2})(1-t_{d}^{2})\ ,\quad k\ge0
\nonumber \\[2mm]
f_{k,q_{\parallel}}^{(2)}(x,z) & = &
f_{k,q_{\parallel}}^{(2a)}(x,z)+f_{k,q_{\parallel}}^{(2b)}(x,z)+\left(2k+1\right)\,f_{k,q_{\parallel}}^{(2c)}(x,z)
\nonumber \\[2mm]
f_{k,q_{\parallel}}^{(3)}(x,z) & = &
\frac{1}{2}\,(1-x^{2})(1-t_{u}t_{d})\ ,\quad k\ge0
\nonumber \\[2mm]
f_{k,q_{\parallel}}^{(4)}(x,z) & = & \frac{1}{B_{\rho}}\,\frac{\alpha_{+}-\alpha_{-}}{\alpha_{+}\alpha_{-}}\,
(1-t_{u}^{2})(1-t_{d}^{2})\ ,\quad k\ge1
\nonumber \\[2mm]
f_{k,q_{\parallel}}^{(5)}(x,z) & = &
f_{k,q_{\parallel}}^{(5a)}(x,z)+f_{k,q_{\parallel}}^{(5b)}(x,z)\ ,
\nonumber \\[2mm]
f_{k,q_{\parallel}}^{(6)}(x,z) & = &
f_{k,q_{\parallel}}^{(2a)}(x,z)-f_{k,q_{\parallel}}^{(2b)}(x,z)+f_{k,q_{\parallel}}^{(2c)}(x,z)\ ,
\nonumber \\[2mm]
f_{k,q_{\parallel}}^{(7)}(x,z) & = &
-f_{k,q_{\parallel}}^{(5a)}(x,z)+f_{k,q_{\parallel}}^{(5b)}(x,z)\ ,
\label{efes}
\end{eqnarray}
where
\begin{eqnarray}
f_{k,q_{\parallel}}^{(2a)}(x,z) & = &
-\frac{1}{2}\,\frac{\alpha_{-}}{\alpha_{+}}\,(1+t_{u})\,(1+t_{d})\,\bigg[m_{u}m_{d}+\frac{1}{z}+(1-x^{2})\frac{q_{\parallel}^{2}}{4}\bigg]\;,\qquad
k\ge-1 \ ,
\nonumber \\[2mm]
f_{k,q_{\parallel}}^{(2b)}(x,z) & = & -\frac{1}{2}\,\frac{\alpha_{+}}{\alpha_{-}}\,(1-t_{u})\,(1-t_{d})\,\bigg[m_{u}m_{d}+\frac{1}{z}+(1-x^{2})\frac{q_{\parallel}^{2}}{4}\bigg]\;,\qquad k\ge1\ ,
\nonumber \\[2mm]
f_{k,q_{\parallel}}^{(2c)}(x,z) & = & \frac{\alpha_{+}-\alpha_{-}}{2\alpha_{+}\alpha_{-}}\,(1-t_{u}^{2})(1-t_{d}^{2})\ ,\qquad k\ge1 \ ,
\nonumber\\[2mm]
f_{k,q_{\parallel}}^{(5a)}(x,z) & = & \frac{1}{4\alpha_{+}}\,\bigg[(1+x)\frac{t_{u}\,(1+t_{u})\,(1-t_{d}^{2})}{B_{u}}
+(1-x)\frac{t_{d}\,(1+t_{d})\,(1-t_{u}^{2})}{B_{d}}\bigg]\;,\qquad k\ge0 \ ,
\nonumber \\[2mm]
f_{k,q_{\parallel}}^{(5b)}(x,z) & = & \frac{1}{4\alpha_{-}}\,\bigg[(1+x)\frac{t_{u}\,(1-t_{u})\,(1-t_{d}^{2})}{B_{u}}
+(1-x)\frac{t_{d}\,(1-t_{d})\,(1-t_{u}^{2})}{B_{d}}\bigg]\;,\qquad k\ge1 \ . \nonumber
\end{eqnarray}
Here, as in the main text, we have used the definitions
$t_{u}=\tanh[(1-x)zB_{u}/2]$, $t_{d}=\tanh[(1+x)zB_{d}/2]$, with
$B_{f}=|Q_{f}B|$ for $f=u,d$. For $k=0$ and $k=-1$ some of the above
functions vanish; therefore, for each expression we have explicitly
indicated the range of values of $k$ to be taken into account.

\end{document}